\documentclass[manuscript]{aastex}

\begin{document}


\title{Galaxy Morphology\altaffilmark{1}}


\author{Ronald J. Buta}
\affil{Department of Physics and Astronomy, University of Alabama, Box 870324, Tuscaloosa, AL 35487}

\altaffiltext{1}{To be published in Planets,
Stars, and Stellar Systems, Vol. 6, Series Editor T. D. Oswalt, Volume
editor W. C. Keel, Springer Reference, 2011}

%\email{aastex-help@aas.org}


\begin{abstract}
Galaxy morphology is a product of how galaxies formed, how they
interacted with their environment, how they were influenced by internal
perturbations, AGN, and dark matter, and of their varied star formation
histories. This article reviews the phenomenology of galaxy morphology
and classification with a view to delineating as many types as possible
and how they relate to physical interpretations.  The old
classification systems are refined, and new types introduced, as the
explosion in available morphological data has modified our views on the
structure and evolution of galaxies.  
\end{abstract}



\noindent
{\it Index terms}: Hubble-Sandage-de Vaucouleurs classification; spiral
galaxies; elliptical galaxies; S0 galaxies; irregular galaxies;
resonance ring galaxies; outer rings; inner rings; nuclear rings;
secondary bars; nuclear bars; pseudorings; grand design spirals;
flocculent spirals; merger morphologies; tidal tails;
collisional ring galaxies; polar ring galaxies; barred galaxies; lens
galaxies; spiral arm classes; counter-winding spiral galaxies;
anemic spiral galaxies; spiral arm multiplicity; dust lanes;
pseudobulges; classical bulges; dust-lane elliptical galaxies;
shell/ripple galaxies; ultra-luminous infrared galaxies; bar ansae;
brightest cluster members; low surface brightness galaxies; dwarf 
galaxies; dwarf spiral galaxies; dwarf spheroidal galaxies; Galaxy
Zoo project; isolated galaxies; blue compact dwarf galaxies;
star-forming galaxy morphologies; warped disks; stellar mass
morphology; luminosity classes; boxy ellipticals; disky ellipticals;
late-type galaxies; early-type galaxies; Magellanic barred galaxies;
color galaxy morphology; atomic and molecular gas galaxy morphology


\tableofcontents

\clearpage

\section{Introduction}

In the nearly 100 years since galaxy morphology became a topic of
research, much has been learned about galactic structure and dynamics.
Known only as ``nebulae" at that time, galaxies were found to have a
wide range of largely inexplicable forms whose relations to one another
were a mystery. As data accumulated, it became clear that galaxies are
fundamental units of matter in space, and an understanding of how they
formed and evolved became one of the major goals of extragalactic
studies. Even in the era of space observations, galaxy morphology
continues to be the backbone of extragalactic research as modern
instruments provide information on galactic structure across a
wide range of distances and lookback times.

In spite of the advances in instrumentation and the explosion of
data, classical galaxy morphology (i. e., the visual morphological
classification in the style of Hubble and others) has not lost its relevance.
The reasons for this are as follows:

\noindent
1. Morphology is still a logical starting point for understanding
galaxies. Sorting galaxies into their morphological categories is similar
to sorting stars into spectral types, and can lead to important 
astrophysical insights.
Any theory of galaxy formation and evolution will have to, at
some point, account for the bewildering array of galactic forms.

\noindent
2. Galaxy morphology is strongly correlated with galactic star
formation history. Galaxies where star formation ceased gigayears ago
tend to look very different from those where star formation continues
at the present time. Classical morphology recognizes these differences
in an ordered way.

\noindent
3. Information on galaxy morphology, in the form of new types of
galaxies, multi-wavelength views of previously known galaxy types,
and higher resolution views of all or part of some galaxies,
has exploded as modern instrumentation has superceded the old
photographic plates that were once used exclusively for galaxy
classification. 

\noindent
4. Galaxy classification has gone beyond the
realm of a few thousand galaxies to that of a {\it million} galaxies
through the Galaxy Zoo project. Not only this, but GalaxyZoo has taken
morphology from the exclusive practice of a few experts to the public 
at large, thus
facilitating citizen science at its best. Galaxy Zoo images are also in
{\it color}, thus allowing the recognition of special galaxy types
and features based on stellar populations or gaseous emission.

\noindent
5. Finally, deep surveys with the Hubble Space Telescope
have extended morphological studies well
beyond the realm of the nearby galaxies that dominated early
catalogues, allowing detailed morphology to be distinguished at
unprecedented redshifts. 

Now, more than ever, galaxy morphology is a vibrant subject that
continues to provide surprises as more galaxies are studied for their
morphological characteristics across the electromagnetic spectrum. It
is clear that a variety of effects are behind observed morphologies,
including environmental density and merger/interaction history,
internal perturbations, gas accretions, nuclear activity, secular
evolution, as well as the diversity in star formation histories, and
that a global perspective based on large numbers of galaxies will
improve theoretical models and give a more reliable picture of galactic
evolution.

The goal of this article is to present the phenomenology of galaxy
morphology in an organized way, and highlight recent advances in
understanding what factors influence morphology and how various
galaxy types are interpreted. The article is a natural follow-up
to the excellent review of galaxy morphology and classification
by Sandage (1975)
in Volume IX of the classic {\it Stars and Stellar Systems} series.
It also complements the recently published {\it de Vaucouleurs
Atlas of Galaxies} (Buta, Corwin, \& Odewahn 2007, hereafter the
dVA), which provided a
detailed review of the state and technique of galaxy classification up
to about the year 2005. Illustrations are very important in a review
of this nature, and the article draws on a large number of sources
of images. For this purpose, the Sloan Digital Sky Survey (SDSS),
the NASA/IPAC Extragalactic Database (NED), and the dVA have been 
most useful. 

\section{Overview}

\begin{figure}
\plotone{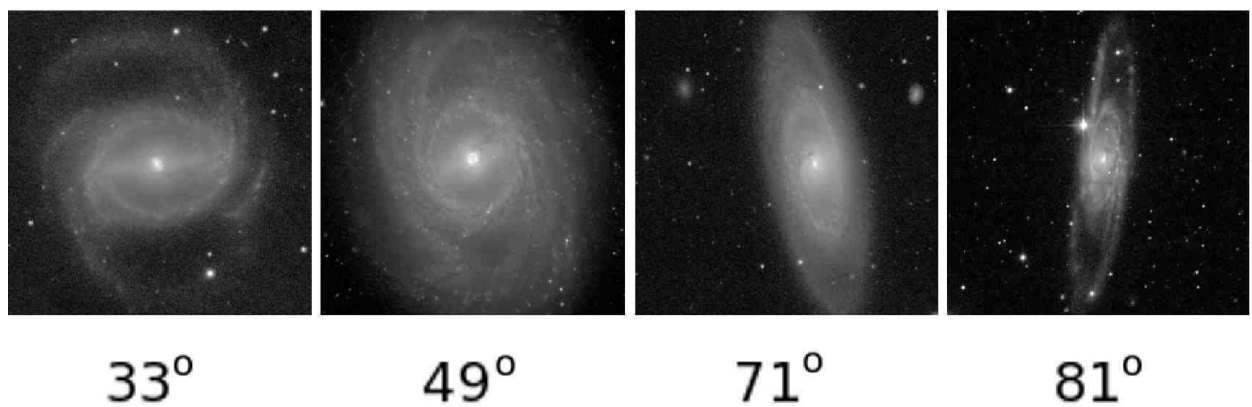}
\caption{Four galaxies of likely similar face-on morphology viewed at 
different inclinations (number below each image). The
galaxies are (left to right): NGC 1433, NGC 3351, NGC 4274, and NGC 5792. 
Images are from the dVA (filters $B$ and $g$).}
\label{difftilts}
\end{figure}

As extended objects rather than point sources, galaxies show a wide
variety of forms, some due to intrinsic structures, others due to the
way the galaxy is oriented to the line of sight.  The random
orientations, and the wide spread of distances, are the principal
factors that can complicate interpretations of galaxy morphology. If we
could view every galaxy along its principal axis of rotation, and from
the same distance, then fairer comparisons would be possible.
Nevertheless, morphologies seen in face-on galaxies can also often be
recognized in more inclined galaxies (Figure~\ref{difftilts}).  It is
only for the highest inclinations that morphology switches from face-on
radial structure to vertical structure. In general we either know the
planar structure in a galaxy, or we know its vertical structure, but we
usually cannot know both well from analysis of images alone.

Galaxy morphology began to get interesting when the ``Leviathan of
Parsonstown", the 72-inch meridian-based telescope built in the 1840s
by William Parsons, Third Earl of Rosse, on the grounds of Birr Castle
in Ireland, revealed spiral patterns in many of the brighter Herschel
and Messier ``nebulae." The nature of these nebulae as galaxies wasn't
fully known at the time, but the general suspicion was that they were
star systems (``island universes") like the Milky Way, only too distant
to be easily resolved into their individual stars. In fact, one of
Parsons' motivations for building the ``Leviathan" was to try and
resolve the nebulae to prove this idea. The telescope did not
convincingly do this, but the discovery of spiral structure itself was
very important because such structure added to the mystique of the
nebulae. The spiral form was not a random pattern and had to be
significant in what it meant. The telescope was not capable of
photography, and observers were only able to render what they saw with
it in the form of sketches. The most famous sketch, that of M51 and its
companion NGC 5195, has been widely reproduced in introductory
astronomy textbooks.

While visual observations could reveal some important aspects
of galaxy morphology, early galaxy classification was based on
photographic plates taken in the blue region of the spectrum. Silver
bromide dry emulsion plates were the staple of astronomy beginning in
the 1870s and were relatively more sensitive to blue light than to red
light. Later, photographs taken with Kodak 103a-O and IIa-O plates
became the standard for galaxy classification. In this part of the
spectrum, massive star clusters, dominated by spectral class O and B
stars, are prominent and often seen to line the spiral arms of
galaxies. These clusters, together with extinction due to interstellar
dust, can give blue light images a great deal of detailed structure for
classification. It is these types of photographs which led to the
galaxy classification systems in use today. 

In such photographs, we see many galaxies as a mix of structures.
Inclined galaxies reveal the ubiquitous {\it disk} shape, the most
highly flattened subcomponent of any galaxy. Studies of Doppler
wavelength shifts in the spectra of disk objects (like HII regions and
integrated star light) reveal that disks rotate differentially. If a
galaxy is spiral, the disk is usually where the arms are found, and
also where the bulk of interstellar matter is found. The radial
luminosity profile of a disk is usually {\it exponential}, with
departures from an exponential being due to the presence of other
structures.

In the central area of a disk-shaped galaxy, there is also often a
bright and sometimes less flattened mass concentration in the form of a {\it
bulge}. The nature of bulges and how they form has been a 
topic of much recent research, and is discussed further in section
9. Disk galaxies range from virtually bulge-less to bulge-dominated.
In the center there may also be a conspicuous {\it nucleus}, a bright
central concentration that was usually lost to overexposure in
photographs. Nuclei may be dominated by ordinary star light, or may be
{\it active}, meaning their spectra show evidence of violent gas
motions.

{\it Bars} are the most important internal perturbations seen in
disk-shaped galaxies. A bar is an elongated mass often made of old
stars crossing the center. If spiral structure is present, the arms
usually begin near the ends of the bar. Although most easily recognized
in the face-on view, bars have generated great interest recently in the
unique ways they can also be detected in the edge-on view. 
Not all bars are made exclusively of old stars. In some
bulge-less galaxies, the bar has considerable gas and recent star
formation. 

Related to bars are elongated disk features known as {\it ovals}.
Ovals usually differ from bars in lacking higher order Fourier components
(i.e., have azimuthal intensity distributions that vary mainly
as 2$\theta$), but nevertheless can be major perturbations in a galactic
disk. The entire disk of a galaxy may be oval, or a part of it may
be oval. Oval disks are most easily detected if there is considerable
light or structure at larger radii.

{\it Rings} are prominent features in some galaxies. Often defined by
recent star formation, rings may be completely closed features or may
be partial or open, the latter called ``pseudorings." Rings can be
narrow and sharp or broad and diffuse. It is particularly interesting
that several kinds of rings are seen, and that some galaxies can have
as many as four recognizeable ring features. {\it Nuclear rings} are
the smallest rings and are typically seen in the centers of barred
galaxies.  {\it Inner rings} are intermediate-scale features that often
envelop the bar in a barred galaxy. Outer rings are large, low surface
brightness features that typically lie at about twice the radius of a
bar. Other kinds of rings, called accretion rings, polar rings, and
collisional rings, are also known but are much rarer than the inner,
outer, and nuclear rings of barred galaxies. The latter kinds of
rings are also not exclusive to barred galaxies, but may be found also
in nonbarred galaxies.

{\it Lenses} are features, made usually of old stars, that have a
shallow brightness gradient interior to a sharp edge. They are commonly
seen in Hubble's disk-shaped S0 class (section 5.2). If a bar is
present, the bar may fill a lens in one dimension.  Lenses may be round
or slightly elliptical in shape.  If elliptical in shape they would
also be considered ovals.

{\it Nuclear bars} are the small bars occasionally seen in the centers
of barred galaxies, often lying within a nuclear ring. When present in
a barred galaxy, the main bar is called the ``primary bar" and the
nuclear bar is called the ``secondary bar." It is possible for a
nuclear bar to exist in the absence of a primary bar.

{\it Dust lanes} are often seen in optical images of spiral galaxies,
and may appear extremely regular and organized. They are most readily
evident in edge-on or highly inclined disk galaxies, but are still
detectable in the face-on view, often on the leading edges of bars
or the concave sides of strong inner spiral arms.

Spiral arms may also show considerable morphological variation.
Spirals may be regular 1, 2, 3, or 4-armed patterns, and may also be
higher order multi-armed patterns. Spirals may be tightly wrapped (low
pitch angle) or very open (high pitch angle.) A {\it grand-design}
spiral is a well-defined global pattern, often detectable as smooth
variations in the stellar density of old disk stars. A {\it flocculent}
spiral is made of small pieces of spiral structure that appear sheared
by differential rotation. Their appearance can be strongly affected by
dust, such that at longer wavelengths a flocculent spiral may appear
more grand-design.  Pseudorings can be thought of as variable pitch
angle spirals which close on themselves, as opposed to continuously
opening, constant pitch angle, logarithmic spirals.

There are also numerous structures outside the scope of traditional
galaxy classification, often connected with strong interactions
between galaxies. Plus, the above described features are not 
necessarily applicable or relevant to what we see in very distant
galaxies. Accounting for all of the observed features of nearby
galaxies, and attempting to connect what we see nearby to what
is seen at high redshift, is a major goal of morphological studies.

\section{Galaxy Classification}

As noted by Sandage (1975), the first step in studying any class of
objects is a classification of those objects. Classification built
around small numbers of shared characteristics can be used for sorting 
galaxies into fundamental categories, which can then be the basis
for further research. From such research, physical relationships
between identified classes may emerge, and these relationships may
foster a theoretical interpretation that places the whole class of
objects into a global context. There is no doubt that such an approach
greatly contributed to the development of the science of biology, and
this is no less true for galaxies.

The genesis of galaxy classification is to take the complex
combinations of structures described in the previous section and
summarize them with a few type symbols. Sandage (1975) 
describes the earlier classification systems of Wolf,
Reynolds, Lundmark, and Shapley that
fell into dis-use more than 50 years ago. 
The Morgan (1958) spectral type/concentration
classification system, which was based on a connection between
morphology (specifically central concentration)
and the stellar content of the central regions, was used
recently by Bershady, Jangren, \& Conselice (2000) in a mostly
quantitative manner (see also Abraham et al. 2003).
Thus, Morgan's system has in a way survived into the modern era but not
in the purely visual form that he proposed. 
Only one Morgan galaxy type, the supergiant cD type, is still used 
extensively (section 10.6). Van den Bergh's luminosity/arm morphology
classification system is described by van den Bergh (1998; see
section 6.5).

\begin{figure}
\plotone{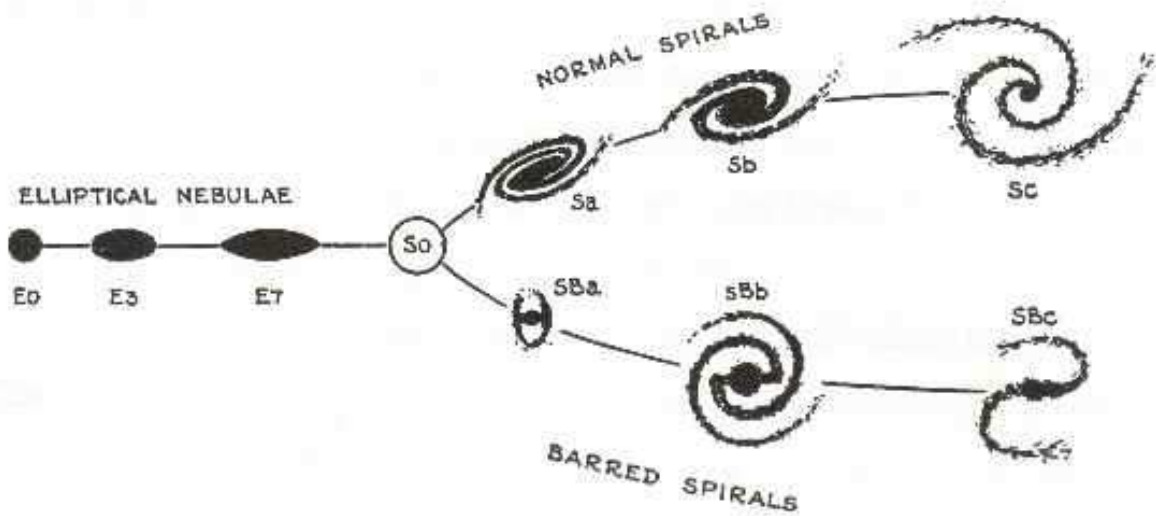}
\caption{Hubble's (1936) ``tuning fork" of galaxy morphologies is the basis for modern
galaxy classification.}
\label{tuning_fork_schematic}
\end{figure}

The big survivor of the early visual classification systems was that of
Hubble (1926, 1936), as later revised and expanded upon by Sandage
(1961) and de Vaucouleurs (1959). Sandage (1975) has argued that one
reason Hubble's view prevailed is that he did not try and account for
every superficial detail, but kept his classes broad enough that the
vast majority of galaxies could be sorted into one of his proposed
bins. These bins were schematically illustrated in Hubble's famous
``tuning fork"\footnote{As recently noted by D. L. Block (Block et al.
2004a), this diagram may have been inspired by a similar schematic by
Jeans (1929).} (Hubble 1936; reproduced in
 Figure~\ref{tuning_fork_schematic}) recognizing a sequence of
progressive flattenings from ellipticals to spirals. Ellipticals had
only two classification details: the smoothly declining brightness
distribution with no inflections, and no evidence for a disk; and the
ellipticity of the isophotes, indicated by a number after the ``E"
symbol. (For example, E3 means the ellipticity is 0.3.) Spirals were
systems more flattened than an E7 galaxy that could be subdivided
according to the degree of central concentration, the degree of
openness of the arms, the degree of resolution of the arms into
complexes of star formation (all three criteria determining position
along the fork), and on the presence or absence of a bar (determining
the appropriate prong of the fork).

The S0 class at the juncture of the prongs of the fork was still
hypothetical in 1936. As ``armless disk galaxies," S0s were mysterious
because all examples known in 1936 were barred.  These were classified
as SBa, but this was a troubling inconsistency because nonbarred Sa
galaxies had full spiral patterns. Hubble predicted the existence of
nonbarred S0s to fill the gap between type E7\footnote{ van den Bergh
(2009a) shows that E0-E4 galaxies are more luminous on average than are
E5-E7 galaxies, suggesting that all E7 galaxies (and not many have been
recognized) are actually S0 galaxies.  Van den Bergh argues that
genuine E galaxies may be no more flattened than E6.} and Sa and cure
what he felt was a ``cataclysmic" transition.

It was not long before Hubble himself realized that the tuning fork
could not adequately represent the full diversity of galaxy
morphologies, and after 1936 he worked on a revision that included real
examples of the sought group: nonbarred S0 galaxies. Based on
fragmentary notes he left behind, Sandage (1961) prepared the {\it
Hubble Atlas of Galaxies} to illustrate Hubble's revision, and also
added a third dimension: the presence or absence of a ring. This was
the first major galaxy atlas illustrating a classification system in a
detailed, sophisticated way with beautifully produced photographs.
Hubble's revision, with van den Bergh luminosity classes (Sandage \&
Tammann 1981), was updated and extended to types later than Sc by
Sandage \& Bedke (1994).

Because Sandage (1961) and Sandage \& Bedke (1994) describe the
Hubble-Sandage revision so thoroughly, the details will not be repeated
here. Instead, the focus of the next section will be on the de
Vaucouleurs revision as outlined in the dVA. The reasons for this are:
(1) the de Vaucouleurs classification provides the most familar galaxy
types to extragalactic researchers, mostly because of extensive
continuing use of the Third Reference Catalogue of Bright Galaxies
(RC3, de Vaucouleurs et al. 1991); and (2) the de Vaucouleurs
classification is still evolving to cover more details of galaxy
morphology considered significant at this time. It should be 
noted that both the de Vaucouleurs and Hubble-Sandage revisions
are strictly applicable only to $z$ $\approx$ 0 galaxies and that it is
often difficult to fit objects having $z$ $>$ 0.5 neatly into
the categories defining these classification systems. High redshift
galaxy morphology is described in section 13.

\section{A Continuum of Galactic Forms}

The Hubble tuning fork is useful because it provides a visual
representation of information Hubble (1926) had only stated in
words. The fork contains an implication of continuity. For example,
it does not rule out that there might be galaxies intermediate
in characteristics between an "Sa" or "Sb" spiral, or between a
normal "S" spiral and a barred "SB" spiral.  Continuity along the
elliptical galaxy sequence was always implied as a smooth variation
from round ellipticals (E0) to the most flattened ellipticals (E7).
Sandage (1961) describes the modifications that made the Hubble
system more three-dimensional: the introduction of the (r) (inner
ring) and (s) (pure spiral) subtypes. Continuity even with this
characteristic was possible, using the combined subtype (rs). Thus,
already by 1961, the Hubble classification system had become much
more complicated than it was in 1926 or 1936. The addition of the 
S0 class was one reason for this, but the (r) and (s) subtypes were
another.

In the Hubble-Sandage classification, it became common to denote
galaxies on the left part of the Hubble sequence as ``early-type"
galaxies and those on the right part as ``late-type" galaxies. 
By the same token, Sa and SBa spirals became ``early-type spirals"
while Sc and SBc spirals became ``late-type spirals." Sb and SBb
types became known as ``intermediate-type spirals." The reason
for these terminologies was convenience and borrows terminology
often used for stars. Young, massive stars of spectral classes O and B
were known as ``early-type stars" while older stars of cooler spectral
types were known as ``late-type stars." Hubble stated that his use of
these temporal descriptions for galaxies had no evolutionary
implications. An irony in this is that it eventually became clear that
early-type galaxies are dominated by late-type stars, while
late-type galaxies often have significant numbers of early-type
stars.

\begin{figure}
\plotone{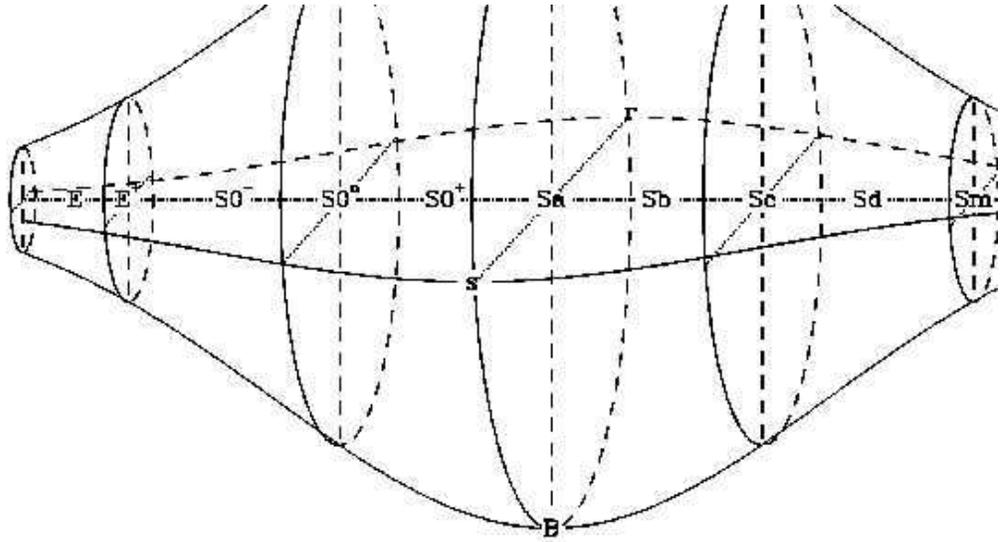}
\caption{de Vaucouleurs's (1959) classification volume, a revision and extension of the Hubble
tuning fork. The three dimensions are the stage (Hubble type), the family (apparent bar strength),
and the variety (presence or absence of an inner ring).}
\label{classification_volume}
\end{figure}

De Vaucouleurs (1959) took the idea of continuity of galaxy
morphology a step further by developing what he referred to as the {\it
classification volume} (Figure~\ref{classification_volume}). 
In this revision of the Hubble-Sandage (1961) classification, galaxy morphology
represents a continuous sequence of forms in a three-dimensional volume
with a long axis and circular cross-sections of varying size. The long
axis of the volume is the {\it stage}, or type, and it represents the
long axis of the original Hubble tuning fork. The short axes are the
family and the variety, which refer to apparent bar strength and the
presence or absence of an inner ring, respectively. In addition to Hubble's
original stages E, Sa, Sb, and Sc, the classification volume includes
new stages: late ellipticals: E$^+$, ``very late" spirals: Sd,
``Magellanic spirals": Sm, and ``Magellanic irregulars": Im. The S0
class is included in the same position along the sequence, between E's
and spirals, but is subdivided into three stages.
The characteristics defining individual stages are described
further in section 5.2.

The stage is considered the most fundamental dimension of the
classification volume because measured physical parameters, such
as integrated color indices, mean surface brightnesses, and neutral
hydrogen content correlate well with position along the sequence
(e. g., Buta et al. 1994).
Early-type galaxies tend to have redder colors, higher average
surface brightnesses, and lower neutral hydrogen content than
late-type galaxies. The family and variety axes of the classification volume 
indicate the considerable variations in morphology at a given stage. 
A famous sketch of families
and varieties near stage Sb, drawn by de Vaucouleurs himself during a
cloudy night at McDonald Observatory circa 1962, is shown in 
Sandage (1975) and in Figure 1.13 of the dVA.
The classification volume is broader in the middle compared to the
ends because the diversity of galaxy morphology is largest at stages
like S0/a and Sa. Bars and rings are often most distinct and most
recognizable at these stages. Such features are not characteristic
of E galaxies, so the volume must be narrow at that end. Along the
S0 sequence, bars and rings are barely developed among the earliest S0s
(S0$^-$) and well-developed among the late S0s (S0$^+$), thus the
volume begins to broaden. Among very late-type galaxies, Sd, Sdm, Sm,
and Im, bars are actually very frequent, but closed rings (r) are not.
Thus, the volume narrows at that end as well.

For the purposes of illustrating morphology, blue light digital images
converted to units of magnitudes per square arcsecond are used when
available.  This approach is described in the dVA, and requires
calibration of the images, usually based on published photoelectric
multi-aperture photometry. In addition, some of the illustrations used
(especially in section 15) are from the Sloan Digital Sky Survey or
from other sources.  These are not in the same units but still provide
excellent illustrations of morphology.  SDSS color images differ from
dVA images mainly in the central regions, where SDSS images sometimes
lose detail.

Unlike the dVA, the scope of this article extends beyond the
traditional $UBVRI$ wavebands.
It is only during the past 20 years that significant morphological
information has been obtained for galaxies outside these
bands, mostly at mid- and far-ultraviolet and mid-IR wavelengths from space-based
observatories capable of imaging in these wavelengths to unprecedented
depths, providing a
new view of galaxy morphology that is only beginning to be explored.
A useful review of many issues in morphology is provided by van
den Bergh (1998).

\section{Galaxy Types: Stage, Family, and Variety}

\subsection{Elliptical and Spheroidal Galaxies}

\begin{figure}
\plotone{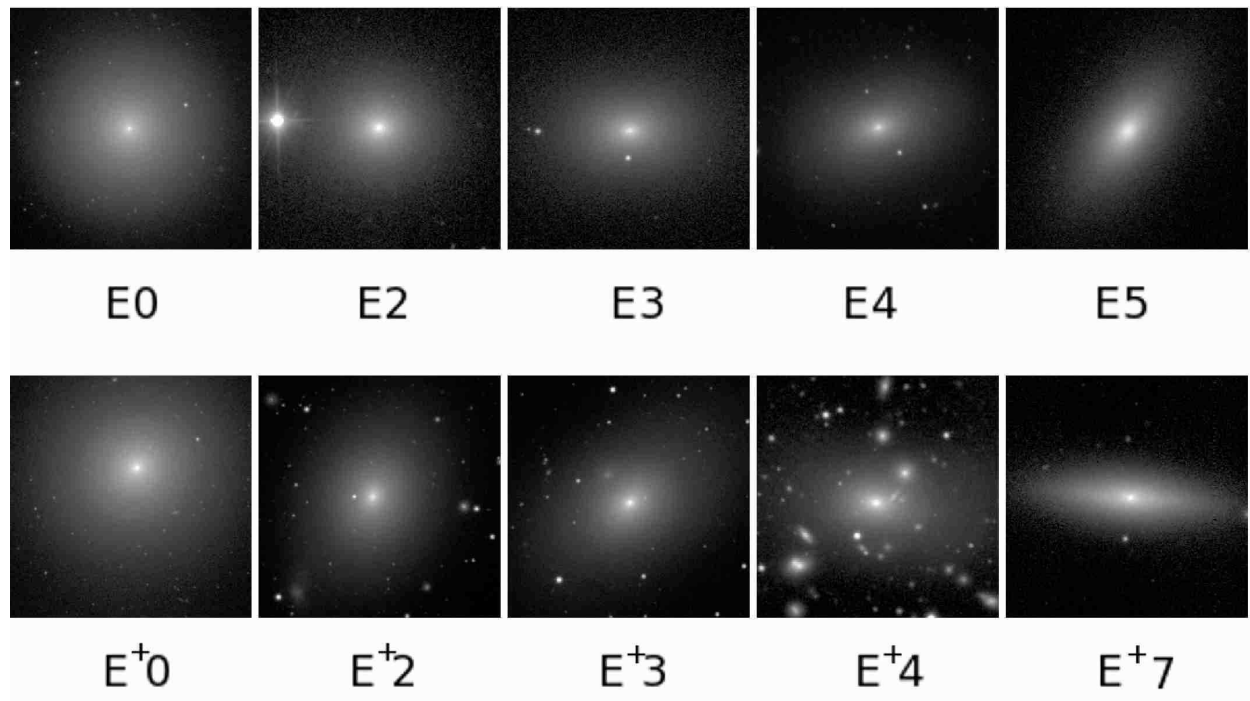}
\caption{Examples of elliptical galaxies of different projected shapes.
Type E galaxies are normal ellipticals with no structural details. 
From left to right the galaxies shown are NGC 1379, 3193, 5322, 1426, and
720.  Type E$^+$ galaxies are ``late" ellipticals, which may include faint
extended envelopes typical of large cluster ellipticals, or simple
transition types to S0$^-$. The examples shown
are (left to right): NGC 1374, 4472, 4406, 4889, and 4623. All of these
images are from the dVA (filters $B$, $V$, and $g$).}
\label{egals}
\end{figure}

Elliptical galaxies are smooth, amorphous systems with a continuously
declining brightness distribution and no breaks, inflections, zones, or
structures, as well as no sign of a disk.  Figure~\ref{egals} shows
some good examples.  Because ellipticals are dominated by old stars and
are relatively dust-free, they look much the same at different
wavelengths. Hubble's subclassification of ellipticals according to
apparent ellipticity (En, where n=10(1-b/a), b/a being the apparent
flattening) was useful but virtually no physical characteristics of
ellipticals correlate with this parameter (Kormendy \& Djorgovski
1989). The n value in the En classification is simply the projected
ellipticity and not easily interpreted in terms of a true flattening
without direct knowledge of the orientation of the symmetry planes.
Luminous ellipticals are thought to be triaxial in structure with an
anisotropic velocity dispersion tensor, while lower luminosity
ellipticals are more isotropic oblate rotators (Davies et al. 1983).
Studies of rotational to random kinetic energy ($V$/$\sigma$) versus
apparent flattening ($\epsilon$) show that massive ellipticals are slow
anisotropic rotators.  Ellipticals follow a fundamental plane
relationship between the effective radius $r_e$ of the light
distribution, the central velocity dispersion $\sigma_0$, and mean
effective surface brightness $<I_e>$ (see review by Kormendy \&
Djorgovski 1989).  Dwarf elliptical galaxies may not follow the same
relation as massive ellipticals; this is discussed by Ferguson \&
Binggeli (1994).

The lack of fundamental significance of Hubble's En classification led
some authors to seek an alternative, more physically useful approach.
Kormendy \& Bender (1996; see also the review by Schweizer 1998)
proposed a revision to Hubble's tuning fork handle that orders
ellipticals according to their velocity anisotropy, since this is a
significant determinant of E galaxy intrinsic shapes. Velocity
anisotropy correlates with the {\it deviations} of E galaxy isophotes
from pure elliptical shapes, measured by the parameter $a_4/a$, the
relative amplitude of the $\cos 4\theta$ Fourier term of these
deviations. If this relative amplitude is positive, then the isophotes
are pointy, disky ovals, while if negative, the isophotes are boxy ovals. A
boxy elliptical is classified as E(b), while a disky elliptical is
classified as E(d).  The correlation with anisotropy is such that E(b)
galaxies have less rotation on average and more velocity anistropy than
E(d) galaxies.

\begin{figure}
\plotone{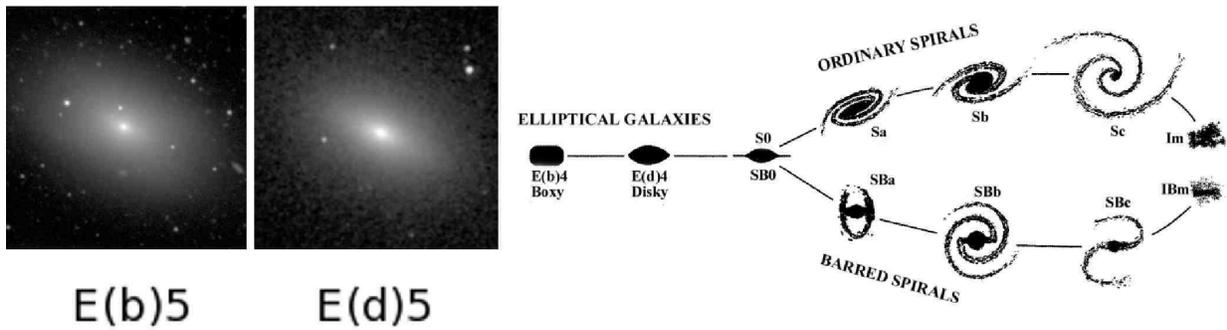}
\caption{Revised classification of elliptical galaxies from
Kormendy \& Bender (1996), as schematically incorporated into
Hubble's (1936) ``tuning fork." At left are two examples of
boxy and disky ellipticals: NGC 7029 (left, $B$-band) and NGC 4697 ($JHK_s$
composite, 2MASS image from NED).}
\label{boxydisky}
\end{figure}

The Kormendy \& Bender proposed classification is shown in
Figure~\ref{boxydisky}, together with two exceptional examples that
show the characteristic isophote shapes. The leftmost of these two, NGC
7029, is an unusually obvious boxy elliptical at larger radii. This
boxiness is the basis for the classification E(b)5. However, NGC 7029
is not boxy throughout: it shows evidence of a small inner disk and
hence is disky at small radii. This is not necessarily taken into
account in the classification. The other image shown in
Figure~\ref{boxydisky} is NGC 4697, a galaxy whose isophotes are
visually disky. The idea with the Kormendy \& Bender classification is
that it is the disky ellipticals which connect to S0 galaxies, and not
the boxy ellipticals. However, NGC 7029 demonstrates that diskiness and
boxiness can be a function of radius, thus perhaps a smooth connection
between E(b) and E(d) types [i.e., type E(b,d)] is possible and the
order shown in the Kormendy \& Bender revision to the Hubble tuning
fork, with boxy Es blending into the disky Es, could be reasonable.

Note that while the classical En classifications of ellipticals were
designed by Hubble to be estimated by eye, this is not easily done for
the E(b) and E(d) classifications, which are most favored to be 
seen only when the disk is nearly edge-on.  Face-on Es with imbedded
disks will not show disky isophotes.  For example, NGC 7029 and 4697
are extreme cases where the isophotal deviations are obvious by eye.
But for most E galaxies, the E(b)n and E(d)n classifications can only
be judged reliably with measurements of the $a_4/a$ parameter.

The de Vaucouleurs classification of ellipticals includes a slightly
more advanced type called E$^+$, or ``late" ellipticals. It was
originally intended to describe ``the first stage of the transition to
the S0 class" (de Vaucouleurs 1959). Five examples of E$^+$ galaxies
are shown in the second row of Figure~\ref{egals}. Galaxies classified
as E$^+$ can be the most subtle S0s, but many of the E$^+$ cases listed
in RC3 are the brightest members of clusters that have shallow enough
brightness profiles to appear to have an extended envelope (see section
10.6). Of the five E$^+$ galaxies shown, only NGC 4623 (rare type
E$^+$7) seems consistent with de Vaucouleurs's original view. It is
also a much lower luminosity system than the other four cases shown.
While S0$^-$ is the type most often confused with ellipticals in visual
classification, the bin has a wide spread from the most obvious to the
least obvious cases. Thus, a type like E$^+$ is still useful for
distinguishing transitions from E to S0 galaxies.

\begin{figure}
\plotone{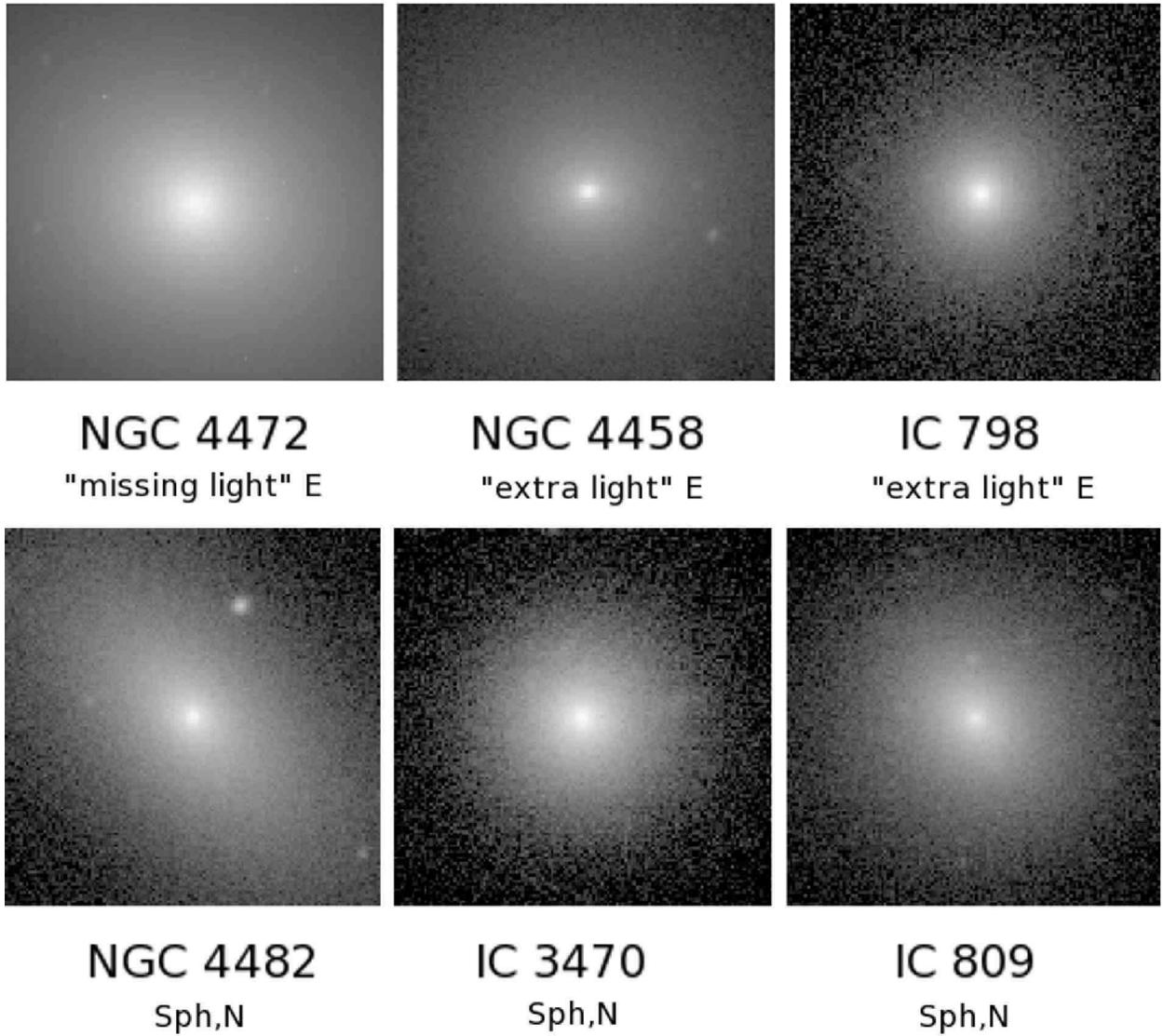}
\caption{Illustrations of 6 early-type galaxies in the Virgo Cluster
with photometric classifications from Kormendy et al. (2009):
NGC 4472 (core E, $M_V$ = $-$23.2);
NGC 4458 (coreless E, $M_V$ = $-$19.0);
IC 798 (VCC 1440; `coreless E, $M_V$ = $-$16.9);
NGC 4482 (nucleated spheroidal, $M_V$ = $-$18.4);
IC 3470 (VCC 1431; nucleated spheroidal, $M_V$ = $-$17.4);
IC 809 (VCC 1910; nucleated spheroidal, $M_V$ = $-$17.4). The images shown
are all based on SDSS $g$-band single or mosaic images, and are
in units of mag arcsec$^{-2}$.}
\label{cores}
\end{figure}

The photometric properties of ellipticals depend on luminosity. In
terms of Sersic $r^{1 \over n}$ profile fits, large, luminous
ellipticals tend to have profiles described better by $n$$\gtrsim$4,
while smaller, lower luminosity ellipticals tend to have $n$$<$4, with
values as low as 1 (e. g., Caon, Capaccioli, \& D'Onofrio 1993). Graham \&
Guzm\'an (2003) discuss the implications of this correlation on
proposed dichotomies of elliptical galaxies (e.g., Kormendy 1985; see
also Ferrarese et al. 2006). These studies received a major impetus
from the massive photometric analysis of Virgo Cluster elliptical
galaxies by Kormendy et al. (2009). Two issues were considered by these
authors (Figure~\ref{cores}). The first was whether galaxies classified
as dwarf ellipticals (``dE"; see section 15.2) in the Virgo Cluster
really are the low luminosity extension of more massive, conventional
ellipticals, or something different altogether.  Based on parameter
correlations, such as $r_{10\%}$, the major axis radius of the isophote
containing 10\% of the total visual luminosity, versus $\mu_{10\%}$,
the surface brightness of this isophote, Kormendy et al. show that even
the most elliptical-like and luminous dE galaxies lie on a distinct
sequence from normal elliptical galaxies, which tend to lie on a higher
density sequence. In a graph of $B$-band central surface brightness
versus absolute $B$-band magnitude, the dE galaxies lie in a region
occupied by Magellanic irregular galaxies, suggesting a link between
the groups and consistent with the earlier conclusions of Kormendy
(1985). Kormendy et al. (2009) suggest reclassifying Binggeli, Sandage,
\& Tammann's (1985) "dE" galaxies and related objects (like dwarf S0,
or dS0 galaxies) as "spheroidal" (Sph) galaxies, including the type
"Sph,N", meaning "nucleated spheroidal" galaxy (section 15.2).
Figure~\ref{cores} shows several examples of Sph,N galaxies as compared
with several genuine elliptical galaxies. The morphological appearance
alone does not necessarily distinguish the two classes. The
classification is physical, being based mainly on parameter
correlations. Kormendy et al.  suggest that Sph galaxies are formed
from late-type systems by environmental effects and supernovae.

The second issue considered was the physical distinction between
``core" elliptical galaxies, those where the surface brightness profile
approaches either a constant level or a slightly sloped level with
radius approaching zero, and ``coreless" ellipticals (also known as
``power law" Es) where the inner profile steepens with decreasing
radius (Kormendy 1999). Kormendy et al. (2009) illustrated both types
relative to a Sersic $r^{1\over n}$ fit to the outer regions of the
luminosity profiles. In this representation, core Es are ``missing
light" relative to the fit while coreless Es have ``extra light." The
top row of Figure~\ref{cores} shows one core E (NGC 4472) and two
coreless Es [NGC 4458 and IC 798 (VCC 1440), the latter a low-luminosity
dwarf]. The subtle distinctions are evident in these images, with
NGC 4472 showing a soft center and NGC 4458 showing a strong center.
The terminology for both types is mostly historical (Kormendy 1999)
and somewhat counter to the visual impression (i.e., NGC 4472 lacks
a bright core while NGC 4458 has one, yet the latter is technically
coreless). Kormendy et al. show that core and coreless E galaxies 
have different Sersic indices, velocity dispersion anisotropy,
isophote shapes, and rotational character, with the core Es being
of the boxy type and the coreless Es of the disky type in
Figure~\ref{boxydisky}. The distinction may be tied to the number
of mergers that formed the system.

\subsection{S0 and Spiral Galaxies}

The full classification of spiral and S0 galaxies involves the recognition
of the stage, family, and variety.
In de Vaucouleurs's classification approach, the implication for bars,
inner rings, and stages is a continuum of forms (de Vaucouleurs 1959),
so that there are no sharp edges to any category or ``cell" apart from the
obvious ones (for example, there are no galaxies less ``barred" than
a nonbarred galaxy, nor are there galaxies more ringed than those with
a perfectly closed ring). 

\begin{figure}
\plotone{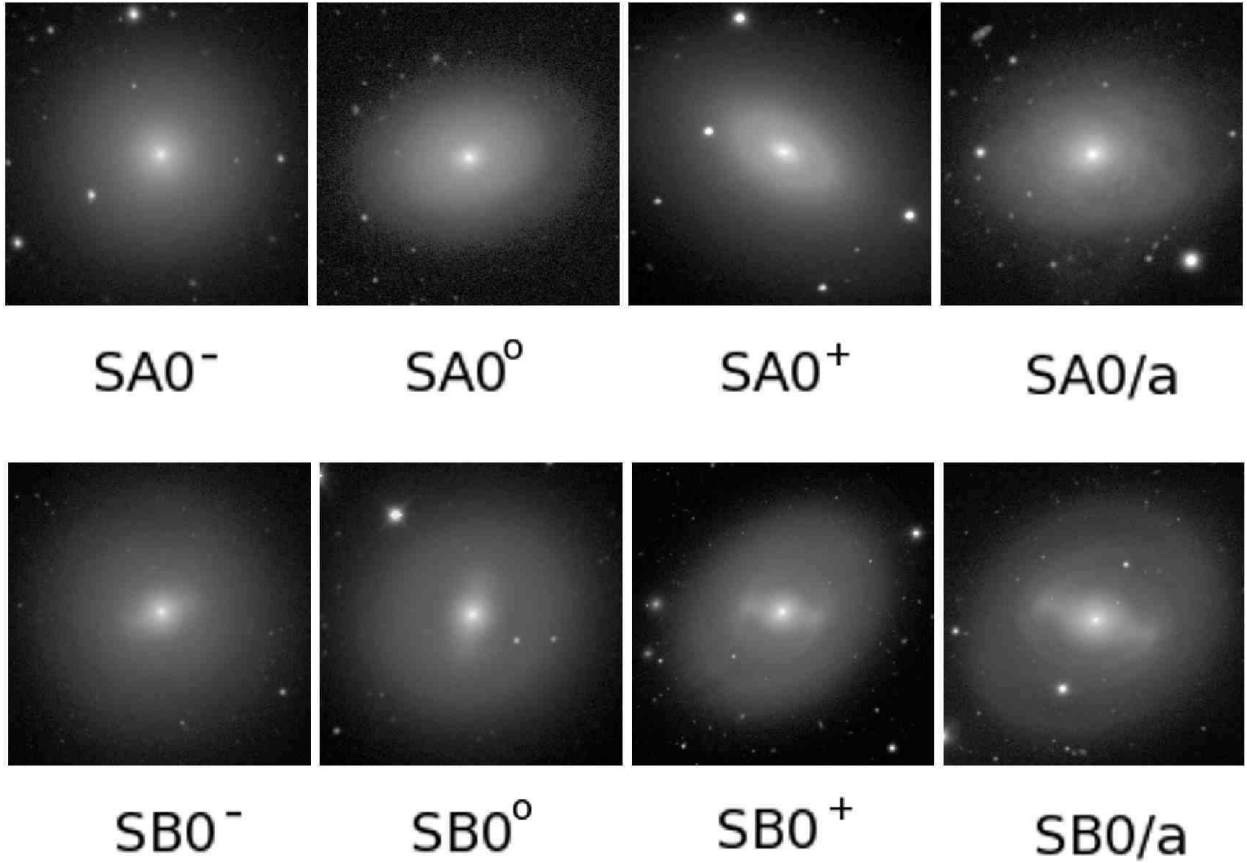}
\caption{Examples of barred and nonbarred S0 galaxies of different stages
from ``early" (S0$^-$), to ``intermediate" (S0$^o$), to ``late" (S0$^+$),
including the transition stage to spirals, S0/a.  
The galaxies shown are (left to right): Row 1 - NGC 7192,
1411, 1553, and 7377; Row 2 - NGC 1387, 1533, 936, and 4596.
All images are from the dVA (filters $B$ and $V$).}
\label{S0gals}
\end{figure}

The classification of S0 galaxies depends on recognizing the presence
of a disk and a bulge at minimum, and usually a lens as well, and no
spiral arms.  Examples are shown in Figure~\ref{S0gals}.  The display
of galaxy images in units of mag arcsec$^{-2}$ makes lenses especially
easy to detect, as noted in the dVA.  Even if a lens isn't obvious, a
galaxy could still be an S0 if it shows evidence of an exponential
disk.  (Lenses are also not exclusive to S0s.) The ``no spiral arms"
characteristic is much stricter in the Hubble-Sandage classification
than in the de Vaucouleurs interpretation, because varieties (r, rs,
and s) are carried into the de Vaucouleurs classifications of S0s. This
allows the possibility of a classification like SA(s)0$^-$, which would
be very difficult to recognize. Bars enter in the classification of S0s
in a similar manner as for spirals.  Figure~\ref{S0gals} shows mainly
stage differences among nonbarred and barred S0s. The stage for S0s
ranges from early (S0$^-$), to intermediate (S0$^o$), and finally to
late (S0$^+$), in a succession of increasing detail. The earliest
nonbarred S0s may be mistaken for elliptical galaxies on photographic
images, and indeed Sandage \& Bedke (1994) note cases where they
believe an S0 galaxy has been misclassified as an elliptical by de
Vaucouleurs in his reference catalogues (see also {\it The Revised
Shapley-Ames Catalogue}, RSA, Sandage \& Tammann 1981). This kind of
misinterpretation is less likely for types S0$^o$ and S0$^+$, because
these will tend to show more obvious structure.

The morphological distinction between E and S0 galaxies has been
considered from a quantitative kinematic point of view by Emsellem et
al. (2007). These authors argue that the division of early-type
galaxies into E and S0 types is ``contrived", and that it is more
meaningful to divide them according to a quantitative kinematic
parameter called $\lambda_R$, the specific angular momentum of the
stellar component, which is derived from a two-dimensional velocity
field obtained with the SAURON integral field spectrograph (Bacon et
al. 2001). Based on this parameter, early-type galaxies are divided
into slow and fast rotators, i. e., whether they are characterized by
large-scale rotation or not. In a sample of 48 early-types, most were
found to be fast rotators classified as a mix of E and S0 types, while
the remainder were found to be slow rotators classified as Es. This
kind of approach, which provides a more physical distinction among
early-types, does not negate completely the value of the E and S0
subdivisions, but highlights again the persistent difficulty of
distinguishing the earliest S0s from Es by morphology alone.

The transition type S0/a shows the beginnings of spiral structure.  Two
examples are included in Figure~\ref{S0gals}, one nonbarred and the
other barred.  Type S0/a is a well-defined stage characterized in the
de Vaucouleurs 3D classification volume as having a high diversity in
family and variety characteristics. The type received a negative
characterization as the ``garbage bin" of the Hubble sequence at one
time because troublesome dusty irregulars, those originally classified
as ``Irr II" by Holmberg (1950) and later as ``I0" by de Vaucouleurs,
seemed to fit better in that part of the sequence. [In fact, de
Vaucouleurs, de Vaucouleurs, \& Corwin (1976) assigned the numerical
stage index $T$=0 to both S0/a and I0 galaxies.] However, this problem
is only a problem at optical wavelengths. At longer wavelengths (e.g.,
3.6 microns), types such as Irr II or I0 are less needed as they are
defined mainly by dust (Buta et al. 2010a).

\begin{figure}
\plotone{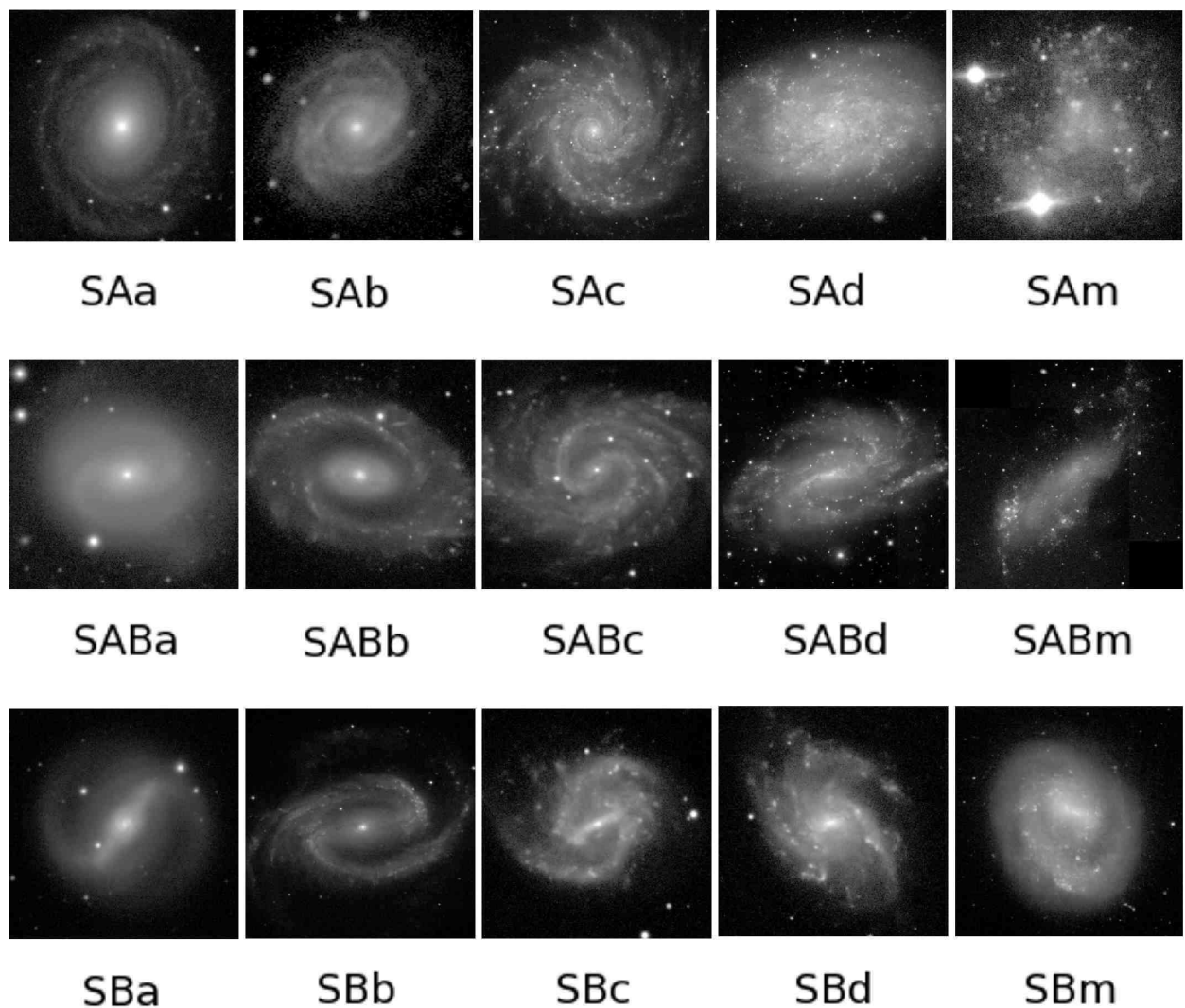}
\caption{Stage classifications for spirals, divided according to bar 
classifications into parallel sequences. The galaxies illustrated are
(left to right):
Row 1 - NGC 4378, 7042, 628, 7793, and IC 4182; Row 2 -
NGC 7743, 210, 4535, 925, and IC 2574; Row 3 - NGC 4314, 1300, 3513,
4519, and 4618. All images are $B$-band from the dVA.}
\label{tuning_fork}
\end{figure}

\begin{figure}
\plotone{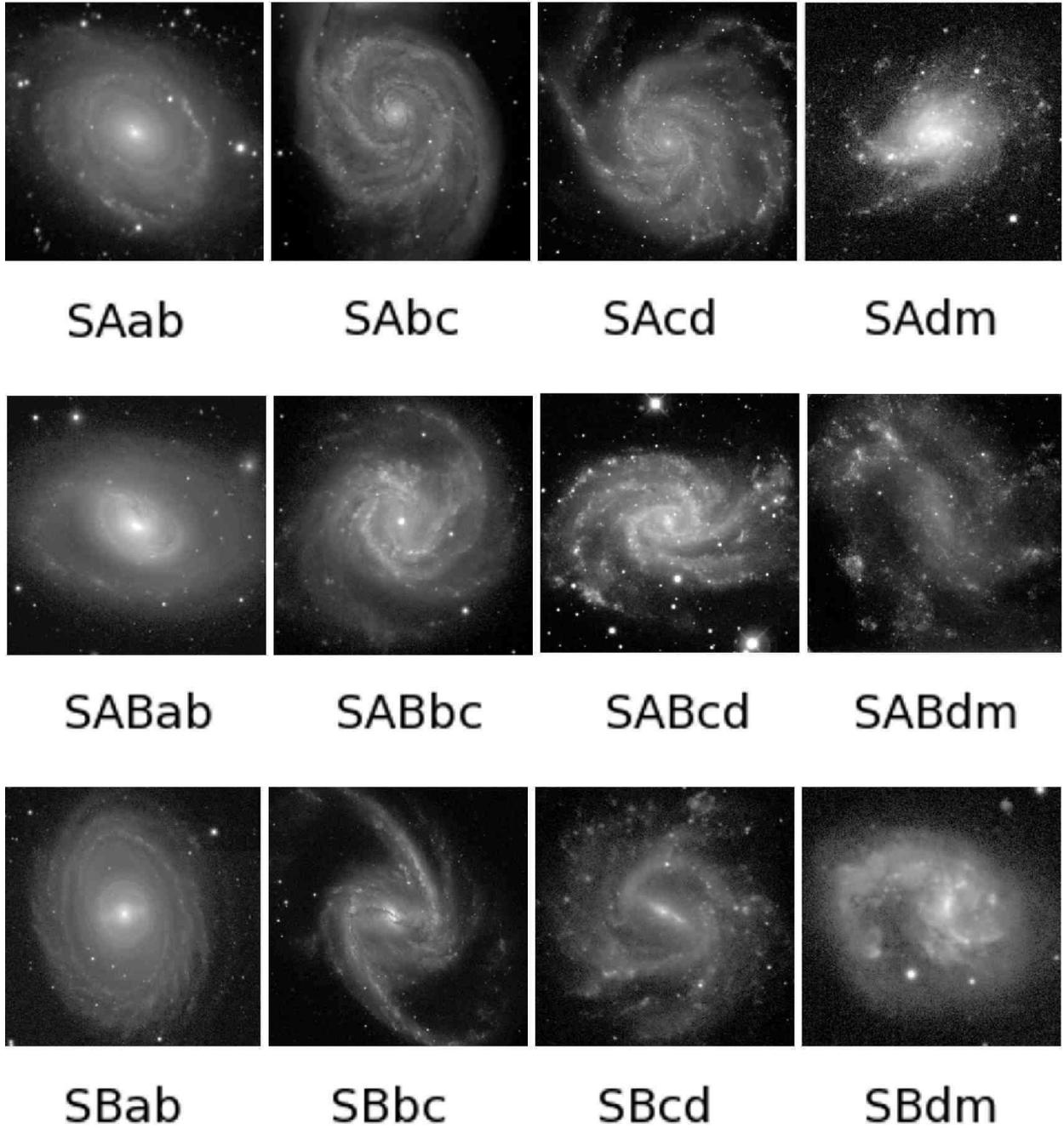}
\caption{Sequences of stages intermediate between the main stages
illustrated in Figure~\ref{tuning_fork}. The galaxies are
(left to right): Row 1
- NGC 2196, 5194, 5457, and 4534; Row 2 - NGC 3368,
4303, 2835, and 4395; Row 3 - NGC 1398, 1365, 1073, and 
4027. All images are $B$-band from the dVA, except for NGC 4534, which
is SDSS $g$-band.}
\label{intermediate_types}
\end{figure}

In general, the
stage for spirals is based on the appearance of the spiral arms (degree
of openness and resolution) and also on the relative prominence of the
bulge or central concentration. These are the usual criteria
originally applied by Hubble (1926, 1936). Figure~\ref{tuning_fork}
shows the stage sequence for spirals divided according to bar
classification (SA, SAB, SB), and as modified and extended 
by de Vaucouleurs (1959) to include Sd and Sm types. Intermediate
stages, such as Sab, Sbc, Scd, and Sdm, are shown in 
Figure~\ref{intermediate_types}. As noted by de Vaucouleurs (1963),
these latter stages are almost as common as the basic ones.

The three Hubble criteria are basically seen in the illustrated
galaxies.  Sa galaxies tend to have significant bulges, and
tightly-wrapped and relatively smooth spiral arms. Sab galaxies are
similar to Sa galaxies, but show more obvious resolution of the arms.
Sb galaxies have more resolution and more open arms, and generally
smaller bulges than Sab galaxies. Sbc galaxies have considerable
resolution and openness of the arms, and also usually significant
bulges. In Sc galaxies, the bulge tends to be very small and the arms
patchy and open. Scd galaxies tend to be relatively bulgeless, patchy
armed Sc galaxies. Stage Sd is distinctive mainly as almost completely
bulgeless late-type spirals with often ill-defined spiral structure.

Stages Sdm and Sm are the most characteristically asymmetric stages,
the latest spiral types along the de Vaucouleurs revised Hubble
sequence. They are described in detail by de Vaucouleurs \& Freeman
(1972) and by Odewahn (1991). Sm is generally characterized by
virtually no bulge and a single principle spiral arm. If a bar is
present, it is usually not at the center of the disk isophotes, unlike
what is normally seen in earlier type barred spirals. This leads to the
concept of an {\it offset barred galaxy}. The single spiral arm
emanates from one end of the bar. As noted by Freeman (1975), this is a
basic and characteristic asymmetry of the mass distribution of
Magellanic barred spirals. Sdm galaxies are similar, but may show a
weaker or shorter second arm. In Figure~\ref{tuning_fork}, NGC 4618 is
an especially good example of an SBm type (Odewahn 1991), while in
Figure~\ref{intermediate_types}, NGC 4027 is illustrated as type SBdm.

An important issue regarding these galaxies is whether the optically
offset bar is also offset from the dynamical rotation center of the
disk. In a detailed HI study of the interacting galaxy pair NGC 4618
and 4625, Bush \& Wilcots (2004) found very regular velocity fields
and extended HI disks, but no strong offset of the rotation center from
the center of the bar. This is similar to what Pence et al.  (1988)
found for the offset barred galaxy NGC 4027, based on optical
Fabry-Perot interferometry. In contrast, both Magellanic Clouds, which
are also offset barred galaxies, were found to have HI rotation centers
significantly offset from the center of the bar (Kerr \& de
Vaucouleurs 1955).

In general, the application of Hubble's three spiral criteria allows
consistent classification of spiral types. Nevertheless, sometimes the
criteria are inconsistent. For example, small bulge Sa galaxies
are described by Sandage (1961) and Sandage \& Bedke (1994). Barred
galaxies with nuclear rings can have spiral arms like those of an
earlier Hubble type and very small bulges. In such conflicting cases,
the emphasis is usually placed on the appearance of the arms. Also,
while late-type Sdm and Sm
galaxies are characteristically asymmetric, other types may be
asymmetric as well. On average, the bulge-to-total luminosity ratio is
related to Hubble type, but the result is sensitive to how
galaxies are decomposed (e. g., Laurikainen et al. 2005).
Asymmetry has been quantified by Conselice (1997).

\begin{figure}
\plotone{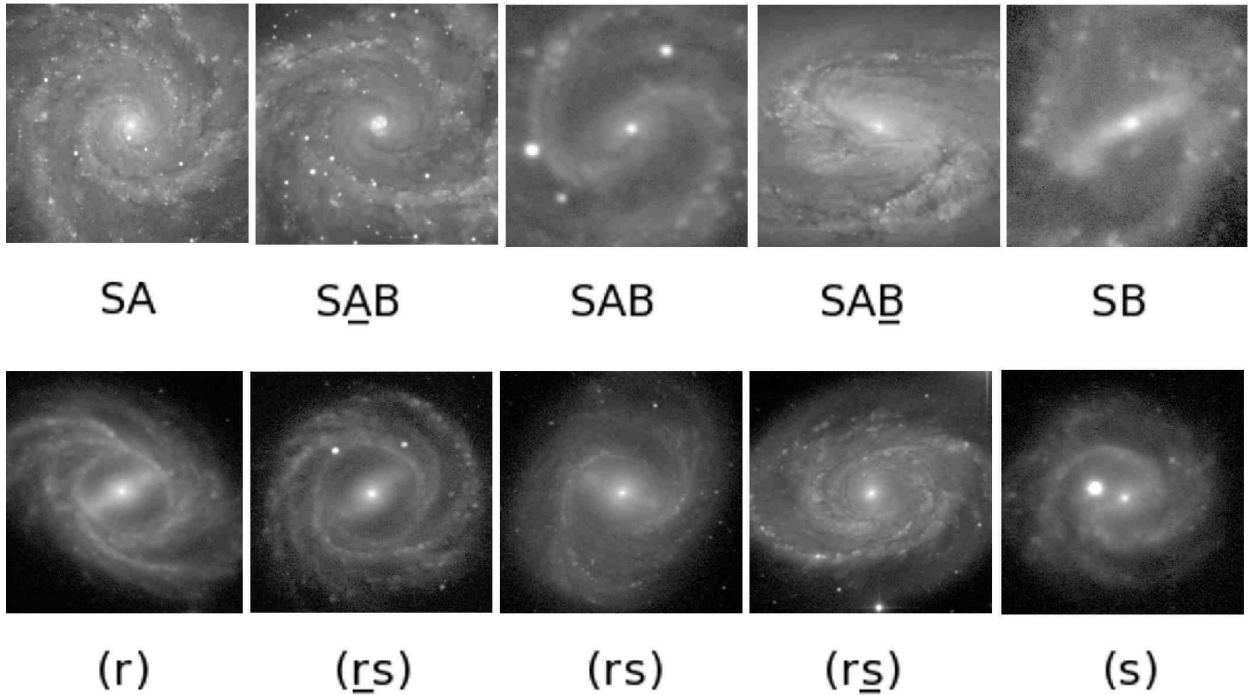}
\caption{The continuity of family and variety characteristics among
spiral galaxies, including underline classifications used by
de Vaucouleurs (1963). The galaxies are (left to right): Row 1 - NGC 628, 2997,
4535, 3627, and 3513; Row 2 - NGC 2523, 3450, 4548, 5371, and 3507.
All images are $B$-band from the dVA.}
\label{continuity}
\end{figure}

The family classifications SA, SAB, and SB are purely visual
estimates of bar strength, for both spirals and S0s. 
They are highlighted already
in Figures~\ref{S0gals}--~\ref{intermediate_types}, but the 
continuity of this characteristic 
is better illustrated in Figure~\ref{continuity}, where de Vaucouleurs
(1963) underline classifications (S$\underline{\rm A}$B and SA$\underline{\rm B}$)
are also shown. An SA galaxy has no evident bar in general,
although high inclination can cause a mistaken SA classification if a
bar is highly foreshortened. Also, internal dust may obscure a
bar (see, e. g., Eskridge et al. 2000). An SB galaxy should have a clear,
well-defined bar. The intermediate bar classification SAB is one of
the hallmarks of the de Vaucouleurs system, and is used to recognize
galaxies having characteristics intermediate between barred and
nonbarred galaxies. It is used for well-defined ovals or
simply weaker-looking normal bars. The weakest primary bars
are denoted S$\underline{\rm A}$B while the classification
SA$\underline{\rm B}$ is usually assigned to more classical bars that
appear only somewhat weaker than conventional bars. Most of the time,
galaxies which should be classified as SA$\underline{\rm B}$
are simply classified as SB.  

Variety is also treated as a continuous classification
parameter (Figure~\ref{continuity}, second row). 
A spiral galaxy having a completely closed or very nearly
completely closed inner ring is denoted (r). The spiral arms usually
break from the ring. If the spiral arms break directly from the central
region or the ends of a bar, 
forming a continuously winding, open pattern, the variety is
(s). The intermediate variety (rs) is also well-defined. Inner rings
which appear broken or partial are in this category. The
``dash-dot-dash in brackets" morphology: (-o-), where a bar with a
bulge is bracketted by spiral arcs overshooting the bar axis, is very
typical of variety (rs). The example of this shown in Figure~\ref{continuity}
is NGC 4548. We use the notation $\underline{\rm r}$s to
denote an inner ring made up of tightly wrapped spiral arms that do not
quite close, while the notation r$\underline{\rm s}$ is used for very open,
barely recognizable, inner pseudorings. A good example of the former is
NGC 3450, while an example of the latter is NGC 5371.

\begin{figure}
\plotone{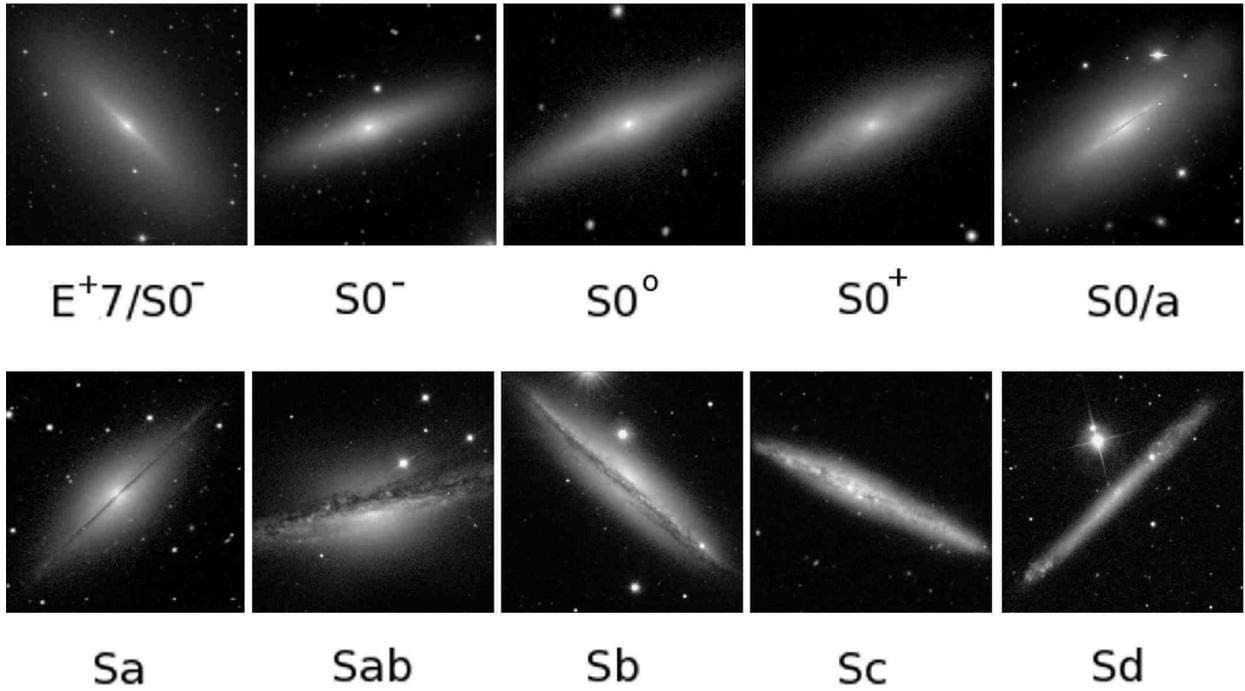}
\caption{Classification of edge-on galaxies by stage. The galaxies are
(left to right):
Row 1 - NGC 3115, 1596, 7332, 4425, and 5866; Row 2 -
NGC 7814, 1055, 4217, 4010, and IC 2233. All images are from the dVA
($B$ and $V$ filters).}
\label{edge-ons}
\end{figure}

A {\it spindle} is a highly inclined disk galaxy. For blue-light
images, usually an ``sp" after the classification 
automatically implies considerable uncertainty in the interpretation,
because family and variety are not easily distinguished when
the inclination is high. Figure~\ref{edge-ons} shows, however, that stages
can be judged reasonably reliably for edge-on galaxies. One important
development in the classification of edge-on galaxies has been the ability
to recognize edge-on bars through boxy/peanut and ``X"-shapes.
Boxy/peanut bulges in edge-on galaxies were proven to be bars
seen edge-on from kinematic considerations (e. g., 
Kuijken \& Merrifield 1995). This shape
is evident in NGC 4425 (Row 1, column 4 of Figure~\ref{edge-ons}; see
also section 9).

For spiral and S0 galaxies that are not too highly inclined (i.e., not
spindles), once the stage, family, and variety are determined these are
combined in the order family, variety, stage for a final full type. For
example, NGC 1300 is of the family SB, variety (s), and stage b, thus
its full type is SB(s)b. The S0$^+$ galaxy NGC 4340 has both a bar and
inner ring and its full type is SB(r)0$^+$.  The classification is
flexible enough that if, for example, the family and variety of a
galaxy cannot be reliably determined owing to high inclination, while
the stage can still be assessed, then the symbols can be dropped and a
type such as ``Sb" or ``S0" can still be noted.

\subsection{Irregular Galaxies}

\begin{figure}
\plotone{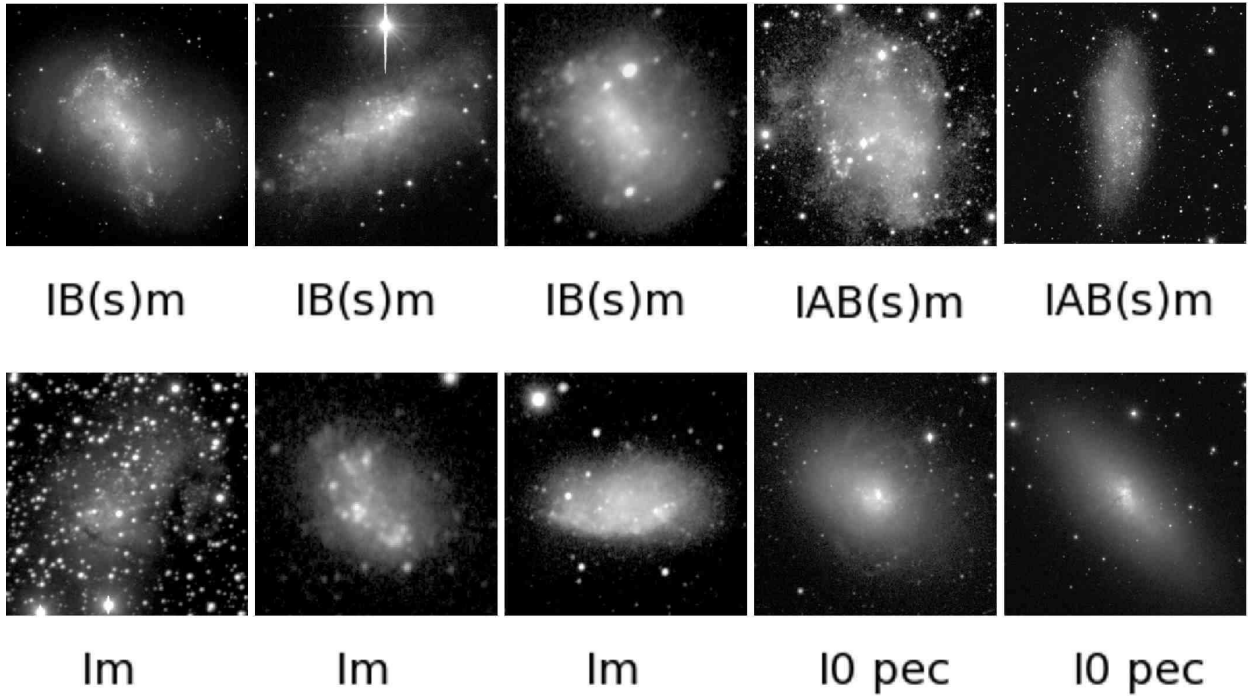}
\caption{Examples of irregular galaxies ranging in absolute blue magnitude from
$-$14 to $-$18. The galaxies are (left to right): Row 1 - NGC 4449, 1569, 1156, DDO 50,
and A2359-15 (WLM galaxy); Row 2 - IC 10, DDO 155, DDO 165, NGC 1705,
NGC 5253. The ``pec" stands for peculiar. All images are $B$-band
from the dVA.}
\label{irregulars}
\end{figure}

Magellanic irregular galaxies represent the last normal stage of the
de Vaucouleurs revised Hubble sequence. Several examples are shown in
Figure~\ref{irregulars}. The objects illustrated in the top row are
all examples of (s)-variety irregulars with bars or some trace of a
bar. Nevertheless, not all Magellanic irregulars have bars. Irregulars 
of the lowest luminosities are usually classified simply as Im since
the sophistication of structure needed to distinguish something like
``family" may not exist for such galaxies.

Irregular galaxies are important for their star formation characteristics.
As noted by Hunter (1997), irregulars are similar to spirals in having
both old and young stars, as well as dust, atomic, molecular, and ionized
gas, but lack the spiral structure that might trigger star formation. 
Thus, they are useful laboratories for examining how star formation occurs
in the absence of spiral arms.

Although irregulars are largely defined by a lack of well-organized
structure like spiral arms, the two lower right galaxies in 
Figure~\ref{irregulars} are not so disorganized looking and seem different
from the other cases shown. NGC 5253 looks almost like a tilted S0 
galaxy, yet it has no bulge at its center nor any obvious lenses. Instead,
the central area is an irregular zone of active star formation. The
central zone was interpreted by van den Bergh (1980a) as ``fossil evidence"
for a burst of star formation, possibly triggered by an interaction with
neighboring M83. This is a case where the de Vaucouleurs classification
of I0 seems reasonable: NGC 5253 is an early-type galaxy with a central
starburst, probably the youngest and closest example known (Vanzi \& Sauvage
2004). It is a Magellanic irregular galaxy imbedded in a smooth S0-like
background known to have an early-type star spectrum. 
NGC 1705, also shown in Figure~\ref{irregulars}, is similar but
has a super star cluster near the center and obvious peculiar filaments.
It is classified as a blue compact dwarf by Gil de Paz  et al. (2003).
Both galaxies are classified as Amorphous by Sandage \& Bedke (1994).

\section{Other Dimensions to Galaxy Morphology}

The de Vaucouleurs classification volume recognizes three principal
aspects of galaxy morphology, but clearly there are many more dimensions
than three. Stage, family, and variety are the dimensions
most clearly highlighted in blue light images and have a wide scope.
Other dimensions may be considered and for some there is explicit
notation in use. 

\subsection{Outer Rings and Pseudorings}

\begin{figure}
\plotone{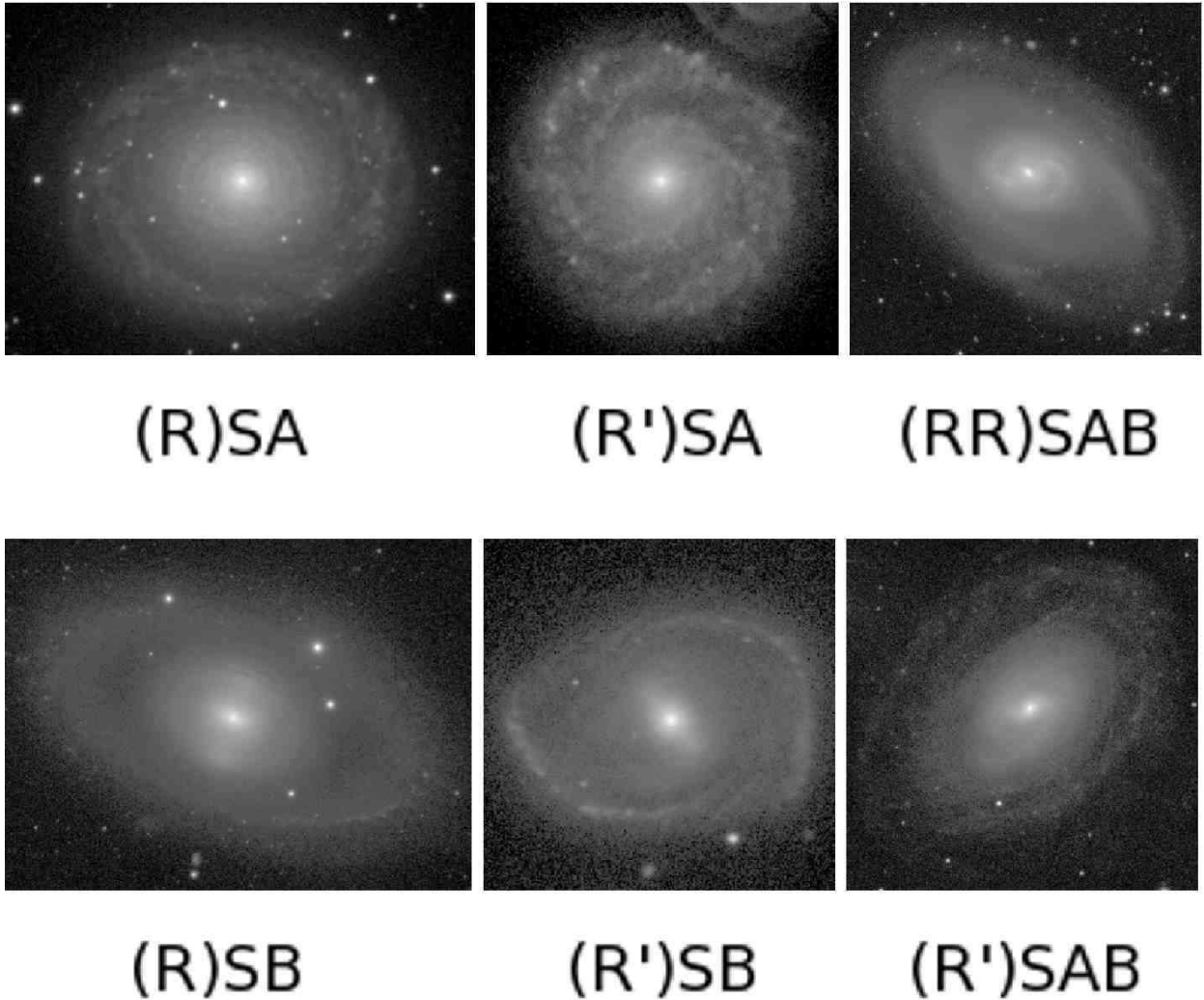}
\caption{Examples of outer rings (R) and outer pseudorings (R$^{\prime}$)
in barred and nonbarred galaxies. Also shown is a rare example with two
largely detached outer rings (RR). The galaxies are (left to right): Row 1 -
NGC 7217, IC 1993, and NGC 2273; Row 2 - NGC 3945, NGC 1358,
and NGC 1371. All images are from the dVA and are $B$-band except for
NGC 2273, which is $r$-band.}
\label{outer_rings}
\end{figure}

Published de Vaucouleurs types include an extra dimension known as the outer
ring/pseudoring classification. 
Several examples of outer rings and pseudorings are shown
in Figure~\ref{outer_rings}. 
An outer ring is a large, often diffuse
structure, typically seen in barred early-type galaxies (stages S0$^+$
to Sa) at a radius approximately twice that of the bar. 
Closed outer rings are recognized
with the type symbol (R) preceding the main part of the classification.
For example, an SB(r)0$^+$ galaxy having an outer ring has a full
classification of (R)SB(r)0$^+$. Interestingly, rare cases of 
double outer ring galaxies, type (RR), are known, where
two detached outer rings are seen; an example 
is NGC 2273 shown in the upper left frame of Figure~\ref{outer_rings}.

In later-type galaxies, a large outer ring-like feature is often seen made of
outer spiral arms whose variable pitch angle causes them to close
together. These features are classified as outer pseudorings,
symbolized by (R$^{\prime}$) preceding
the main type symbols [e.g., as in (R$^{\prime}$)SB(r)ab]. Outer
pseudorings are mainly observed in Sa to Sbc galaxies, and are only
rarely seen in the very late stages Sc-Sm.

Among bright nearby galaxies, outer rings and pseudorings are found at about
the 10\% level (Buta \& Combes 1996). Typically, outer rings
are fainter than 24 mag arcsec$^{-2}$ in blue light. With such low surface
brightnesses, the rings can be easily lost to Galactic extinction.
The division between outer rings and pseudorings is also not sharp.
Some outer pseudorings are only barely distinguishable from outer
rings. Continuity applies to these features as it does for
inner rings although there is no symbol other than ``S" for an
outer spiral pattern which does not close into an outer
pseudoring.

\begin{figure}
\plotone{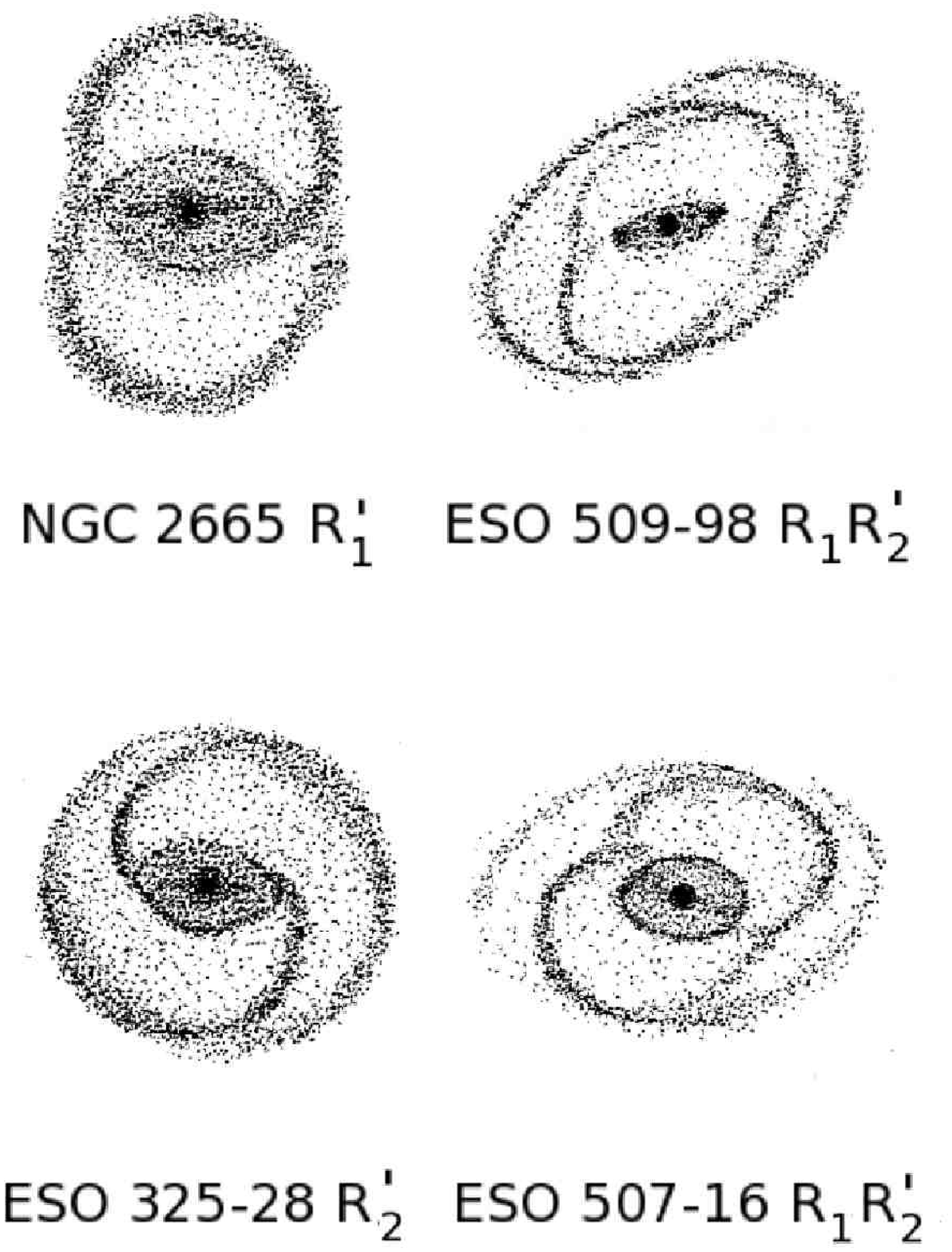}
\caption{Schematic representations of outer Lindblad resonance (OLR)
morphologies (Buta \& Combes 1996).}
\label{olr_schematic}
\end{figure}

\begin{figure}
\plotone{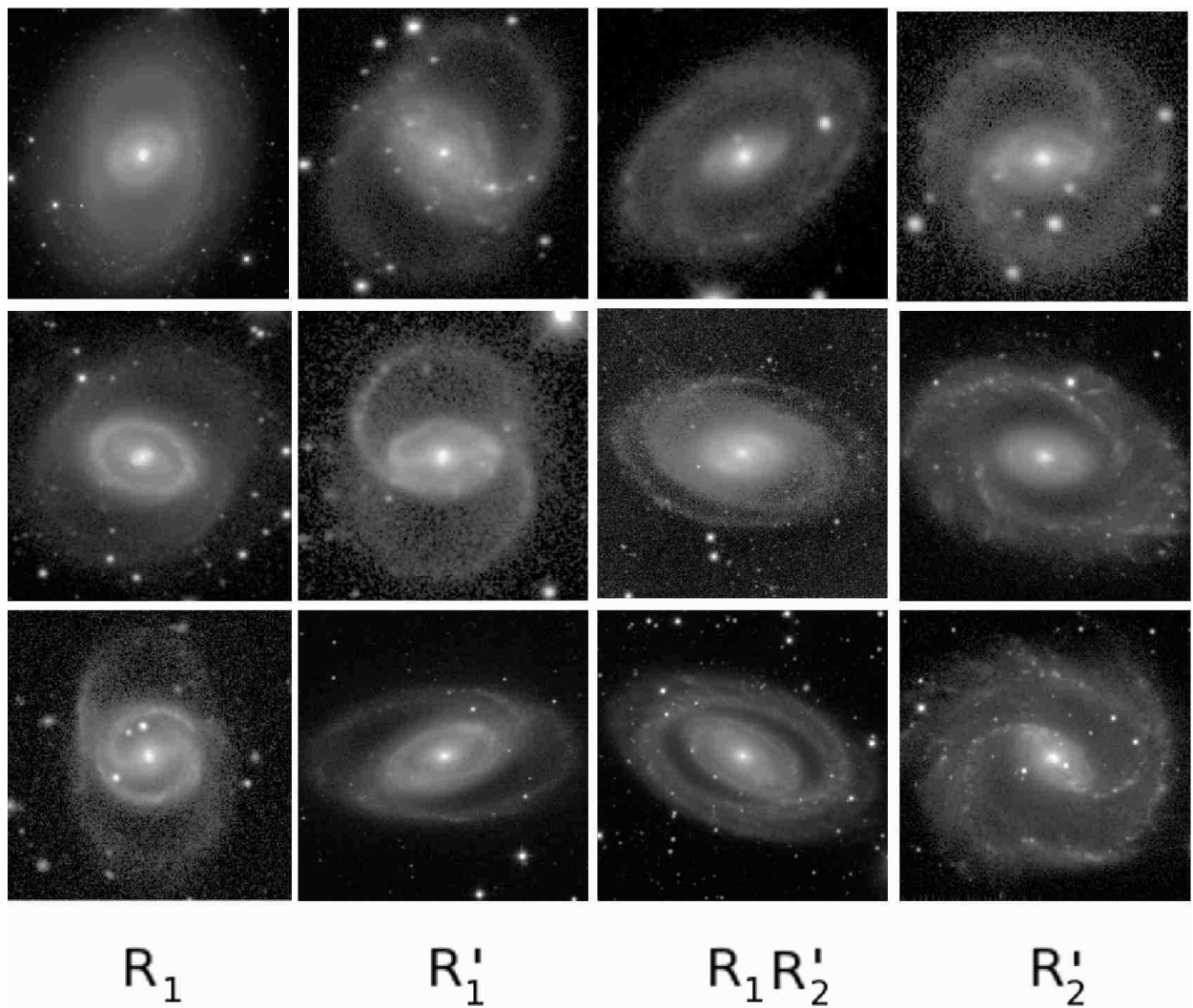}
\caption{Examples of OLR subclasses of outer rings and pseudorings.
The galaxies are (left to right):
Row 1 - NGC 1326, NGC 2665, ESO 509$-$98, and ESO 325$-$28;
Row 2 - NGC 3081, UGC 12646, NGC 1079, and NGC 210;
Row 3 - NGC 5945, 1350, 7098, and 2935. All images are $B$-band from the dVA.}
\label{olr_subclasses}
\end{figure}

Although closed outer rings (R) are equally well-recognized in the RSA
and the Carnegie and Hubble Atlases, outer pseudorings are a unique
feature of the de Vaucouleurs revision. The value of recognizing these
features is that many show morphologies consistent with the theoretical
expectations of the outer Lindblad resonance (OLR, Schwarz 1981), one
of the major low-order resonances that can play a role in disk
evolution.  Resonance rings are discussed further in section 10.1, but
Figure~\ref{olr_schematic} shows schematics of the morphologies
generally linked to this resonance. The schematics are designed to
highlight the subtle but well-defined aspects of these features, while
Figure~\ref{olr_subclasses} shows images of several examples of each
morphology, including the ``models" used for the schematic.  Outer
rings of type R$_1$ are closed rings that are slightly dimpled towards
the bar axis, a shape which connects directly to one of the main
periodic orbit families near the OLR as shown in Schwarz (1981) and in
the dVA.  Outer pseudorings of type R$_1^{\prime}$ are similar to type
R$_1$ but are made of two spiral arms that wind approximately 180$^o$
with respect to the ends of the bar. Even these will usually show a
dimpled shape. Outer pseudorings of type R$_2^{\prime}$ are different
from this in that two spiral arms wind 270$^o$ with respect to the ends
of the bar, such that the arms are doubled in the quadrants immediately
trailing the bar.

The shapes R$_1$, R$_1^{\prime}$, and R$_2^{\prime}$ were predicted by
Schwarz (1981) based on ``sticky-particle" numerical simulations. Not
predicted by those simulations (but later shown in extensions of those
simulations by Byrd et al. 1994 and Rautiainen \& Salo 2000) is an
interesting combined ring morphology called R$_1$R$_2^{\prime}$, which
consists of a closed R$_1$ ring and an R$_2^{\prime}$ pseudoring. This
combination is especially important because it demonstrates not only a
continuity of morphologies among outer rings and pseudorings different
from the continuity between outer rings and pseudorings in general, but
also it is a morphology that can be linked directly to the dynamics of
barred galaxies.

Note that the classification shown in Figures~\ref{olr_schematic} and
~\ref{olr_subclasses} does not depend on whether the rings are in fact
linked to the OLR. The illustrated morphologies are abundant and easily
recognized regardless of how they are interpreted. (Section 10.1
discusses other interpretations that have been proposed.) Although the
Schwarz models guided the search for these morphologies, Rautiainen,
Salo, \& Buta (2004) and Treuthardt et al. (2008) showed that some
outer pseudorings classified as R$_1^{\prime}$ are more likely related
to the outer 4:1 resonance and not the OLR. These cases are generally
recognizable by the presence of secondary spiral arcs in a four-armed
pattern in the area of the bar (NGC 1433 in Figure~\ref{difftilts} and
ESO 566$-$24 in Figure~\ref{spiraltypes} are examples).

The OLR subclassifications are used in the same manner as the
plain outer ring and pseudoring classifications. For example,
NGC 3081 has the full type (R$_1$R$_2^{\prime}$)SAB(r)0/a. 

\subsection{Inner and Outer Lenses}

\begin{figure}
\vspace{-1.0truein}
\plotone{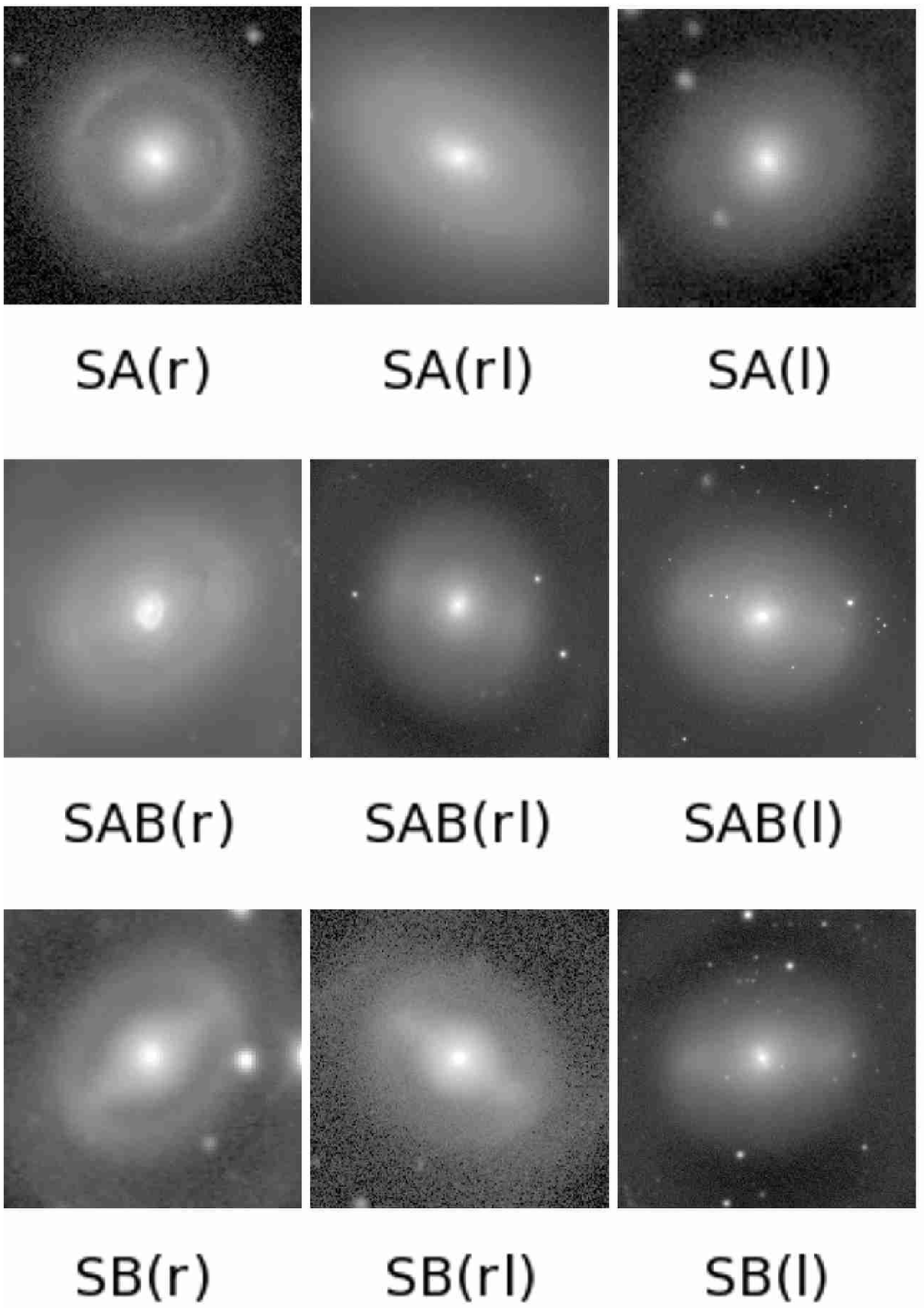}
\caption{}
\label{lenses}
\end{figure}

The value of recognizing lenses as significant morphological components
was first emphasized by Kormendy (1979), who suggested a dynamical
link between inner lenses, which are often filled by a bar in one
dimension, and dissolved or dissolving bars. Kormendy noted that
lenses can be of the inner or outer type, in a manner
analogous to inner and outer rings. He suggested the notation
(l) for inner lenses and (L) for outer lenses to be used in
the same position of the classification as inner and outer
rings. For example, the galaxy NGC 1543 is type (R)SB(l)0/a
while galaxy NGC 2983 is type (L)SB(s)0$^+$. Figure~\ref{lenses}
demonstrates the continuity between rings and lenses, which is
evident not only among barred galaxies but among nonbarred ones
as well. This continuity is recognized by the type symbol (rl),
also used by Kormendy (1979). This type refers to a low
contrast inner ring at the edge of a clear lens. Even underline
classifications ($\underline{\rm r}$l) and (r$\underline{\rm l}$) may 
be recognized.  A rare classification, (r$^{\prime}$l),
refers to an inner pseudoring/lens, a type of feature that 
is seen in NGC 4314 and recognized as such in the dVA.

Similarly, Figure~\ref{outer_lenses} shows a continuity between outer
rings and outer lenses through the type classification (RL), referring
to an outer lens with a weak ring-like enhancement. Underline types
$\underline{\rm R}$L and R$\underline{\rm L}$ may also be recognized.
The origin of outer lenses could be in highly evolved outer rings.

\subsection{Nuclear Rings and Bars}

The central regions of barred galaxies often include distinct
morphological features in the form of small rings and secondary
bars. The rings, known as nuclear rings because of their proximity
to the nucleus well inside the ends of the primary bar, are
sites of some of the most spectacular starbursts known in normal
galaxies. The rings are typically $\approx$1.5 kpc in linear
diameter and intrinsically circular in shape. Figure~\ref{nrnb}
(top row) shows three examples: NGC 1097, 3351, and 4314. These
images highlight the small bulges that seem characteristic of
nuclear-ringed barred galaxies. The three galaxies illustrated
have types ranging from Sa to Sb, but based on the bulge size
the types would be considerably later. For example, NGC 3351 has
the bulge of an Sd galaxy.

Comer\'on et al. (2010) carried out an extensive statistical study
of nuclear ring radii, and identified a subclass known as ``ultra-compact"
nuclear rings (UCNRs). Such rings were recognized mainly in Hubble
Space Telescope images and are defined to be less than 200pc in diameter.
(See Figure~\ref{bulges} for an example in NGC 3177.)
Comer\'on et al. showed that UCNRs are the low size tail of the
global nuclear ring population. This study also showed that bar
strength impacts the sizes of nuclear rings, with stronger bars 
generally hosting smaller nuclear rings than weaker bars.

Comer\'on et al. (2010) were also able to derive a reliable estimate of
the relative frequency of nuclear rings as 20\%$\pm$2\% over the type
range S0$^-$ to Sd, confirming with smaller error bars the previous
result of Knapen (2005). Assuming that nuclear rings are a normal part
of galaxy evolution, these authors argue that the rings may survive for
2-3 Gyr. Interestingly, it was also found that 19\%$\pm$4\% of nuclear
rings occur in nonbarred galaxies, implying either that the rings may
have formed when a bar was stronger (evidence of bar evolution) or that
ovals or other mechanisms can lead to their formation. Mazzuca et al.
(2009; see also Knapen 2010) connect some of the properties of nuclear
rings to the rate at which the rotation curve rises in the inner
regions.

The most extreme nuclear ring known is found in the SBa galaxy ESO
565$-$11 (see also section 7). At 3.5kpc in diameter, not only is it one of
the largest known nuclear rings, but also the ring has an extreme
elongated shape compared to more typical nuclear rings.

Nuclear bars lie in the same radial zone as nuclear rings and sometimes
lie inside a nuclear ring. Three examples are shown in the second row
of Figure~\ref{nrnb}. These average about one-tenth the size of a
primary bar. There is no preferred angle between the axis of the
nuclear bar and the primary bar, suggesting that the two features have
different pattern speeds (Buta \& Combes 1996; dVA).

Neither nuclear rings nor nuclear bars were recognized in the original
Hubble-Sandage-de Vaucouleurs classifications, presumably in part
because the use of small-scale photographic plates for extensive galaxy
classification limited the detectability of the features in the
(typically) overexposed centers. Modern multi-band digital imaging
greatly facilitates the detection of the small rings and bars, allowing
their inclusion in the classification. Buta \& Combes (1996) and Buta
et al. (2010a) suggested the notation nr for nuclear rings and
nb\footnote{In a study of galactic nuclei, van den Bergh (1995)
proposed the notation ``NB" for nuclear bars, although what he refers
to are not the same as the features described here.} for nuclear bars,
respectively, to be used as part of the variety classification as in,
for example, SB(r,nr)b for NGC 3351, or SAB(l,nb)0/a for NGC 1291.
Continuity may exist for these features like other rings and primary
bars.  [For example, nuclear lenses (nl) may also be recognized.] In
blue light images, the appearance of the central region of a barred
galaxy can be strongly affected by dust.  For example, NGC 1365 shows a
nuclear spiral in blue light, while in the infrared, the morphology is
that of a nuclear ring (Buta et al.  2010a). The morphologies of some
galaxies have a full complement of classifiable features. For example,
accounting for all the rings and bars seen in NGC 3081, the
classification is (R$_1$R$_2^{\prime}$)SAB(r,nr,nb)0/a.

Lisker et al. (2006) use the terminology ``S2B" for double-barred galaxies,
a reasonable alternative approach to classifying these objects.
Lisker et al. successfully identified nuclear bars in galaxies
at redshifts $z$=0.10-0.15 (from HST ACS observations), the most distant 
ones recognized thus far.

\subsection{Spiral Arm Morphologies}

A classification such as ``Sb" tells one that a galaxy is a spiral
of moderate pitch angle and degree of resolution of the arms, and that
a significant bulge may be present. The type does not directly tell:
(1) the multiplicity of the spiral pattern; (2) the character of
the arms (massive, filamentary, grand design, or flocculent); 
or (3) the sense of winding
of the arms (leading or trailing the direction of rotation). These
are nevertheless additional dimensions to galaxy morphology. 

The multiplicity of the spiral pattern refers to the actual number
of spiral arms, usually denoted by the integer $m$. Examples of
spirals having $m$=1 to 5 are illustrated in Figure~\ref{spiraltypes}.
The multiplicity is not necessarily straightforward to determine and
may be a function of radius. For example, a spiral may be two-armed
in the inner regions and multi-armed in the outer regions. Spirals
of low $m$ are usually {\it grand design}, a term 
referring to a well-defined global (meaning galaxy-wide)
pattern of strong arms. The typical
grand design spiral has two main arms, as in NGC 5364 (lower left 
frame of Figure~\ref{spiraltypes}). In contrast, a flocculent
spiral has piecewise continuous arms but no coherent global pattern
(Elmegreen 1981). NGC
5055 is an example shown in the middle left frame of Figure~\ref{spiraltypes}.
This category is relevant mainly to optical wavebands. In the infrared,
an optically flocculent spiral like NGC 5055 reveals a more coherent global
grand design spiral (Thornley 1996; see also Figure~\ref{nir03}), 
indicating that dust is partly responsible for the flocculent appearance.

The terms ``massive" and ``filamentary" arms are due to Reynolds (1927)
and are discussed by Sandage (1961, 1975).
Massive arms are broad, diffuse, and of relatively low contrast, as in
M33, while filamentary arms are relatively thin in comparison, and lined
by knots or filaments, as in NGC 5457 (M101). De Vaucouleurs (1956)
originally used these distinctions as part of his classification, but
later dropped the references to spiral arm character probably because
of the complexity it added to his types.

Elmegreen \& Elmegreen (1987) used a different approach to spiral arm
character by recognizing a series of spiral {\it arm classes} based on
arm continuity and length (but not necessarily contrast). Ten classes
ranging from flocculent (ACs 1-4) to grand design (5-12; numbers 10 and
11 were later dropped). Examples of each are illustrated in
Figure~\ref{arm_classes} (see Elmegreen \& Elmegreen 1987 for a
description of each class).  Thus, spiral character is a well-developed
additional dimension to galaxy classification.  A simpler approach
advocated by Elmegreen \& Elmegreen is ``G" for grand-design, ``F" for
flocculent, and ``M" for multiple-armed. The arms of grand design
spirals are in general thought to be density waves and may in fact
represent quasi-steady wave modes (e.g., Bertin et al. 1989; Zhang
1996, 1998, 1999), although there is also some evidence that spirals may be
transient (see review by Sellwood 2010). Flocculent spirals may be
sheared self-propogating star formation regions (Seiden \& Gerola
1982).

Figure~\ref{spiraltypes} also shows two examples of a new class of spirals,
called {\it counter-winding} spirals. In these cases, an inner spiral
pattern winds outward in the opposite sense to an outer spiral pattern.
In the case of NGC 4622 (row 2, middle), the inner pattern has only
a single arm and the outer pattern has two arms, while in NGC 3124
(row 2, middle right), the inner pattern is two-armed while the outer
pattern is at least four-armed.
The two cases are very different because NGC 4622 is essentially nonbarred
while in NGC 3124, the inner spiral is classified as a bar. The
presence of oppositely winding spiral patterns in the same galaxy
means that one set of arms is trailing (opening opposite the direction
of rotation) while the other set is
leading (opening into the direction of rotation). 
In general, studies of the dust distribution as well as the
rotation of spirals has shown that trailing arms are the rule (de
Vaucouleurs 1958). Surprisingly, straightforward analysis of a velocity
field and the dust pattern in NGC 4622 led Buta, Byrd, \& Freeman
(2003) to conclude that the strong outer two-armed pattern in this galaxy
is leading, while the inner single arm is trailing. This led to
the characterization of NGC 4622 as a ``backwards spiral galaxy,"
apparently rotating the wrong way.
An additional nonbarred counter-winding spiral has been identified
in ESO 297$-$27 by Grouchy et al. (2008). In this case, the same
kind of analysis showed that an inner single arm leads while a
3-armed outer pattern trails. No comparable analysis has yet been
made for NGC 3124.

Vaisanen et al. (2008) have shown that a two-armed (but not
counter-winding) spiral in the strongly interacting galaxy IRAS
18293-3413 is leading.  Even with this, the number of known leading
spirals is very small (dVA). Leading spirals are not expected to be as
long-lived as trailing spirals since they do not transfer angular
momentum outwards and this is needed for the long-term maintenance of a
spiral wave (Lynden-Bell \& Kalnajs 1972).

An interesting example of leading ``armlets" was described by Knapen et
al. (1995a), who used $K$-band imaging of the center of the grand
design spiral M100 to reveal details of its nuclear bar and well-known
nuclear ring/spiral. The nuclear bar has a leading twist that connects
it to two bright $K$-band ``knots" of star formation. This morphology
was interpreted in terms of the expectations of gas orbits in the vicinity
of an inner inner Lindblad resonance (IILR; Knapen et al. 1995b).

The final galaxy in Figure~\ref{spiraltypes} is NGC 4921, an example of
an {\it anemic} spiral.  This is a type of spiral that is deficient in
neutral atomic hydrogen gas, and as a consequence it has a lower amount
of dust and star formation activity. The arms of NGC 4921 resemble
those of an Sb or Sbc galaxy in pitch angle and extensiveness, but are
as smooth as those typically seen in Sa galaxies. Anemic spirals were
first recognized as galaxies with ``fuzzy" arms (see van den Bergh
1998) where star formation has been suppressed due to ram-pressure
stripping in the cluster environment. In the case of NGC 4921, the
environment is the Coma Cluster. The idea is that such galaxies will
eventually turn into S0 galaxies (van den Bergh 2009a). Anemic spiral
galaxies are further discussed in section 10.2.

Seigar et al. (2008) have demonstrated the existence of a correlation
between spiral arm pitch angles and supermassive central black hole
masses. The sense of the correlation is such that black hole mass is
highest for the most tightly wound spirals and lowest for the most open
spirals. The correlation is expected because black hole mass is tightly
correlated to bulge mass and central mass concentration, and spiral arm
pitch angle is tied to shear in galactic disks, which itself depends on
mass concentration (Seigar et al. 2005).

\subsection{Luminosity Effects}

Luminosity effects are evident in the morphology of galaxies through surface
brightness differences between giants and dwarfs, and through the 
sophistication of structure such as spiral arms. van den Bergh (1998)
describes his classification system which takes luminosity effects into
account using a set of luminosity classes that are analogous to those
used for stars. The largest, most massive spirals have long and
well-developed arms, while less massive spirals have less well-defined arms.

The nomenclature for the classes parallels that for stars: I (supergiant 
galaxies), II (bright giant galaxies), 
III (giant galaxies), IV (subgiant galaxies) and V (dwarf galaxies). Intermediate cases I-II, II-III, III-IV, 
and IV-V, are also recognized.

Figure~\ref{lclasses} shows galaxies which van den Bergh (1998) considers
primary luminosity standards of his classification system. 
The original van den Bergh standards for these classes were based on the small scale paper prints of the 
Palomar Sky Survey. Sandage and Tammann (1981) adopted the precepts of the van den Bergh classes but 
revised the standards based on large-scale plates. 
In general, luminosity class I galaxies have the longest, most well-developed 
arms, luminosity class III galaxies have short, patchy arms extending from the 
main body, while luminosity 
class V galaxies have very low surface brightness and only a hint of spiral structure. The classes are
separated by type in Figure~\ref{lclasses} because among Sb galaxies, 
few are of luminosity class III or fainter, while 
among Sc and later type galaxies, the full range of luminosity classes
is found. van den Bergh does not use types like Sd or Sm for conventional
de Vaucouleurs late-types, but instead uses S$^-$ and S$^+$ to denote
``early" (smoother) and ``late" (more patchy) subgiant spirals. Similarly,
van den Bergh uses Sb$^-$ and Sb$^+$ to denote
``early" and ``late"  Sb spirals, respectively. (Some of 
these would be classified as Sab and Sbc by de Vaucouleurs.) 
According to the standards listed by van den Bergh (1998), an Sb I galaxy is 
2-3 mag more luminous than an Sb III galaxy, while an Sc I galaxy is
more than 4 mag more luminous than an S V galaxy.

\section{The Morphology of Galactic Bars and Ovals}

Bars are among the most common morphological features of disk-shaped
galaxies. Unlike spiral arms, bars cross the ``spiral-S0 divide" in the
Hubble sequence and are abundant among spirals
(at the 50-70\% level) when both SAB and SB types are
considered (de Vaucouleurs 1963; Sellwood \& Wilkinson 1993). The
bar fraction has cosmological significance (Sheth et al. 2008), and
many estimates of the nearby galaxy bar fraction have been made from
both optical and IR studies (see Buta et al. 2010a for a summary of
recent work). 

Bars are fairly well-understood features of galaxy morphology that have
been tied to a natural instability in a rotationally supported stellar
disk (see review by Sellwood \& Wilkinson 1993). The long-term maintenance
of a bar in a mostly isolated galaxy is thought to depend on how effectively
it transfers angular momentum to other galaxy components, such as the halo
(Athanassoula 2003). Bars are thought to be transient features that,
in spiral galaxies, may dissolve and regenerate several times over a Hubble
time (Bournaud \& Combes 2002). Alternatively, bars may be long-lived
density wave modes that drive secular evolution of both the stellar
and gaseous distributions (Zhang \& Buta 2007; Buta \& Zhang 2009).
The possible secular evolution of bars in
S0 galaxies is discussed by Buta et al. (2010b).
Bars are also thought to drive spiral density waves (Kormendy \& Norman
1979; Buta et al. 2009; Salo et al. 2010).

The actual morphology of bars shows interesting variations that merit
further study.  The family classifications SAB and SB indicate some
measure of bar strength, but do not allude to the varied appearances of
bars even among those only classified as SB. Regular bars, such as
those illustrated in Figure~\ref{tuning_fork}, are the conventional
types that define the SB class. Figure~\ref{ansae} shows ``ansae"-type
bars, referring to bars which have ``handles" or bright enhancements at
the ends. Martinez-Valpuesta et al. (2007) carried out a statistical
study and found that ansae are present in $\approx$40\% of early-type
barred galaxies and are very rare for types later than stage Sb. Ansae
are usually detectable in direct images, but their visibility can be
enhanced using unsharp-masking (all the right frames for each galaxy in
Figure~\ref{ansae}).  Morphologically, ansae may be small round
enhancements like those seen in NGC 5375 and 7020, but in some cases,
ansae are approximately linear enhancements, giving the bar a parallelogram
appearance as in NGC 7098, or curved arcs, giving the bar a partial
ring appearance as in NGC 1079. Color index maps in the dVA show that
ansae are generally as red as the rest of the bar, indicating the
features are stellar dynamical in origin, rather than gas-dynamical.
Nevertheless, ansae made of star-forming regions are known.
Martinez-Valpuesta et al.  (2007) illustrate the case of NGC 4151, a
well-known Seyfert 1 galaxy with a strong bar-like inner oval. The
appearance of this galaxy's ansae in the 1.65$\mu$m $H$-band is shown
in the lower right frames of Figure~\ref{ansae}, where the ansae are
seen to have irregular shapes compared to the others shown.

Another subtlety about bars is their general boxy character. Athanassoula
et al. (1990) showed that generalized ellipses fit the projected isophotes of
bars better than do normal ellipses. For 11 or 12 SB0 galaxies examined
in this study, a significant degree of boxiness was found near the bar
semi-major axis radius.

The unsharp-masked image of NGC 7020 in Figure~\ref{ansae} shows an X-shaped
pattern in the inner regions that is the likely signature of a significantly
three-dimensional bar. NGC 1079 and 5375 shows hints of similar structure.
The X-pattern is expected to be especially evident in edge-on galaxies
which show the extended vertical structure of bars. Many examples have
been analyzed (Bureau et al. 2006; see also the dVA).
Buta et al. (2010a) have suggested that edge-on bars recognized from the
X-pattern be denoted ``SB$_x$."

The cause of bar ansae is uncertain. In simulations, Martinez-Valpuesta
et al. (2006) found that ansae form late, after a second bar-buckling episode
in a disk model with a live axisymmetric halo,
and appear as density enhancements in both the face-on and edge-on
views. 

Figure~\ref{ovals} shows three examples of galaxies having oval inner
disks. These features are described in detail by Kormendy and
Kennicutt (2004=KK04), who present both photometric and kinematic criteria
for recognizing them. The images in Figure~\ref{ovals} are optical and
have been cleaned of foreground and background objects, and have also
been deprojected based on the mean shape and major axis position angle of
faint outer isophotes. In all three cases, the presence of an outer
ring allows clear recognition of the oval shape, assuming that the inner and
outer structures are in the same plane. The upper panels of Figure~\ref{ovals}
show the full morpholopy with the outer rings, while the lower panels
show the bright oval inner disks. The shapes of the oval disks are
varied and range from axis ratio 0.84 for NGC 4736 to 0.55 for NGC 1808.
The most striking example is NGC 4941, whose oval disk includes a 
bright, normal-looking spiral pattern with isophotal axis ratio 0.68.
Many other examples are provided by KK04.

The ovals appear to play a bar-like role in these galaxies. The outer
rings may be resonant responses to the nonaxisymmetric potential of the
ovals, which clearly harbor a great deal of mass in spite of the
mildness of their departures from axisymmetry. As noted by KK04, oval
disk galaxies are expected to evolve secularly in much the same manner
as typical barred galaxies. On the other hand, ovals themselves could
be products of bar secular evolution. Laurikainen et al. (2009) found
that the near-IR bar fraction in S0 galaxies is significantly less than
that in S0/a or early-type spiral galaxies (also found by Aguerri et
al. 2009 in the optical), while the oval/lens fraction is higher,
suggesting that some ovals/lenses might be dissolved bars. Further
evidence that bars might be dissolving in some galaxies is the
detection of extremely weak bars in residual images of visually
nonbarred S0 galaxies where a two-dimensional decomposition model has
been subtracted. Such a bar is detected in the SA0$^o$ galaxy NGC 3998
(Laurikainen et al. 2009). Aguerri et al. (2009) suggest that central
concentration is a key factor in bar evolution, and that a unimodal
distribution of bar strengths argues against the idea that bars
dissolve and reform (Bournaud \& Combes 2002).

Regular barred galaxies also often include an oval bounded by an
elongated inner ring. The deprojected blue light images of two examples
are shown in Figure~\ref{inner_rings}. NGC 1433 has a very strong
normal bar and one of the most intrinsically elongated inner rings
known. The inner ring lies at the edge of an oval which is more conspicuous
at longer wavelengths. In NGC 1433, the inner ring, the oval, and the
bar are all aligned parallel to each other. The situation is different 
in ESO 565$-$11, whose bright oval is strongly misaligned with a 
prominent bar, but similar to NGC 1433, the oval is bounded by an
inner pseudoring. The suggestion in this case is that the bar and
the oval are independent patterns.

\section{Dust Morphologies}

Dust lanes are often the most prominent part of the interstellar
medium detectable in an optical image of a galaxy. 
Figure~\ref{dust_types} shows different classes of
dust lanes, using direct optical images on the left for each galaxy,
and a color index map on the right. The color index maps are coded
such that blue features are dark while red features (like dust lanes)
are light. 

The bars of intermediate (mainly Sab to Sbc) spirals
often show {\it leading} dust lanes, that is, well-defined lanes
that lie on the leading edges of the bars, assuming the spiral arms
trail. The example shown in Figure~\ref{dust_types}, NGC 1530, has
an exceptionally strong bar and the lanes are very straight,
regular, and well-defined. These dust lanes are a dynamical effect
associated with the bar. The lanes may be curved or straight.
Athanassoula (1992) derived models of bar dust lanes and tied
the curvature to the strength of the bar in the sense 
that models with stronger bars
developed straighter dust lanes. Comer\'on et al. (2009) recently
tested this idea by measuring the curvature of actual dust lanes
as well as quantitative values of the bar strengths for 55
galaxies. They found
that strong bars can only have straight dust lanes, while weaker
bars can have straight or curved lanes.

In the same manner as bars, a strong spiral often has dust lanes on the
concave sides of the inner arms.  This is shown for NGC 1566 in the
upper right panels of Figure~\ref{dust_types}.  Both bar and spiral
dust lanes are face-on patterns. Another type of face-on pattern is the
{\it dust ring}. The inner dust ring of NGC 7217 is shown in the right,
middle frames of Figure~\ref{dust_types}, and it appears as the dark,
inner edge of a stellar ring having the same shape. Dust rings can also
be detected in more inclined galaxies.

An inclined galaxy with a significant bulge also can show another
dust effect: in such a case, the bulge is viewed through the dust
layer on the near side of the disk, while the dust is viewed through
the bulge on the far side of the disk. This leads to a reddening
and extinction asymmetry across the minor axis such that the near
side of the minor axis is more reddened and extinguished than the
far side. In conjunction with rotation data, this near side/far 
side asymmetry was used by Hubble (1943) and de Vaucouleurs (1958)
to show that most spirals trail the direction of rotation.

The lower frames of Figure~\ref{dust_types} show the planar
dust lanes seen in edge-on spiral galaxies. The lane in NGC 7814
(lower left frames) is red which indicates that the galaxy is
probably no later in type than Sa. This is consistent with the
large bulge seen in the galaxy. However, the planar dust lane
seen in NGC 891, type Sb, has a thin blue section in the middle
of a wider red section. The blue color is likely due to outer
star formation that suffers relatively low extinction. Some
individual star forming regions can be seen along the dust lane.
In spite of the blue color, we are only seeing the outer edge
of the disk in the plane. 

Also related to galactic dust distributions are observations of
{\it occulting galaxy pairs}, where a foreground spiral galaxy
partly occults a background galaxy, ideally an elliptical
(White \& Keel 1992). With such pairs, one can estimate the optical
depth of the foreground dust, often in areas where it might not
be seen easily in an isolated spiral. An excellent example is
described by Holwerda et al. (2009), who are able to trace the
dust distribution in an occulting galaxy to 1.5 times than the
standard isophotal radius.

Another way of illustrating the dust distribution in galaxies is
with {\it Spitzer Space Telescope} Infrared Array Camera (IRAC)
images at 8.0$\mu$m wavelength. This is discussed further in 
section 12.

\section{The Morphologies of Galactic Bulges}

A bulge is a very important component of a disk galaxy. In the context
of structure formation in a cold dark matter (CDM) cosmology, bulges
may form by hierarchical merging of disk galaxies, a process thought to
lead to elliptical galaxies if the disks have approximately equal mass.
Bulges formed in this way should, then, resemble elliptical galaxies,
especially for early-type spirals.  The bulges of later-type spirals,
however, can be very different from the expectations of a merger-built
bulge (also known as a ``classical" bulge).  In many cases, the bulge
appears to be made of material associated with the disk.

KK04 reviewed the concept of ``pseudobulges," referring to galaxy
bulges that may have formed by slow secular movement of disk gas to the
central regions (also known as disk-like bulges; Athanassoula 2005).
The main driving agent for movement of the gas is thought to be bars,
which are widespread among spiral galaxies and which exert gravity
torques that can move material by redistributing the angular momentum.
Inside the corotation resonance, where the bar pattern speed equals the
disk rotation rate, gas may be driven into the center to provide the
raw material for building up a pseudobulge. KK04 review the evidence
for such processes and argue that pseudobulges are a strong indication
that secular evolution is an important process in disk-shaped
galaxies.

Figure~\ref{bulges} shows the morphologies of both classical bulges and
pseudobulges.  The four classical bulge galaxies shown in the lower
row, M31, NGC 2841, M81, and M104, have bright smooth centers and no
evidence for spiral structure or star formation. Classical bulges also
tend to have rounder shapes than disks, and can have significant
bulge-to-total luminosity ratios as illustrated by M104. Classical
bulges are also more supported by random motions than by rotation. Many
references to classical bulge studies are given by KK04. Formation
mechanisms of such bulges are discussed in detail by Athanassoula (2005).

The two upper rows of Figure~\ref{bulges} are all pseudobulges as
recognized by KK04. The first row shows HST wide $V$-band (filter
F606W) images of the inner 1-1.3 kpc of four galaxies, NGC 3177, 4030,
5377, and 1353, in the type range Sa-Sbc.  The areas shown account for
much of the rise in surface brightness above the inward extrapolation
of the outer disk light in these galaxies, and would be considered
bulges just on this basis.  The HST images show, however, considerable
spiral structure, dust, small rings, and likely star formation in these
regions, characteristics not expected for a classical bulge. KK04 argue
that instead these are pseudobulges that are highly flattened, have a
projected shape similar to the outer disk light, have approximately
exponential brightness profiles (Sersic index $n$$\approx$1-2), and
have a high ratio of ordered rotation to random motions.  KK04 argue
that a low Sersic index compared to $n$=4 appears to be the hallmark of
these pseudobulges, and a signature of secular evolution.

The second row in Figure~\ref{bulges} shows other kinds of pseudobulges
discussed by KK04. NGC 6782 and 3081 (two left frames)
have secondary bars, and KK04 considered that such features indicate 
the presence of a pseudobulge because bars are always disk features.
In each case, the secondary bar lies inside a nuclear ring. 

The other two galaxies in the second row of Figure~\ref{bulges}, NGC
128 and 1381, are examples of boxy or box-peanut bulges. These features
have been linked to the vertical heating of bars, and if this is what
they actually are, then KK04 argue that boxy and box-peanut bulges are
also examples of pseudobulges.  However, boxy and box/peanut bulges
would {\it not} necessarily be the result of slow movement of gas by
bar torques, and subsequent star formation in the central regions, but
instead would be related to the orbital structure of the bar itself
(Athanassoula 2005).

Recent studies have shown that pseudobulges are the dominant type of
central component in disk galaxies. Although originally thought to be
important only for late-type galaxies, Laurikainen et al. (2007) showed
that pseudobulges are found throughout the Hubble sequence, including
among S0-S0/a galaxies, based on sophisticated two-dimensional 
photometric decompositions. Such galaxies frequently have nuclear bars,
nuclear disks, or nuclear rings. Laurikainen et al. also found that
bulge-to-total ($B/T$) flux ratios are much less than indicated by
earlier studies, especially for early Hubble types, and that the
Sersic index averages $\lesssim$ 2 across all types. The lack of gas
in S0 and S0/a galaxies complicates the interpretation of their
pseudobulges in terms of bar-driven gas flow and subsequent star formation.
Instead, Laurikainen et al. link the pseudobulges in early-type galaxies
to the evolution of bars. Laurikainen et al. (2010) also showed that
S0s can have pseudobulges if they are stripped spirals, without
invoking any bar-induced evolution. 

\section{Effects of Interactions and Mergers}

Galaxy morphology is replete with evidence for gravitational
interactions, ranging from minor, distant encounters, to violent
collisions and major/minor mergers. Many of the most puzzling and
exotic morphologies can be explained by interactions, and even
sublimely normal galaxies, like ordinary ellipticals, have been
connected to catastrophic encounters. Up to 4\% of bright nearby
galaxies are involved in a major interaction (Knapen \& James 2009).
In clusters, other types of interactions may occur, such as gas
stripping and truncation of the star-forming disk. In this section, a
variety of the types of morphologies that may be considered the results
of external interactions are described and illustrated.

\subsection{Normal versus Catastrophic Rings}

The three types of rings described so far, nuclear, inner, and outer
rings, are aspects of the morphology of relatively normal, undisturbed
galaxies. Inner rings and pseudorings are found in more than 50\% of
normal disk galaxies (Buta \& Combes 1996), while outer rings and
pseudorings are found at the 10\% level. The latter rings could be more
frequent because their faintness may cause them to go undetected, which
is less likely to occur for inner rings. As has been noted, nuclear
rings are found at the 20\% level (Comer\'on et al. 2010). The high
abundance of these types of ring features suggests that they are mainly
products of {\it internal} dynamics, and in fact all three ring types
have been interpreted in terms of internal processes in barred
galaxies. The main interpretation of these kinds of rings has been in
terms of orbital resonances with the pattern speed of a bar, oval, or
spiral density wave. Resonances are special places where a bar can
secularly gather gas owing to the properties of periodic orbits (Buta
\& Combes 1996).  Buta (1995) showed that the intrinsic shapes and
relative bar orientations of inner and outer rings and pseudorings
supports the resonance interpretation of the features. Schwarz (1981)
suggested the outer Lindblad resonance for outer rings and pseudorings,
while Schwarz (1984) suggested the inner 4:1 ultraharmonic resonance
for inner rings and pseudorings, and the inner Lindblad resonance for
nuclear rings.  Knapen et al. (1995b) and Buta \& Combes (1996) provide
further insight into these interpretations.

The resonance idea may only be valid in the case of weak
perturbations.  In the presence of a strong perturbation, the concept
of a specific {\it resonance radius} can break down, although the idea
of a broad {\it resonance region} could still hold (Contopoulos 1996).
Regan and Teuben (2003, 2004) argue that nuclear rings and inner rings
are better interpreted in terms of orbit transitions, that is, regions
where periodic orbits transition from one major orbit family to
another, as in the transition from the perpendicularly-aligned $x_2$
family to the bar-aligned $x_1$ family (Contopoulos \& Grosbol 1989).

Normal rings have also been interpreted in terms of ``invariant
manifolds" which emanate from the unstable $L_1$ and $L_2$ Lagrangian
points in the bar potential (Romero-G\'omez et al. 2006, 2007;
Athanassoula et al. 2009a,b). This approach has also had some success
in predicting the shapes and orientations of inner and outer rings,
such as the R$_1$, R$_1^{\prime}$, R$_1$R$_2^{\prime}$, and
R$_2^{\prime}$ morphologies shown in Figure~\ref{olr_schematic} and
Figure~\ref{olr_subclasses}. A morphology called ``rR$_1$", which
includes an  oval inner ring and a figure eight-shaped R$_1$ ring (see
NGC 1326 in Figure~\ref{olr_subclasses}), is especially well-represented
by this kind of model. The manifolds are tubes which guide orbits
escaping the $L_1$ and $L_2$ regions. Note that in this interpretation,
outer rings are not necessarily associated with the OLR (Romero-G\'omez
et al. 2006).

Although the vast majority of the ring-like patterns seen in galaxies
are probably of the resonance/orbit-transition/invariant-manifold
type, other classes of rings are known
that are likely the result of more catastrophic processes, such as
galaxy collisions. Figure~\ref{ringtypes} shows resonance rings in
comparison to three other types: accretion rings, polar rings, and
collisional rings (the latter commonly referred to as ``ring
galaxies"). The three accretion rings shown, in Hoag's Object
(Schweizer et al. 1987), IC 2006 (Schweizer et al. 1989), and NGC 7742
(de Zeeuw et al. 2002) are thought to be made of material from an
accreted satellite galaxy. For IC 2006 and NGC 7742, the evidence for
this is found in {\it counter-rotation}:  the material in the rings
counter-rotates with respect to the material in the rest of the
galaxies. In Hoag's Object and IC 2006, the accreting galaxy is a
normal E system.

Polar rings are also accreted features except that the accreting galaxy
is usually a disk-shaped system, most often an S0
(Whitmore et al. 1990). In these cases, the accreted material
comes in at a high angle to the plane of the disk. The configuration
is most stable if the accretion angle is close to 90$^o$, or over the poles
of the disk system. This limits the ability of differential precession
to cause the ring material to quickly settle into the main disk.
The polar feature can be a ring or simply an inclined and extended
disk. Whitmore et al. (1990) presented an extensive catalogue of
probable and possible polar ring galaxies.

The main example illustrated in Figure~\ref{ringtypes} is NGC 4650A,
where the inner disk component is an S0. Galaxies like NGC 4650A
have generated considerable research because polar rings probe
the shape of the dark halo potential (e. g., Sackett et al. 1994).
The galaxies are also special because the merging
objects have retained their distinct identities, when most mergers
lead to a single object. Brook et al. (2008) link the misaligned disks
of polar ring galaxies to the process of hierarchical structure formation 
in a cold dark matter scenario.

While polar rings are most easily recognizable when both disks are
nearly edge-on to us, cases where one or the other disk is nearly
face-on have also been recognized. One example, ESO 235$-$58 (Buta
\& Crocker 1993)
is shown in the middle-right panel of Figure~\ref{ringtypes}.
In this case, the inner component is almost exactly edge-on and shows
a planar dust lane, and is likely a spiral rather than an S0. 
The ring component is inclined significantly
to the plane of this inner disk but may not be polar. The faint outer
arms in this component caused ESO 235$-$58 to be misclassified as
a late-type barred spiral in RC3. Spiral structure in polar disks
has been shown to be excitable by the potential of the inner disk,
which acts something like a bar (Theis, Sparke, \& Gallagher 2006).

An example where the main disk is seen nearly face-on is NGC 2655
(Sparke et al. 2008). In this case, the polar ring material is seen
as silhouetted dust lanes at an uncharacteristic angle to the inner
isophotes. NGC 2655 also shows evidence of faint shells/ripples,
indicative of a recent merger (section 10.3.3). Sil'chenko \& Afanasiev (2004)
have discussed NGC 2655 and other similar examples of inner polar
rings in terms of the triaxiality of the potential.

Also illustrated in Figure~\ref{ringtypes} is NGC 660, which was
listed as a possible polar ring galaxy by Whitmore et al. (1990).
Like ESO 235$-$58, NGC 660 has an aligned dust lane in its inner
disk component, which thus is likely to be a spiral, not an S0. 
The extraplanar disk is actually far from polar,
being inclined only 55$^o$ (van Driel et al. 1995). A recent study
of massive stars in the ring is given by Karataeva et al. (2004).

Collisional ring galaxies (Arp 1966; Appleton \& Struck-Marcell 1996)
are thought to be cases where a larger
galaxy suffers a head-on collision with a smaller galaxy
down its polar axis. The collision causes an expanding density wave of massive
star formation, and multiple rings are possible. Three examples
are shown in Figure~\ref{ringtypes}. Theys \& 
Spiegel (1976) have discussed various classes of ring galaxies.
Arp 147 (Arp 1966) is an example of 
type ``RE", referring to a sharp elliptical ring with an empty
interior.  The Cartwheel (Higdon 1995) and the Lindsay-Shapley
ring (Arp \& Madore 1987)
are examples of type ``RN", meaning an elliptical ring with an
off-center nucleus. Not shown in Figure~\ref{ringtypes} is a third
category called ``RK", where a single, large knot lies on one side of the
ring, making the system very asymmetric.

Madore, Nelson, \& Petrillo (2009) have published a comprehensive atlas of all
known likely collisional ring galaxies, many taken from the catalogue
of Arp \& Madore (1987).
Based on this study, only 1 in 1000 galaxies is a collisional ring galaxy. 
For entry, the Madore et
al. atlas requires at least two objects in the immediate vicinity of
the ring that might plausibly be the intruder galaxy. Most of the rings
are not in
cluster environments, however. The atlas also brings attention to 
several double-ring collisional systems, which have been predicted by numerical 
simulations (see Struck 2010 for a review). The unusual radial ``spokes"
in the Cartwheel, a feature not seen in any other collisional ring galaxy,
could be related to interactive accretion streams (Struck et al. 1996).

Romano, Mayya, \& Vorobyov (2008) present images of several ring
galaxies that show the pre-collision stellar disk. They also show that
rings are generally delineated by blue knots and that the off-centered
nuclei are usually more yellow in color. In addition, some of the
companion galaxies show diffuse asymmetric outer light suggesting that
they are being stripped.

Figure~\ref{ringtypes} shows that accretion rings can account for some
of the rings seen in nonbarred galaxies. Buta \& Combes (1996) argue
that a bar is an essential element in resonance ring formation. ESO
235$-$58 shows that a polar ring-related system can resemble a ringed,
barred galaxy. The three collisional rings are all very distinctive
from the others.

\subsection{Environmental Effects on Star-Forming Disks}

Galaxy clusters are excellent laboratories for detecting the effects of
environment on galaxy morphology and structure. Frequent mergers and
environmental conversion of spirals into S0s are thought to be at the
heart of the morphology-density relation, where early-type galaxies
dominate cluster cores, and spirals and irregulars are found mainly
in the outer regions (Dressler 1980; van der Wel et al. 2010).

The issue of environmental effects has a direct bearing on how we might
interpret the Hubble sequence. For example, the continuity of galaxy
morphology certainly seems apparent from the discussions in previous
sections of this review. The Hubble sequence E-S0-Sa-Sb-Sc-I appears
physically significant when total colors, mean surface brightnesses,
and HI mass-to-blue light ratios are considered, and the way features
are recognized in the classification systems also favors the
continuity. Morphological continuity does {\it not}, however,
automatically imply that the galaxy types are in fact ordered
correctly. For example, although Hubble placed S0s as a transition type
between elliptical galaxies and spirals, this placement has been
questioned by van den Bergh (1998, 2009a).  Based on a statistical
analysis of types given in the RSA, van den Bergh showed that S0
galaxies are typically 0.8-1.0 mag less luminous than E and Sa
galaxies, implying that S0 galaxies, on the whole, cannot really be
considered intermediate between E and Sa galaxies.\footnote{ In
contrast to van den Bergh's study of RSA S0 galaxies, Laurikainen et
al. (2010) found that the absolute $K_s$-band magnitudes of a
well-defined sample of S0s are similar to those of early-type spirals
in the OSUBSGS sample. The sample was mostly drawn from RC3 and
includes some galaxies classified as ellipticals in RC3 and as S0s in
the RSA (see section 12).} The preponderance of S0 galaxies in clusters
led to the early suggestion (e.g., Spitzer \& Baade 1951; Gunn \&
Gott 1972; Moore et al. 1996) that some type of external environmental
interaction was responsible for stripping a spiral galaxy of its
interstellar medium.  If this actually occurred, then, as suggested by
van den Bergh (2009a), this could imply that S0 galaxies have lost a
substantial fraction of their spiral mass due to interactions.
Alternatively, van den Bergh (2009b) argues that stripping of a lower
luminosity, late-type spiral should be easier than stripping of a
higher luminosity, early-type spiral, which could also account for the
luminosity difference. In an examination of the environment of S0
galaxies, van den Bergh (2009b) found no significant difference in the
average luminosities, flattenings, or distribution of S0 subtypes in
clusters, groups, or the field, indicating that some S0s develop as a
result of internal effects, such as the influence of an active galactic
nucleus.

Barway et al. (2011) noted that lower luminosity S0s have a higher bar
fraction than higher luminosity S0s (83\% versus 17\%), suggesting that
the two groups form in different ways (see also Barway et al. 2007).
These authors suggest that faint S0s are stripped late-type spirals,
which are known to have a high bar fraction (Barraza et al. 2008).

Environmental effects in clusters do not always have to involve drastic
transformations in morphology. Sometimes the effects are more subtle.
Figure~\ref{virgotypes} shows several spiral galaxies that are also
members of the Virgo Cluster. These galaxies highlight processes that
affect the star-forming disk while leaving the older stellar disk
relatively unaffected. NGC 4580 and 4689 are galaxies having a patchy
inner disk and a smooth outer disk, called ``Virgo types" by van den
Bergh et al. (1990). These objects suggest that the environment of such
galaxies has somehow truncated the star-forming disk, with a greater
concentration of truncated disks toward the cluster core.  Similar
results are obtained from observations of the HI gas disks of Virgo
cluster galaxies (e.g., Giovanelli \& Haynes 1985; Cayatte et al.  1994;
Chung et al. 2009).

Koopmann \& Kenney (2004) summarize the results of an extensive survey
of H$\alpha$ emission from Virgo Cluster galaxies, and identify
different categories of environmentally-influenced star formation
characteristics based on H$\alpha$ imaging. The blue light images of
examples of each category are included in Figure~\ref{virgotypes}, and
show how the subtleties are manifested in regular morphology. Using a
sample of isolated spiral galaxies to define ``normal" star formation,
Koopmann and Kenney defined several categories of Virgo spiral galaxy
star-forming disks: Category ``N" refers to disks whose star formation
is within a factor of three of the normal levels. ``E" cases have star
formation enhanced by more than a factor of three compared to normal.
``A" cases are ``anemic" spirals (section 6.4) having star formation
reduced by more than a factor of 3 compared to normal.  ``T/N" refers
to galaxies where the star-forming disk is sharply cut off, but inside
the cutoff, the star formation levels are normal (the [s] means
truncation is severe).  One of these, NGC 4580, is so unusual that
Sandage \& Bedke (1994) classify it as Sc(s)/Sa, where the Sc part is
the inner disk and the Sa part is the outer disk.  In ``T/A galaxies",
the inner star-formation is at a low level, as in anemic cases, while
in ``T/C" galaxies, most of the star formation is confined within the
inner 1 kpc. Koopmann and Kenney found that the majority of Virgo
Cluster spiral galaxies have truncated star-forming disks.

The idea is that the interstellar medium (ISM) of a cluster galaxy can
interact with the intra-cluster medium (ICM), stripping the ISM (via
ram pressure; Gunn and Gott 1972) but leaving the stellar disk intact.
Truncated gas and star-forming disks result because ram-pressure
stripping is more severe in the outer parts of galaxies (e. g., Book and
Benson 2010 and references therein). In Virgo, most galaxies with
truncated star-forming disks have relatively undisturbed stellar disks
and normal to slightly enhanced inner disk star formation rates,
suggesting that ICM-ISM stripping is the main mechanism in the
reduction of their star formation rates. The cases found to have
relatively normal or enhanced star formation rates are preferentially
located in the outer parts of the cluster and likely have never visited
the core region. Only galaxies which go near the center get
significantly stripped.  However, tidal effects also contribute to
morphological changes. Several galaxies, including many of the T/C
class, display peculiarities consistent with tidal effects, such as
nonaxisymmetric circumnuclear star formation, shell features (e.g., NGC
4424 in Figure~\ref{virgotypes}), and enhanced inner star formation
rates.

A recent study by Yagi et al. (2010) provides dramatic and clear
evidence of disk gas stripping in galaxies thought to be relatively new
arrivals to the core region of the Coma Cluster. Using deep H$\alpha$
imaging, these authors detected ionized gas in clouds that are mostly
outside the main disk of a dozen Coma galaxies. Three distinct
morphologies of the distributions of these clouds were found: (1)
connected clouds that blend with disk star formation; (2) long,
connected lines of clouds that extend from a central gas knot but are
not related to the disk light; and (3) clouds completely detached from
the main disk. Examples of these categories are illustrated in
Figure~\ref{coma}. Yagi et al. interpret them in terms of an
evolutionary gas-stripping sequence where category (1) galaxies are in
an earlier phase of stripping while the category (3) galaxies are in
the most advanced phase. It is likely that large disk galaxies in Coma
would be completely stripped eventually because of the cluster's high
ICM density and broad velocity distribution. The same process seen
in Coma likely occurs in Virgo but is only partial for the large
spirals owing to the lower ICM density and velocities in Virgo
(Koopmann \& Kenney 2004).

\subsection{Interacting and Peculiar Galaxies}

\subsubsection{Tidal Tails, Arms, and Bridges}

It is perhaps fitting that the first major spiral galaxy discovered was
in the interacting pair M51 (section 2). Numerical simulations (Salo 
\& Laurikainen 2000a,b) have shown that both parabolic and bound passages
of the companion, NGC 5195, can explain the observed morphology and
other characteristics of the system. It turns out
that M51 defines a class of interacting systems known as M51-type
pairs. Figure~\ref{pec_Es} shows an example in the pair NGC 2535-6.
In each case, the larger component has a strong two-armed spiral,
with one arm appearing ``drawn" to the smaller companion. An
extensive catalogue of M51-type pairs is provided by Jokimaki et
al. (2008).

Other distant encounters can produce tidal tails or bridges of material
between galaxies (top row, Figure~\ref{pec_Es}). 
NGC 4676, also known as the "Mice," is a pair of
strongly interacting galaxies where a very extended tidal tail has formed
in one component. The strongly-interacting pair NGC 5216/18 has developed
a bright connecting bridge of material, and each component shows tidal
tails. The evolution of this system, and the role of encounters on bar
formation, is described by Cullen et al. (2007). 

\subsubsection{Dust-Lane Ellipticals}

Figure~\ref{pec_Es} also shows several examples of morphologies that may
result from minor mergers of a small galaxy with a more massive,
pre-existing elliptical galaxy. Bertola (1987) brought attention
to the unusual class of {\it dust-lane ellipticals}, where an otherwise
normal elliptical galaxy shows peculiar lanes of obscuring dust. It was
de Vaucouleurs's personal view that ``if an elliptical shows dust, then 
it's not an elliptical!" However, Bertola showed that an unusual case like
the radio elliptical galaxy NGC 5128, where a strong dust lane lies along the
{\it minor axis} of the outer light distribution, is simply the nearest
example of a distinct class of objects. Further study showed that dust-lane
ellipticals come in several varieties. The minor axis dust lane type appears
most common, but cases of alignment along the major axis of the outer
isophotes (major axis dust lanes) as well as cases of misalignment are
also known (see the upper left panels of Figure~\ref{pec_Es}). The origin
of these very regular dust lanes is thought to be a merger of a gas-rich
companion (e.g., Oosterloo et al. 2002). The regularity of the dust lanes 
suggests that the mergers are in advanced states.

\subsubsection{Shell/Ripple Galaxies}

The two lower left frames of Figure~\ref{pec_Es} show examples of
galaxies having ``shells,", or faint, arc-shaped brightness enhancements of
varying morphology. They were first discovered on deep photographs
by Malin (Malin \& Carter 1980), and appeared to be
associated mainly with elliptical galaxies. In fact, the first examples,
NGC 1344 and 3923, are classified in catalogues as ordinary ellipticals
because the shells are not detectable on photographs of average depth. 
Once the class was recognized, a detailed search led to other examples
which were listed by Malin \& Carter (1980, 1983). The term ``shells"
implies a particular three-dimensional geometry that Schweizer
\& Seitzer (1988) argued imposes a prejudice on the interpretation
of the structures. They proposed instead the alternate term ``ripples,"
which implies less of a restrictive geometry. 

The explanation of shell/ripple galaxies is one of the great success
stories in galactic dynamics (see review by Athanassoula \& Bosma 1985). 
Shells are thought to be
remnants of a minor merger between a massive elliptical and a lower mass disk-like galaxy. The main requirements 
are that the disk-shaped galaxy be ``cold", or lack any random motions, and that the potential of the elliptical 
galaxy should be rigid, meaning the elliptical is much more massive than its companion. The smaller galaxy's stars 
fall into the center of the galaxy and phase wrap, or form alternating outward-moving density waves made of the disk 
galaxy's particles near the maximum excursions of their largely radial orbits in the rigid potential. Many, but not all, of 
the main properties of shell Es can be explained by this model. 
Other issues concerning shell galaxies 
are reviewed by Kormendy and Djorgovski (1989).

Taylor-Mager et al. (2007; see their Figure 2) 
have proposed a simple classification of
interacting systems that highlights different interaction classes.
A pre-merger (type pM) includes two interacting galaxies that are
sufficiently far apart to suffer little apparent distortion. A minor
merger (mM) is two galaxies showing evidence of merging, but one component
is much smaller than the other. A major merger (M) has two comparable
brightness galaxies in the process of merging, while a merger remnant
(MR) is a state sufficiently advanced that the merging components are
no longer distinct. 

\subsubsection{Ultra-Luminous Infrared Galaxies}

Related to interacting systems are the infrared-bright galaxies
first identified by Rieke \& Low (1972) based on 10$\mu$m photometry.
From studies based on the Infrared Astronomical Satellite (IRAS),
Sanders \& Mirabel (1996) classified a galaxy as a ``luminous infrared
galaxy" (LIRG) if its luminosity in the 8-1000$\mu$m range is between
10$^{11}$ and 10$^{12}$ L$_{\odot}$. If the luminosity in the same
wavelength range exceeds 10$^{12}$ $L_{\odot}$, then the object is
called an ``ultra-luminous infrared galaxy" (ULIRG). Detailed studies
have shown that at high redshifts, LIRGS and ULIRGS are a dominant
population of objects (see discussion in Pereira-Santaella et al. 2010).

The morphologies of nine ULIRGS were studied using HST $B$ and $I$-band
images by Surace et al. (1998). Their montage of six of these objects is
shown in Figure~\ref{ulirgs}. In every case there are clear signs of
interactions, and all are likely linked to mergers or mergers in progress.
Several, like Mk 231, have bright Seyfert nuclei. Arribas et al. (2004)
obtained extensive imaging of local LIRGS, and found a similar high
proportion of strongly interacting and merging systems.

The merger rate is considered one of the most important parameters
for understanding galaxy evolution. It has been difficult to estimate,
and issues connected with it are discussed by Jogee et al. (2009; see also
Conselice 2009).
A merger is considered major if it involves a companion ranging
from 1/4 to approximately equal mass to the main galaxy. A major
merger of two spiral galaxies can destroy both disks and lead to an
$r^{1\over 4}$ law profile remnant through violent relaxation.
Minor mergers involve companions having 1/10 to 1/4 the mass of the
main galaxy. Both types of mergers, while in progress, can lead to 
many specific morphological features such as highly distorted shapes,
tidal tails and bridges, shells and ripples, and warps. Even some bars
and spiral patterns are thought to be connected to interactions,
and especially galaxies with a double nucleus are thought to be
mergers. As discussed in section 10.1, 
mergers or collisions may also be at the heart of rare
morphologies such as ring 
and polar ring galaxies. Using visual classifications
of merger types, Jogee et al. (2009) estimate that 16\% of high
mass galaxies have experienced a major merger, while 45\% have
experienced a minor merger, during the past 3-7Gyr ($z$=0.24-0.80).

\subsection{Warps}

A warp is an apparent bend or slight twist in the shape of the disk of
a spiral galaxy (see Sellwood 2010 and references therein). 
In a warp, stars and gas clouds move in roughly circular
orbits, but the orientation of these orbits relative to the inner disk
plane changes with increasing radius. Warps are most easily detected
in edge-on galaxies because the bending of the outer orbits makes the
galaxy look like an integral sign. 
Although often most pronounced in an HI map, 
warps can be seen in ordinary optical images of edge-on galaxies.
Figure~\ref{warps} shows three galaxies having strong optical
warping of the disk
plane. In two of the galaxies, the bright inner disk is unwarped, while
a fainter and thicker outer disk zone is twisted relative to the inner 
disk. In UGC 3697, the warping is exceptionally visible. In general,
optical warping is less severe than HI warping. 

Warping is a very common aspect of spiral galaxies (e.g., Binney 1992)
and has been interpreted in terms of perturbations (gaseous infall or
interactions) that trigger bending instabilities
(e.g., Revaz \& Pfenniger
2007). Garcia-Ruiz, Sancisi, \& Kuijken (2002) estimated HI warp angles,
the angle between the inner disk plane and the assumed linear warping
zone, to range from nearly 0$^o$ to more than 30$^o$. A useful
summary of previous warp studies is provided by Saha et al. (2009),
who examine warp onset radii in mid-IR images. The theory of warps
is reviewed by Sellwood (2010).

\subsection{The Morphology of Active Galaxies}

The morphology of active galaxies (also called ``excited" galaxies by
van den Bergh 1998) is important to consider because of a possible link
between morphological features and the fueling of the active nucleus.
Early studies showed a preponderance of ring, pseudoring, and bar
features in Seyfert galaxies that suggested the link was bar-driven gas
flow (Simkin, Su, \& Schwarz 1980). Several examples of the morphology
of Seyfert and other active galaxies are shown in Figure~\ref{actives}.
The activity classifications are based mainly on spectroscopy, not on
morphology, and are described by Veron-Cetty \& Veron (2006).

A detailed study of active galaxy morphologies by Hunt \& Malkan (1999)
provided similar results to the early studies. These authors examined
the morphologies of a large sample of galaxies selected on the basis of
their 12$\mu$m emission, and found that outer rings and inner/outer
ring combinations are 3-4 times higher in Seyfert galaxies than in
normal spirals. In contrast, bars were found to occur with the same
frequency ($\approx$69\%) in Seyferts as in normal spirals, while for
HII/starburst galaxies, the frequency was much higher ($>$80\%).
Although outer rings are found mostly in barred galaxies, bars do not
promote the nuclear activity of Seyfert galaxies. Hunt \& Malkan (1999)
interpret this inconsistency in terms of timescales: it takes roughly
3$\times$10$^9$ yr for a closed outer ring to form, a timescale during
which a bar may weaken or dissolve. Because of this, a high ring
frequency in Seyferts would indicate an advanced evolutionary state.
Related to the same issue, Comer\'on et al. (2010) found that
nuclear rings do not correlate with the presence of nuclear activity.

The study of Hunt \& Malkan (1999) used mostly RC3 classifications to
deduce the bar fraction in active galaxies. These visual
classifications are based on blue light images and hence dust could
effectively obscure some bars. Knapen et al. (2000) used high
resolution near-IR images of well-defined samples and quantitative bar
detection methods to deduce that bars are more frequent in Seyfert
galaxies than in a control sample of non-active galaxies: 79\% $\pm$
7.5\% versus 59\% $\pm$ 9\% (see also Laine et al. 2002). Laurikainen,
Salo, \& Buta (2004) came to a similar conclusion for 180 galaxies
in the OSUBSGS, based on near-IR $H$-band images. The former studies
used ellipse fits to identify bars, while Laurikainen et al. used
Fourier analysis.

McKernan, Ford, \& Reynolds (2010) also consider outer rings and pseudorings
as probes of models of AGN fueling from interactions and mergers. The idea
is that a closed outer ring takes a long time to form and is very fragile,
being sensitive to interactions and changes in the bar pattern speed (e.g.,
Bagley et al. 2009). An interaction can change a closed outer ring into
a pseudoring and could possibly destroy the ring. Thus, rings are probes
of the interaction history of active galaxies. McKernan et al. found
no difference between the AGN found in ringed galaxies and those found
in galaxies without rings. But in those with rings, recent interactions
can be ruled out and activity may be tied to short-term internal
secular evolutionary processes.

Bahcall et al. (1997) presented HST images of 20 luminous, low redshift
quasars observed with a wide $V$-band filter.
Figure~\ref{quasars} is reproduced from their paper and shows images
of the host galaxies after removal of most of the quasar light. The images show
a variety of morphologies, including ellipticals, interacting pairs,
systems with obvious tidal disturbances, and normal-looking spirals.
An example of the latter is PG1402+261 ($z$=0.164), 
which is type (R$_1^{\prime}$)SB(rs)a based partly also on the image 
shown in Figure 7 of Bahcall et al. Based on the
number of hosts showing signs of interactions, as well as the number of
companions, Bahcall et al. conclude that interactions may trigger the
quasar phenomenon.

\subsection{The Morphology of Brightest Cluster Members}

Matthews, Morgan, \& Schmidt (1964) observed the optical morphologies
of the radio sources identified with the brightest members of rich
galaxy clusters.
They found that the most common form was what Morgan (1958) called a
``D" galaxy, meaning a galaxy having an elliptical-like inner region
surrounded by an extensive envelope (see discussion in van den Bergh
1998). Although these superficially
resembled Hubble's S0s, none were found having a highly-elongated shape, 
implying that the galaxies are not as highly flattened as typical S0s. Another
characteristic of the cluster D galaxies was their very large linear size
and exceptional luminosity, much larger than a typical cluster member. 
To denote these extreme objects in the Morgan system, the prefix
c was added as in the old classification of supergiant stars. Even today,
Morgan's notation ``cD" is used to describe these supergiant galaxies which 
are generally considered outside the scope of the Hubble system. 

The most detailed study of the photometric properties of brightest cluster
members (BCMs) was made by Schombert (1986, 1987, 1988; see also various
references therein). The main BCM types Schombert
considered were gE (giant ellipticals), D, and cD, distinguished mainly
from the appearance of profile shape. D galaxies are larger and more diffuse
with shallower profiles than gE galaxies, while a cD galaxy is the same as
a D galaxy but with a large extended envelope (Schombert 1987). cD envelopes
can extend to 500 kpc or more. Kormendy \& Djorgovski (1989) argue that
only cD galaxies are sufficiently physically distinct from ellipticals
to merit being a separate class, and recommended that the ``D" class
not be used.

Two cD galaxies and two gE galaxies are shown in Figure~\ref{bcms}. To give
an idea of the scale of these objects, the vertical dimension of the frames
corresponds to 201, 232, and 132 kpc for (left to right)
UGC 10143 (A2152), NGC 4874/89 (A1656), and NGC 6041 (A2151), respectively.
The cD classification of NGC 4874 is due to Schombert (1988), and one
can see in Figure~\ref{bcms} that it is much larger and has a shallower
brightness profile than nearby NGC 4889. 
The cD envelope is detected as an excess of light
in the outer regions relative to a generalized brightness profile, and may
not even be the light that leads to the visual classification of cD. 

Based on structural deviations such as the large radii, shallow profile
slopes, and bright inner regions, Schombert (1987) concluded that BCMs
fit well with the predictions of merger simulations, including accretion
and ``cannibalism" of smaller cluster members. Properties of cD envelopes
(as separated photometrically from the parent galaxy) may suggest a 
stripping process for their formation (Schombert 1988). 

As noted in section 5.1, many BCMs in RC3 received the classification
E$^+$, suggesting that the characteristic brightness profiles give
a hint of an envelope interpreted as an incipient disk. The distribution
of axial ratios of cDs actually is flatter on average than normal
ellipticals (Schombert 1986), but it is not clear that the perceived 
envelopes in BCM E$^+$ galaxies are actually as flattened as a typical disk.
A local example of a gE galaxy is M87, classified as type E$^+$0-1
by de Vaucouleurs. 

\section{Star Formation Morphologies}

\subsection{H$\alpha$ Imaging}

The standard waveband for galaxy classification, the $B$-band, is
sensitive enough to the extreme population I component that the degree
of resolution of spiral arms into star-forming complexes is part
of the classification. The $B$-band, however, also includes a
substantial contribution from the older stellar background. One way
to isolate only the star-forming regions in a galaxy is imaging
in H$\alpha$, which traces HII regions. Apart from showing the
distribution of star formation (modified by extinction), H$\alpha$
imaging also traces the rate of global photoionization, which in
turn directly traces the rate of formation of stars more massive
than about 10$M_{\odot}$ (Kennicutt, Tamblyn, \& 
Congdon 1994). The initial mass function (IMF), either assumed or
constrained in some way (using, for example, broadband colors),
allows H$\alpha$ fluxes to be converted to the total star formation
rate over all stellar masses.

H$\alpha$ imaging often shows well-organized patterns of HII regions
that follow structures like spiral arms and especially rings. Images
of six early-to-intermediate-type ringed galaxies are shown in 
Figure~\ref{halpha} (Crocker, Baugus, \& Buta 1996). 
The way H$\alpha$ imaging usually works is 
a galaxy is imaged in or near its redshifted H$\alpha$ wavelength, and then
in a nearby red continuum wavelength. The net H$\alpha$ image is the
difference between the H$\alpha$ image and the scaled red continuum
image. Often, the H$\alpha$ filter used is broad enough to include
emission from [NII] 6548 and 6584. For each of the galaxies shown
in Figure~\ref{halpha}, the left image is the red continuum image, while
the right image is the net H$\alpha$ image. Of the four barred spirals
shown (NGC 1433, 7329, 6782, and 7267), three show no HII regions
associated with the bar. These three (NGC 1433, 7329, and 6782)
all have conspicuous inner rings which are lined with HII regions.
As shown by Crocker, Baugus, \& Buta (1996), the distribution of
HII regions around inner rings is sensitive to the intrinsic shape
of the ring. When the ring is highly elongated, the HII regions
concentrate around the ends of the major axis (which coincide with
the bar axis; NGC 6782), while when the ring is circular, the HII regions are
more uniformly spread around the ring (NGC 7329). 
Inner ring shapes do not correlate well with maximum relative
bar torques in a galaxy (dVA), but Grouchy et al. (2010) have shown that
when local forcing is considered instead, ring shapes and bar
strengths are well-correlated.

The case of NGC 7267 is different in that most of its H$\alpha$ emission
is concentrated in the bar. Martin \& Friedli (1997) have argued that star
formation along galactic bars could provide clues to gas flow in the
inner regions of galaxies and the fueling of starbursts and AGN. These
authors present models which suggest an evolution from a pure
H$\alpha$ bar, to an H$\alpha$ bar with ionized gas in the center, to a
gas-poor bar with strong nuclear or circumnuclear star formation. This
suggests that the bar of NGC 7267 is younger than those in NGC 1433,
7329, and 6782.

The two other galaxies shown in Figure~\ref{halpha} are nonbarred
or only weakly-barred. NGC 6935 is an interesting case where a nonbarred
galaxy has a strong ring of star formation. Grouchy et al. (2010)
found that the star formation properties of inner rings, but not
the distribution of HII regions, are independent of the ring shapes
and bar strengths in a small sample. The case of NGC 7702 is different
from the others. This is a late S0 (type S0$^+$) showing a very strong
and largely stellar inner ring. The ring shows little ionized gas and
appears to be in a quiescent phase of evolution.

\subsection{Ultraviolet Imaging}

The best global imaging of nearby galaxies at ultraviolet wavelengths
has been obtained with the {\it Galaxy Evolution Explorer} (GALEX,
Martin et al. 2005), which provided detailed images of galaxies of all types
at wavelengths of 0.225 and 0.152$\mu$m. These images reveal young
stars generally less than $\approx$100 Myr old but are affected by dust 
extinction. There is a strong correlation between the UV morphology
and the H$\alpha$ morphology (Figure~\ref{nuv-halpha}). Like H$\alpha$, 
UV fluxes from galaxies can be used to estimate star
formation rates once extinction is estimated, and are particularly
sensitive to the ratio of the present to the average past star formation
rate. UV
imaging is an effective way of decoupling the recent star formation
history of a galaxy from its overall, long-term star formation history.

Figure~\ref{nuv} shows near-UV (0.225$\mu$m) images of eight galaxies
over a range of Hubble types. The two earliest types, NGC 1317 and
4314 (both Sa) show a near-UV morphology dominated by a bright
circumnuclear ring of star formation. The SB(r)b galaxy NGC 3351
shows a conspicuous inner ring of  star formation and little emission
from its bar region. In contrast, the SBc galaxy NGC 7479 shows
strong near-UV emission from its bar. The late-type galaxies NGC 628
[type SA(s)c] and NGC 5474 [type SA(s)m] are typical of their types,
but most interesting is NGC 4625.
A key finding of GALEX was extended UV emission well beyond the optical
extent of some galaxies. NGC 4625 is an example where the main optical
part [type SABm] is only a small fraction of the extent of the UV disk
(Gil de Paz et al. 2005).

The final object shown in Figure~\ref{nuv} is NGC 5253, a basic example
of what de Vaucouleurs classified as I0 (section 5.3). The inner region
is a bright boxy zone of star formation, and even the extended disk is
prominent.

\subsection{Atomic and Molecular Gas Morphology}

Related to star formation morphologies are the distributions of atomic
and molecular gas. Far from being randomly-distributed clouds of
interstellar material, atomic and molecular gas morphologies can be
highly organized, well-defined patterns. Recent high quality surveys
have provided some of the best maps of these distributions in normal
galaxies.  Atomic hydrogen is mapped with the 21cm fine structure
emission line, which has the advantage of not being affected by
extinction and for being sufficiently optically thin in general to
allow total HI masses to be derived directly from HI surface brightness
maps. In addition, HI maps provide information on the kinematics and
dynamics of the ISM, as well on the existence and distribution of dark
matter in galaxies.  Molecular hydrogen is generally mapped using the
$^{12}$CO J=1-0 rotational transition at a wavelength of 2.6mm, under
the assumption that CO and hydrogen mix in a roughly fixed proportion.

Although numerous maps have been made of the HI distribution in nearby
galaxies, the most sophisticated and detailed survey made to date is
the ``The HI Nearby Galaxies Survey" (THINGS, Walter et al. 2008). The
earliest surveys had shown that HI is a tracer of spiral structure in
galaxies, and the THINGS provides some of the highest quality maps
revealing this correlation as well as other characteristics.  From a
morphological point of view, HI maps tend to reveal: (1) enhanced
surface brightness in star-forming features such as spiral arms, rings,
and pseudorings; (2) extended gaseous disks, such that the HI extent
can exceed the optical extent by several times; and (3) supernova-driven
and star-formation driven, wind-blown holes.

Figure~\ref{himaps} shows the HI morphologies of eight THINGS galaxies
ranging from the Sab galaxy NGC 4736 to the Sm galaxy DDO 50 (Holmberg
II).  Bright Sc galaxies like NGC 628 (M74) and NGC 5236 (M83) show HI
distributions that extend well beyond the optical disks. These extended
patterns can include large spirals as in NGC 628. In M81 and M83, the
HI traces the optical spiral structure well. Large rings or pseudorings
are seen in NGC 2841 and NGC 4258, while M81 shows an
intermediate-scale ring of gas that is much less evident optically.
The bright star-forming inner ring in NGC 4736 is well-defined and
easily distinguished in HI, but the galaxy's diffuse stellar outer ring
is more of an asymmetric spiral zone.

Most interesting in HI maps are the obvious holes where there appears
to be a deficiency of neutral gas compared to surrounding regions.
Especially large holes are seen in the HI morphology of the late-type
dwarf DDO 50. These holes are thought to be regions cleared by the
stellar winds and explosions of massive stars contained or once
contained within them. The holes are 100pc to 2kpc in size, and have
different systematic properties in early and late-type galaxies in the
sense that holes may last longer in late-type dwarfs owing to the lack
of serious shear due to strong differential rotation (Bagetakos et al.
2009). Holes may also be found outside the standard isophotal optical
angular diameter.

The dwarf galaxy DDO 154 shown in Figure~\ref{himaps} has one of the
largest ratios of HI to optical diameter, a factor of 6 at least
according to Carignan \& Purton (1998), who also estimated that 90\%
of the mass of the galaxy is in the form of dark matter. The HI and
optical morphologies bear little resemblance to one another. Another
example of strongly uncorrelated HI and optical morphologies is NGC
2915, a very gas-rich galaxy whose optical morphology is a blue compact
dwarf while its at least 5 times larger HI morphology includes a
prominent outer spiral with no optical counterpart (Meurer et al.
1996), leading to the concept of a purely ``dark spiral." Bertin \&
Amorisco (2010) consider a general intepretation of such large outer HI
spirals, especially those seen in galaxies like NGC 628 and NGC 6946:
the spirals represent short trailing waves that carry angular momentum
outwards from corotation and, at least in the gaseous component,
penetrate a normal barrier at the OLR of the stellar disk pattern and
go far out into the HI disk. The short trailing waves are thought to
excite the global spiral arms seen in the main optical body of the
galaxy (where the prominent optical spirals of NGC 628 and NGC 6946 are
found). How NGC 2915 fits into this picture is unclear since the main
stellar body of this galaxy is not a grand-design spiral.

Galaxies whose HI disks do not extend much beyond the optical light
distribution are also of interest. NGC 4736 in Figure~\ref{himaps} is
an example. The large, nearby ringed barred spiral NGC 1433 was found
by Ryder et al. (1996) to have neutral hydrogen gas concentrated in its
central area, its inner ring, and in its outer pseudoring, with no gas
outside the visible disk light and a lower amount of gas in the bar
region compared to the ring regions.  Higdon et al. (1998) showed that
in the ringed, barred spiral NGC 5850, neutral gas is concentrated in
the inner ring and in the asymmetric outer arm pattern, with little or
no emission detected outside this pattern. The asymmetry led Higdon et
al. to propose that NGC 5850 has possibly experienced a high speed
collision with nearby NGC 5846.

In galaxy clusters, it is well-known that environmental effects can
truncate an HI disk so that it is {\it smaller} than the optical disk
light. This is dramatically illustrated in the high resolution VLA HI
maps of Virgo Cluster galaxies by Chung et al. (2009), who found that
galaxies within 0.5 Mpc of the cluster core have severely truncated HI
disks typically less than half the size of the optical standard
isohotal diameter, $D_{25}$. A variety of earlier studies had already
shown these galaxies to be HI-deficient compared to field galaxies of
similar types. As noted in section 10.2, an interaction between a
cluster galaxy's ISM and the intra-cluster medium can account for
these unsual modifications of HI morphology. Chung et al. also
provide evidence for this interaction in some morphologies that
appear to be gas stripping in progress.

The Berkeley-Illinois-Maryland Survey of Nearby Galaxies (BIMA SONG;
Regan et al. 2001; Helfer et al. 2003) provided some of the highest
quality CO maps of normal galaxies. CO emission is often seen in
intermediate (Sab-Sd) galaxies, which have a high enough gas content
and metallicity to allow the $^{12}$CO J=1-0 2.6mm emission line to be
detectable.  The CO distributions of eight such galaxies from this
survey are shown in Figure~\ref{comaps}. These display some of the
range of molecular gas morphologies seen. CO traces the inner spiral
arms of NGC 628, 1068, and 4535, and is seen along the bar of NGC 2903.
A common CO morphology is a large-diameter ring of giant molecular
clouds (GMCs), without a central CO concentration, as seen in NGC 2841
and 7331. The rings are the peaks of exponentially-declining
distributions. The coherent inner molecular gas ring in NGC 7331
appears more like a typical resonance ring, and has an estimated
molecular gas mass of 3.4$\times$10$^9$ $M_{\odot}$ (Regan et al.
2004). The CO distribution in NGC 2403 appears to be concentrated in
individual GMCs, with little diffuse emission, while that in NGC 3351
is characterized by a small central bar aligned nearly perpendicular to
the galaxy's primary bar.  Helfer et al. (2003) show that the Milky
Way, M31, and M33 have CO morphologies that are consistent with the
range of morphologies found by BIMA SONG.

\section{Infrared Observations: Galactic Stellar Mass Morphology}

Infrared observations have considerable advantages over optical
observations of galaxies. While traditional $B$-band images are
sensitive to dust and extinction (both internal and external), the
effects of extinction in the near- and mid-IR are much less, and become
virtually negligible at 3.6$\mu$m. Spiral galaxies imaged at
wavelengths successively longer than $B$-band become progressively
smoother-looking, not only due to the reduced effect of extinction, but
also to the de-emphasis of the young blue stellar component.  The
combination of these effects has led to the popular idea that IR
imaging reveals the ``stellar backbone" of galaxies, i. e., the
distribution of actual stellar mass (e. g., Rix \& Rieke 1993; Block et
al. 1994). Thus, infrared imaging has become a staple for studies of
the gravitational potential and stellar mass distribution in galaxies
(e. g., Quillen, Frogel, \& Gonz\'alez 1994), and for the
quantification of bar strength from maximum relative gravitational
torques (e.g., Buta \& Block 2001; Laurikainen \& Salo 2002).

Infrared imaging has also revealed interesting outer structures such as
the large outer red arcs seen in M33, which have been interpreted by
Block et al. (2004b) to be swaths of extremely luminous carbon stars
formed from external accretion of low metallicity gas. Power spectrum
analysis of the IR structures in classical spirals has been used to
detect azimuthal ``star streams" and to evaluate the role of turbulence
on star formation and spiral structure (Block et al. 2009 and
references therein).

Near-infrared imaging from 0.8-2.2$\mu$m can be successfully obtained
from groundbased observatories but with the serious drawback that the
brightness of the sky background increases substantially over this
range. As a result, it has not been possible to achieve a depth of
exposure at, for example, 2.2$\mu$m comparable to the kinds of depths
achievable at optical wavelengths without excessive amounts of
observing time.  The first major near-IR survey designed for
large-scale morphological studies was the Ohio State University Bright
Spiral Galaxy Survey (OSUBSGS, Eskridge et al.  2002), which included
optical $BVRI$ and near-IR $JHK$ images of 205 bright galaxies of types
S0/a to Sm in a statistically well-defined sample selected to have
total blue magnitude $B_T$$\leq$12.0 and isophotal diameter
$D_{25}$$\leq$6\rlap{.}$^{\prime}$5. This survey allowed a direct
demonstration of how galaxy morphology actually changes from optical to
near-IR wavelengths, not merely for a small, selected sample of
galaxies, but for a large sample covering all spiral subtypes.
The main near-IR filter used in this survey was the $H$-band at
1.65$\mu$m.

The OSUBSGS was later complemented by the {\it Near-Infrared S0 Survey}
(NIRS0S, Laurikainen et al. 2005, 2006, 2007, 2009, 2010; Buta et al.
2006), a $K_s$-band imaging survey of 174 early-type galaxies in the
type range S0$^-$ to Sa, but mostly including S0s, some of which were
misclassified as ellipticals in RC3. NIRS0S images are deeper than
OSUBSGS near-IR images owing to the use of larger telescopes and longer
on-source times. Although S0 galaxies are dominated by old stars and
are usually smooth even in blue light images, the $K_s$ band was chosen
to complement the OSUBSGS sample of spirals in order to make a fair
comparison between bar strengths and bulge properties of S0s and
spirals. Also, S0 galaxies are not necessarily dust-free, and near-IR
imaging is still necessary to penetrate what dust they have. NIRS0S has
led to several important findings about S0 galaxies: (1) a class of
S0s, not previously recognized, having prominent lenses but very small
bulges that are more typical of Sc galaxies than of earlier type
spirals (example: NGC 1411, Laurikainen et al. 2006); (2) considerable
evidence that S0 galaxies have pseudobulges just as in many spirals
(Laurikainen et al. 2007).  While the bulges of the latter are likely
to be made of rearranged disk material in many cases (section 9), those
in S0s are likely to be related to the evolution of bars. S0 bulges
tend to be nearly exponential (Sersic index $n$ $\leq$ 2), are
supported against gravity by rotation rather than random motions, and
often include clear inner disks; (3) good correlations between bulge
effective radii, $r_e$, and disk radial scalelength, $h_R$, as well as
between the $K_s$-band absolute magnitudes of the bulge and disk,
suggest that S0 bulges are not formed from hierarchical mergers,
implying that S0s could be stripped spirals, although the lower bar fraction
in S0s suggest that this is in conjunction with evolution due to bars
and ovals; (4) 70\% of S0-S0/a galaxies have ovals or lenses,
suggesting that bars have been weakened in such galaxies over time; and
(5) not only bulges, but also disks of S0s are similar to those in
S0/a-Scd spirals.

The {\it Two-Micron All-Sky Survey} (2MASS, Skrutskie et al. 2006)
provided near-IR $JHK_s$ images of a much larger galaxy sample than
either the OSUBSGS or NIRS0S, although these images lack the depth of
the OSUBSGS and NIRS0S images in general. 2MASS provided considerable
information on near-infrared galaxy morphology, which led to the
extensive {\it 2MASS Large Galaxy Atlas} (Jarrett et al. 2003).

The best imaging of galaxies at mid-IR wavelengths has been obtained
with the {\it Spitzer Space Telescope} using the Infrared Array Camera
(IRAC, Fazio et al. 2004) and 3.6, 4.5, 5.8, and 8.0$\mu$m filters.
The 3.6 and 4.5$\mu$m filters provide the most extinction-free
views of the stellar mass distribution in galaxies, 
while the 5.8 and 8.0$\mu$m filters reveal the interstellar medium (ISM)
(Pahre et al. 2004). The loss of coolant in 2008 prevented further
observations with the 5.8 and 8.0$\mu$m filters, but the 3.6 and 4.5$\mu$m
filters could still be used. This led to the {\it Spitzer
Survey of Stellar Structure in Galaxies} (S$^4$G, Sheth et al. 2010),
a 3.6 and 4.5$\mu$m survey of 2,331 galaxies of all types closer than 40Mpc.
These wavelengths sample the Rayleigh-Jeans
decline of the stellar spectral energy distribution of all stars
hotter than 2000K. S$^4$G images shown here are from Buta et al. (2010a)
and are based on pre-survey archival images processed in the same
manner as survey images. {\it Spitzer} observations
have a very low background compared to groundbased near-IR observations,
and thus IRAC images are the deepest galaxy images
ever obtained in the IR.

Figure~\ref{m51_frames} compares images of M51 at four wavelengths:
the GALEX 0.15$\mu$m band, the $B$-band (0.44$\mu$m), the near-infrared
$K_s$ band
(2.2$\mu$m), and the IRAC 3.6$\mu$m band. Only the $B$ and $K_s$-band
images are from groundbased observations. The GALEX image reveals 
the extensive star formation in the spiral arms, and the complete absence
of star formation in the companion NGC 5195 as well as in the complex
tidal material north of the companion. The star formation in the arms is
more subdued in the $B$-band, and almost completely subdued in the
$K_s$-band. The arms are so smooth in the $K_s$ band that the galaxy
resembles an Sa or Sab  system. (The $B$-band type is Sbc.)
Surprisingly, this is not the case in the 3.6$\mu$m image whose
considerably greater depth compared to the $K_s$-band image is
evident. The spiral arms in the 3.6$\mu$m band are lined by numerous
resolved objects, many of which are correlated with the
star-forming regions seen in the $B$-band. This is dramatically
seen also in the SB(s)cd galaxy NGC 1559 (Figure~\ref{ngc1559}),
where an IRAC 3.6$\mu$m image is compared with a $B$-band image.
These show that resolved features in the deep 3.6$\mu$m image are 
well-correlated with $B$-band star-forming complexes. Thus, mid-IR
3.6$\mu$m images are {\it not} completely free of the effects of
the extreme population I stellar component (see discussions in
Block et al. 2009 and Buta et al. 2010a).

A sampling of S$^4$G images as compared to $B$-band images for the same
galaxies from the dVA is shown in Figure~\ref{nir01}. The four galaxies
shown, NGC 584, 1097, 628, and 428, have dVA types of S0$^-$, SBb, Sc,
and Sm, respectively, thus covering almost the entire Hubble-de Vaucouleurs
sequence. Although the very dusty interacting system NGC 1097 looks
slightly ``earlier" at 3.6$\mu$m, these images show again that on the whole
the morphology in the two wavebands is very similar. The same is seen for
other galaxies described by Buta et al. (2010a), who found that 3.6$\mu$m
types, judged using the same precepts described in the dVA for blue light
images, are well-correlated with blue-light types. On average mid-IR 
classifications for RC3 S0/a-Sc galaxies 
are about 1 stage interval earlier than $B$-band classifications,
with little difference for types outside this range. The
correlation is much better than what was expected from previous near-IR
studies (Eskridge et al. 2002). 
3.6$\mu$m galaxy morphology is sufficiently contaminated
by recent star formation to allow the same criteria defined for blue
light images to be used for galaxy classification, a surprising result.

Drastic differences between 3.6$\mu$m and $B$-band morphology are seen only
for the most dusty galaxies. One example is NGC 5195, shown also in 
Figure~\ref{m51_frames}. This galaxy, classified as Irr II in the Hubble
Atlas and as I0 by de Vaucouleurs, appears as a regular early-type galaxy
of type SAB(r)0/a (see also Block et al. 1994). Other galaxies
that can look very different are flocculent spirals such as NGC 5055
(Figure~\ref{nir03}). The flocculence largely
disappears at 3.6$\mu$m, and a more global pattern is seen (see also
Thornley 1996). The type of NGC 5055, Sbc, remains 
largely unchanged from $B$ to 3.6 $\mu$m.

As noted by Helou et al. (2004), the mid-IR wavelength domain marks
the transition from emission dominated by starlight to emission
dominated by interstellar dust. While images at 3.6$\mu$m show the
stellar bulge and disk almost completely free of dust extinction, 
an image at 8$\mu$m shows very little starlight but considerable
emission from the ISM in the form of glowing dust.

Figure~\ref{m81frames} shows an 8$\mu$m image of the nearby spiral galaxy M81 
as compared to a $B-I$ color index map coded such that blue star-forming
features are dark while red dust lanes are light. 
The 8$\mu$m image of M81 shows that its ISM is closely associated with
its spiral arms. Comparison with the $B-I$ color index shows that both
the star-forming arms as well as near-side dust lanes can be seen at
8$\mu$m. Even the far-side lanes in the bulge region are clear at
8$\mu$m, where no tilt asymmetry is manifested. Willner et al. (2004)
argue that the dust emission from M81's ISM is likely dominated
by polycyclic aromatic hydrocarbons (PAHs; Gillett et al. 1973) 
which have a prominent
emission feature at 7.7$\mu$m. Willner et al. also showed good correspondence
between the nonstellar dust emission in M81 and the distribution of 
near-ultraviolet (NUV) emission. Some regions with bright dust emission and
little NUV emission were attributed to excessive UV extinction, while 
areas with bright NUV and little dust emission were attributed to the effects
of supernovae.

The $B-I$ color index map of M81 shows an additional set of dust lanes that
have no counterpart in the 8$\mu$m map. These lanes are oriented roughly
perpendicularly to the major axis about halfway between the center and the 
northern arm. Sandage \& Bedke (1994) interpret these as foreground dust
associated with high galactic latitude nebulosities in the halo of our
Galaxy. 

\section{Intermediate and High Redshift Galaxy Morphology}

The key to detecting observable evidence for galaxy evolution, that is,
to actually see morphological differences that are likely attributable to 
evolution, is to observe galaxies at high redshift with sufficient resolution 
to reveal significant details of morphology. Butcher \& Oemler (1978)
had already found strong evidence for morphological 
evolution in the excess number of blue galaxies in very distant ($z$=0.4-0.5)
rich galaxy clusters. These authors
suggested that the blue galaxies are spiral galaxies, and that by the
present epoch, these are the galaxies that become the S0s that dominate
nearby rich, relaxed clusters (like the Coma Cluster, Abell 1656).
An excellent summary of the issues connected with high redshift 
morphological studies is provided by van den Bergh (1998).

Progress on galaxy morphology at high redshift could only be achieved
with the resolution and depth of the Hubble Space Telescope.
The Hubble Deep Field North (HDF-N, Williams et al. 1996),
South (HDFS, Volonteri et al. 2000), and Ultra-Deep Field (HUDF, 
Beckwith et al. 2006), and the GOODS (Great Observatories
Origins Deep Survey; Giavalisco et al. 2004), GEMS (Galaxy Evolution from
Morphology and SEDS; Rix et al. 2004), COSMOS (Cosmic Evolution Survey;
Scoville et al. 2007), and other 
surveys (e.g., Cowie et al. 1995), have provided a large body of information to work with. 
For example, studies of galaxies in the redshift range 0.3 $\leq z \leq$ 0.9 show that 
the proportion of irregular-shaped galaxies dramatically increases  (e.g., 
Abraham et al. 1996). 
This means that the Hubble sequence as we know it did not always exist but was
built up over time via mergers or secular evolution or both.
Observations of submillimeter sources (Chapman et al. 
2003) suggest some of these irregulars are extended major mergers. 

Interpretation of evolution in the various deep surveys depends on knowledge of redshifts, which can
be difficult to measure spectroscopically. A very useful technique for isolating galaxies in high redshift
ranges is the UV-drop out method (Steidel \& Hamilton 1992). Galaxies are compared in different filters,
such as $B_{435}$, $V_{606}$, $i_{775}$, and $z_{850}$. If the redshift is high enough to move the Lyman limit at 0.0912$\mu$m out
of any of the first three filters, there will be a significant drop in flux owing to absorption by hydrogen, 
and the galaxy is said to ``drop out." Galaxies found in this way are called
``Lyman break" galaxies because the effect is partly caused by the spectral
characteristics of hot stars, which show such a break. (Another cause
is UV self-absorption.) 

Steidel et al. (1996) used the UV drop-out approach to identify high $z$ galaxies in the
HDF-N, and also used direct spectroscopy to confirm that the method works.
Beckwith et al. (2006) utilized the method to identify galaxies in the HUDF having $z$ from
3.5 to 7. If a galaxy is seen to drop out of a $U$-band filter such as F300W and seen in a $B$-band
filter such as F450W (both used for the HDF-N; Williams et al. 1996), then the redshift range selected
is $z$=2.4 to 3.4.  van den Bergh (1998) argues that most Lyman
break galaxies are young ellipticals or bulges.

Van den Bergh et al. (2000) describe the issues connected with morphological classifications
of galaxies to redshifts
of $z$$\approx$1. Resolution, band-shifting, and selection effects due to the magnitude-limited
nature of surveys all enter into the interpretation of intermediate to high redshift galaxy morphology. 
Resolution is important because,
typically, a nearby galaxy will have 100 times or more pixels in the image than a high $z$ galaxy will
have for classification. This is fewer pixels than sky survey images of nearby galaxies would have.
The bandpass effect is important because the $B$-band, the standard wavelength for historical
galaxy classification, is not sampling the same part of the spectrum as it would for nearby galaxies.
For example, at $z$=1, a $B$-band filter samples mid-ultraviolet light ($\approx$0.22$\mu$m)
and would be much more sensitive to young star-forming regions
than it would be for nearby galaxies.
Ideally, then, for comparison with nearby galaxies we would like to choose a redshifted band as close as
possible to the {\it rest-frame} $B$-band.
Even accounting for all these effects, significant differences
between nearby and distant morphologies do exist. For example, van den Bergh et al. (2000)
discuss the paucity of grand-design spirals and barred galaxies in the HDF-N, and use artifically
redshifted images of nearby galaxies to demonstrate that the deficiencies are
likely to be real.

Figure~\ref{highz} shows several of the different categories of
intermediate and high redshift galaxy morphologies, based on $V$ and
$i$-band images from the GEMS and Hubble UDF. The redshifts range from
0.42 to 3.35 and provide a wide range of look-back times. First, in
such a range, some galaxies look relatively normal, as shown by the
spiral and elliptical galaxies in the two upper left frames of
Figure~\ref{highz}. The $z$=0.59 spiral is classifiable as type SA(s)bc
and the elliptical as type E3.  The $z$=0.99 spiral shown in the middle
right frame of Figure~\ref{highz} has larger clumps, no clear central
object, and more asymmetry than the $z$=0.59 spiral, but is still
recognizable as a spiral. However, other less familiar categories are
found.  In general, high redshift galaxies are smaller than nearby
galaxies on average (e.g., Elmegreen et al.  2007a=EEFM07).

``Chain galaxies" were first identified by Cowie et al. (1995) and
are linear structures with superposed bright knots that have sizes and blue colors similar
to normal late-type galaxies and relatively flat major axis luminosity
profiles. They have the shapes of edge-on disk galaxies but lack clear bulges
or nuclei. A recent study by Elmegreen, Elmegreen, \& Sheets (2004=EES04) of faint
galaxy morphologies (redshifts 0.5-2) in the Advanced Camera for Surveys (ACS)``Tadpole"
galaxy (UGC10214) field showed that chain galaxies are the most
common linear morphology at magnitudes fainter than $I$=22, accounting for more than
40\% of the sample. Their dominance (also found by Cowie et al. 1995) is interpreted by EES04
as a selection effect because relatively optically thin edge-on galaxies are more favored to be
seen near the limit owing to a higher projected surface brightness than
for face-on versions of the same galaxies. EES04 suggest that chain galaxies
are edge-on irregular galaxies that will evolve to late-type disk galaxies.
Chains are the most flattened linear morphology at faint magnitudes.

``Clump clusters" (EES04) are
somewhat irregular collections of blue knots or clumps with very faint emission between
clumps. The clumps have sizes of $\approx$500pc and masses of $\approx$10$^8$--10$^9$ $M_{\odot}$.
Both of the examples
shown in Figure~\ref{highz} have $z$$>$1. Elmegreen, Elmegreen, \& Hirst (2004=EEH04) identified
clump clusters as the face-on counterparts of the linear chain galaxies, based on the similarities
of the properties of the clumps with those seen in chain galaxies, and on the
distribution of axis ratios of the systems as compared with normal disk galaxies.
The lack of a bulge clump is also consistent with this conclusion.  Nevertheless, analysis
of NICMOS IR images in the HUDF led Elmegreen et al. (2009a) to conclude that 30\% of
clump clusters and 50\% of chain galaxies show evidence of young bulges, implying that
at least half of these galaxies might be genuinely bulgeless. 
In a related study, Elmegreen et al. (2009b) show that
the best local analogues of clump clusters are dwarf irregular galaxies like Ho II, 
scaled up by a factor of 10-100 in mass. This study also brought attention to clump clusters
with faint red background disks, as opposed to blue clump clusters which lack such a feature.
Elmegreen et al. argue that the red background clump clusters are part of an evolutionary sequence leading
from the blue clump clusters to spirals with a ``classical" bulge (e.g., KK04). The
clumps, formed by gravitational instabilities in a turbulent disk, are large and few in
number, and thus will eventually coalesce near the galaxy center if they survive
the effects of supernova explosions (e. g., Elmegreen, Bournaud, \& Elmegreen 2008). 

``Tadpoles" (van den Bergh et al. 1996) are asymmetrically-shaped ``head-tail" morphologies 
with a bright off-centered nucleus and a tail, like a tadpole. A rare local example is
NGC 3991. Tadpoles were recognized in 3\% of the galaxies in HDF-N and were found to
be very blue in color. Usually both the head and the tail are blue, but van den Bergh
et al. (1996) show one example where the head is red and the tail is blue. EEH04 showed that
tadpoles have neither exponential major axis profiles nor clear bulges, and in their sample of
linear objects, tadpoles are the least frequently seen.

The bottom frames of Figure~\ref{highz} show bent chains and rings or partial
rings (Elmegreen \& Elmegreen 2006=EE06). The rings and partial rings are
thought to be mostly collisional in nature (i.e., like the conventional
ring galaxies shown in Figure~\ref{ringtypes}) and show the different
morphologies expected when small companions plunge through a larger disk
galaxy in different ways (Appleton \& Struck-Marcell 1996). Although
bent chains resemble the partial rings, they lack offset nuclei and any evidence
of a background, more face-on disk. EE06 suggest that bent chains are simply
warped versions of the more common linear chains that have suffered an
interaction. The ages of the bent chain clumps are younger than those
found in rings and partial rings, and EE06 argue that relative separations and
sizes of the clumps indicate they form by gravitational instabilities.

Elmegreen et al. (2005) show that approximately 1/3 of the ellipticals catalogued
in the HUDF have prominent blue clumps in their centers
(see also Menanteau et al. 2001, 2004). They argued that these
clumps probably imply accretion events based on comparison of their magnitudes
and colors with local field objects. Menanteau et al. (2001) were able
to reproduce the color distributions with a model having a
starburst superposed on a pre-existing older stellar population.

Galaxy morphology at intermediate and high redshifts also includes
obvious interacting cases as well as possible merger morphologies.
Bridges, tidal tails, plumes, and even M51 analogues are seen as in
nearby galaxies, but are smaller in scale than for nearby objects
(EEFM07). The middle frame of the bottom row shows a possible merger
in progress of two bent chains (or, alternatively, two interacting
spirals), called an ``assembly galaxy" by EEFM07 because they appear to
be assembling from smaller objects.  EEH04 and EEFM07 also discuss the
double systems, considered another category of the linear systems.  The
double systems like the $z$=3.35 one shown in Figure~\ref{highz} are
probably merging ellipticals. EEFM07 also describe ``shrimp galaxies",
which appear to be interacting galaxies with a single curved arm or
tail, curling at one end into a ``body."

Other studies of high redshift galaxy morphology have focussed on the specific redshift ranges that
are selected by the UV drop-out technique. Conselice \& Arnold (2009) examined the morphologies
of galaxies in the $z$=4-6 range from the HUDF, and measured quantitatve parameters such as
the concentration-asymmetry-clumpiness (star formation)
parameters (CAS; Conselice 2003) and other related parameters that are
useful for distinguishing mergers from normal galaxies. 
The CAS system is based on simple global parameters that are easily
derived automatically for large numbers of galaxies. Conselice (2003)
tied the $C$ parameter to the past evolutionary history of galaxies
while parameters $A$ and $S$ measure more active evolution from mergers
and star formation. Conselice \& Arnold found that half of the
HUDF drop-out galaxies they studied have significant asymmetries and may be undergoing merging,
while the other half is mainly smooth symmetric systems that may have collapsed quickly into a
temporary, quiescent state. 

Other quantitative approaches to these issues include the
Sersic $n$ index that characterizes radial luminosity profiles
(Ravindranath et al. 2006; Elmegreen et al. 2007b) and the Gini
coefficient (Abraham et al. 2003; Lotz et al. 2006). The Gini
coefficient provides a way of quantifying high redshift morphology
that does not depend on galaxy shape or the existence of a well-defined
center, and is well-suited to the kinds of objects shown in 
Figure~\ref{highz}. Lotz et al. (2006) found in a sample of 82
Lyman break galaxies that 10-25\% are likely mergers, 30\% are
relatively undisturbed spheroids, and the remainder are disks,
minor mergers, or post-mergers.

Given the rise in peculiar and irregular-shaped galaxies with increasing redshift, the question naturally arises: 
when did the Hubble sequence and all its accompanying details fall into place? This question is considered by 
Conselice et al. (2004), who quantitatively analyzed a well-defined high 
redshift sample using the CAS system. Conselice et al. identify 
``luminous diffuse objects" (LDOs) as galaxies having $C$ less than 1$\sigma$ below the average,
and ``luminous asymmetric objects" (LAOs) as galaxies having $A>S$ .
Some of both classes of objects are covered by the Elmegreen et al. categories described above.
All of the LDOs and LAOs have $M_B$ $<$ $-$19, and Conselice  et al.
suggest such objects might be the precursors of modern disk 
and elliptical galaxies. These are found in the redshift range 0.2 $<z<$ 2, 
suggesting the present day Hubble sequence began taking shape in this interval.
Conselice et al. (2008) consider the morphologies of galaxies more massive than 10$^{10}$ $M_{\odot}$
and in the range 1.2$<$ $z$ $<$ 3. To a $z_{850}$ magnitude of 27, the majority of these galaxies
are peculiar. They conclude that such galaxies undergo 4.3$\pm$0.8 mergers to $z$=3.

\section{Giant Low Surface Brightness Galaxies}

The van den Bergh luminosity classes highlight how luminosity and surface
brightness generally go together. Low surface brightness usually means
low luminosity and small size, hence a dwarf classification. However,
the discovery of rare giant low surface brighness (GLSB) galaxies
by Bothun et al. (1987) shows that morphology can sometimes be misleading
for judging absolute luminosity.
The hallmarks of these objects are a relatively normal bulge and an extremely 
low surface brightness, very large disk. Disk radial scalelengths and 
luminosities are unusually large, 
and extrapolated disk central surface brightnesses are unusually faint,
compared to more normal spirals.
The disks tend to be relatively smooth with a few large, 
isolated HII regions. Bothun et al. (1987) point to a model whereby the disks of these galaxies have such a low gas 
surface density that they are largely unevolved, due to the inefficiency of star formation. 

GLSB galaxies can be classified within the Hubble-Sandage
and de Vaucouleurs classification systems although, as noted by McGaugh, Schombert, and 
Bothun (1995), the majority are classified later than stage Sc. Bulges, bars, rings, and spiral patterns are 
evident in some examples, in spite of the low disk surface brightness. 
Figure~\ref{glsbs} shows three of the originally recognized GLSB examples
(McGaugh, Schombert, \& Bothun 1995): Malin 2 (also known as F568$-$6),
UGC 1230, and UGC 6614. In the images, the length of a side is 131, 38,
and 77 kpc, respectively. These can compared to the giant normal spiral NGC 7531
in the far right frame, where the length of a side is also 38 kpc. Malin
2 and UGC 6614 are especially enormous physical objects. UGC 1230 is
also very large for such a late-type morphology. van den Bergh (1998)
likens the size of Malin 2 to the core of a cluster of galaxies.
He considers ``monsters" like Malin 1 and Malin 2 
to be only one of three types of LSB galaxies. Some LSB galaxies are
as big as normal galaxies, like UGC 1230. Most LSB galaxies, however,
are dwarfs: ellipticals, irregulars and, less frequently, spirals.
These are described further in section 15.2.

An example of another object that could be considered a large LSB galaxy, 
but which lacks a bulge or any evident 
recent star formation, is SGC 2311.8$-$4353, the mysterious ghost-like
companion close to the right of NGC 7531 in Figure~\ref{glsbs}. This
peculiar object is 2/3 the size of NGC 7531 (Buta 1987) but has unknown
redshift. If it is associated with NGC 7531, it would be as much as 30 kpc
in diameter at the faintest detectable isophote level and would clearly
not be a dwarf.

A recent study of three GLSBs (Malin 1, UGC 6614, and UGC 9024)
by Rahman et al. (2007) showed that IR emission
from such objects is consistent with their optically-determined low
star formation rates, with the diffuse optical disks being undetected
from two of the three. A dynamical study of two GLSBs (Malin 1 and
NGC 7589) by Lelli, Fraternali,
\& Sancisi (2010) led the authors to conclude that at least in these
cases, the GLSB galaxy can be thought of as an inner high surface brightness
galaxy having a very extended LSB disk. This is based on the steeply
rising rotation curves found for these galaxies, which is very much like
what is seen in early-type high surface brightness galaxies.

Impey \& Bothun (1997) argue that LSB galaxies brighter than $M_B$=$-$14 contribute significantly to the luminosity 
density of the local universe, are dark-matter-dominated at almost all radii, and have an evolutionary history 
involving late collapse of a low amplitude perturbation, a low star formation rate, and very slow changes. Large 
LSBs are greatly underrepresented in galaxy catalogues but are clearly an important class of objects.

\section{Galaxy Morphology in Color}

\subsection{Normal Galaxies}

The Sloan Digital Sky Survey (York et al. 2000; Gunn et al. 1998)
provides the largest body of information on the colors of galaxies. The
multi-wavelength imaging in $ugriz$ filters has allowed the production
of high quality color images for thousands of galaxies.  Although the
Hubble-Sandage-de Vaucouleurs classification systems were based on blue
light images alone, it is still possible to reliably classify galaxies
with SDSS color images, which are based on combined $gri$ images
(Lupton et al. 2004). In such images one can directly see the stellar
population differences that characterize different galaxy types.

Figure~\ref{colortf} shows the color Hubble sequence from E to Sm.
Once galaxy colors were systematically measured using photoelectric
photometry (de Vaucouleurs, de Vaucouleurs, \& Corwin 1976), it was
noted that integrated colors vary smoothly with advancing stage along
the Hubble sequence. The latest stages have corrected total color index
$(B-V)_T^o$$\approx$0.3-0.4 while E and S0 galaxies have
$(B-V)_T^o$$\approx$0.9-1.0 (e.g., Buta et al. 1994). The latter colors
correspond to yellow-orange while the former are bluish-white. The
colors begin to change at S0/a and Sa and become progressively bluer.
Figure~\ref{colortf} shows the reason for the change. Galaxies earlier
than Sa are dominated by old stars having the colors of K giants.  As
stage advances from Sa to Sm, the spiral structure becomes
progressively more important compared to the bulge. Since the arms are
dominated by complexes of massive young stars, this makes the
integrated colors of the galaxies become progressively bluer until by
the end of the sequence, the bluer colors of these stars have overcome
the yellowish light of the background old disk stars.
Figure~\ref{colortf} also shows that the intermediate colors of
intermediate types such as Sb and Sc are due to the yellowish-orange
colors of bulges and bars as combined with the bluer colors of spiral
arms.

The analysis of integrated SDSS galaxy colors for more than 100,000
galaxies led to one of the most dramatic findings of the survey: a
clear bimodality in the distribution of color that correlates with
morphology: a red peak that includes mainly E, S0, and Sa galaxies, and
a blue peak that includes mainly Sb, Sc, and Irr galaxies (Strateva et
al. 2001). Although the correlation of galaxy color with types had been
known for a long time from photoelectric measurements (e. g., de
Vaucouleurs 1961; Buta et al. 1994), the large sample provided by SDSS
allowed the bimodality to be demonstrated to a high degree of
significance. In plots of $u-r$ color index versus absolute $M_r$
magnitude, the galactic equivalent of a stellar H-R diagram, nearby
early-type galaxies follow a narrow band called the red sequence, while
nearby later-type, mostly star-forming galaxies appear as a broad blue
sequence (also sometimes called the ``blue cloud").  Baldry et al.
(2004) showed that the bimodality (in the form of a double Gaussian
number distribution over all types) is detectable from $M_r$ $\approx$
$-$15.5 to $M_r$ $\approx$ $-$23, being undetectable only for the most
luminous galaxies. Wyder et al. (2007) showed that use of GALEX
near-ultraviolet magnitudes and optical $r$-band magnitudes provides
even greater discrimination between the Gaussian components. Bell et
al. (2004) showed that the bimodality is detectable in faint galaxies
to $z$$\approx$1, indicating that this characteristic of the galaxy
population extends to a lookback time of at least 9Gyr. It is thought
that galaxies evolve from the blue sequence to the red sequence as
their star formation is quenched, perhaps through mergers, gas
depletion, or AGN feedback (e.g., Martin et al. 2007). The possibility
of evolution has engendered great interest in the galaxies lying near
the minimum of the bimodal distribution (the so-called ``green valley";
Thilker et al. 2010 and references therein).

\subsection{Dwarf Galaxies}

Virtually all the galaxies shown in Figure~\ref{colortf} are of
relatively high luminosity, with absolute blue magnitudes $M_B^o$
averaging about $-$20. When physical parameters such as $M_B^o$ are
considered, it becomes clear that the peculiar shape of the de
Vaucouleurs classification volume shown in
Figure~\ref{classification_volume} only highlights the morphological
diversity of families and varieties at each stage, but does not tell us
about the {\it physical parameter space} at each stage, which expands
considerably at each end of the volume (McGaugh, Schombert, and Bothun
1995). Most known dwarf galaxies are either early or late-type, but
not intermediate.

\subsubsection{dE, dS0, BCD, and cE Galaxies}

The most extensive study of dwarf galaxy morphologies
was made by Binggeli, Sandage, \& Tammann (1985=BST), who used deep
photographs to probe the low luminosity population of the Virgo Cluster,
using mostly morphology to deduce cluster membership. Examples
of several categories of Virgo Cluster dwarf galaxies are shown in 
Figure~\ref{dwarfs} using SDSS color images and the classifications of BST.
The most common type is the dwarf elliptical, or dE type, which accounts
for 80\% of the galaxies in the BST catalogue. dE galaxies range 
from $M_B$ = $-$18 to $-$8 (Ferguson \& Binggeli 1994). Many dEs
have an unresolved, star-like nucleus whose presence is indicated
by an N attached to the type, as in dE0,N. The top right panels of
Figure~\ref{dwarfs} show three fairly typical examples. Possibly
related to these normal dE systems are the larger, lower surface brightness
ellipticals (``large dE") shown in the two lower right frames of
Figure~\ref{dwarfs}. 

The second row of Figure~\ref{dwarfs} shows examples of the interesting
class of dwarf S0 galaxies. All of the examples shown are distinct from
dEs in showing a smooth structure but with additional features such as
a lens or a weak bar. dS0 galaxies can also be nucleated and are called
dS0,N. In addition to the low surface brightness dEs and dS0s, the Virgo
Cluster includes two high surface brightness classes of dwarfs. The
cE category refers to compact ellipticals that resemble M32.

The blue compact dwarf (BCD) galaxies are a special class of
star-forming dwarf irregulars characterized by a few bright knots
imbedded in a stellar background of low surface brightness (Sandage and
Binggeli 1984). The most extreme cases are nearly stellar (Thuan \&
Martin 1981).  The knots are often super star clusters associated with
30-Doradus-like HII regions, and the faint background can be very blue
(Thuan et al. 1997).  Spectroscopically, BCDs have narrow emission
lines superposed on a blue continuum, and the lines indicate a low
metallicity. Figure~\ref{dwarfs} shows three examples from the BST
Virgo Cluster catalogue.  Gil de Paz et al. (2003) present an atlas of
more than 100 BCDs that highlight their structure.

As described in section 5.1, the dE galaxies shown in
Figure~\ref{dwarfs} are {\it not} the low luminosity extension of more
luminous ellipticals. These together with the dS0 class are labeled
``spheroidal" galaxies by Kormendy et al. (2009; see
Figure~\ref{cores}), who confirmed the finding by Kormendy (1985) that
these galaxies are more related to evolved low luminosity spirals and
irregulars than to genuine ellipticals.  Based on correlations of
well-defined photometric parameters (e.g., central surface brightness
or velocity dispersion versus core radius or absolute magnitude), these
authors link ellipticals like M32 to the actual low luminosity end of
the E galaxy sequence (see also Wirth \& Gallagher 1984). Thus, the cE
galaxies shown in Figure~\ref{dwarfs} are in a sense truer ``dwarf
ellipticals" than the dE galaxies shown. This does not negate the value
of the BST classifications, since these are purely morphological
interpretations.

\subsubsection{Local Group Dwarf Spheroidals and Irregulars}

The lowest luminosity galaxies that we can study in detail are in the
Local Group. Fortunately, a few are in the area covered by the SDSS so
that they can be illustrated in color. These objects, all fainter than
$M_V$=$-$12, are shown in Figure~\ref{lgroup}. Leo I and Leo II are
usually called dwarf spheroidal, or dSph, galaxies (e.g., Ferguson \&
Binggeli 1994).  dSph galaxies tend to have to have absolute visual
magnitudes $M_V$ $>$ $-$15 and a low degree of flattening; they are
believed to be the most abundant type of galaxy in the Universe. Leo I
($M_V$=$-$11.9) and Leo II ($M_V$=$-$9.6) look different in part
because of their different star formation histories. dSph and dwarf
irregular (dI) galaxies are now known to have complex and varied star
formation histories that may involve multiple episodes of star
formation and effects of interactions (Mateo 1998 and references
therein). The other two galaxies in Figure~\ref{lgroup}, Leo A and DDO
155, are dwarf irregulars having $M_V$$\approx$$-$11.5. A detailed HST
study of the stellar content of Leo A (Cole et al. 2007) showed that
90\% of the star formation in the galaxy occurred less than 8 Gyr ago,
with a peak at 1.5-3 Gyr ago. A useful summary of the properties of
dSph galaxies is provided by van den Bergh (1998) and, most
recently, by Tolstoy, Hill, \& Tosi (2009), who also discuss the
star formation histories of Local Group dI galaxies.

The SDSS has facilitated the discovery of many new Local Group dwarf
galaxies (Belokurov et al. 2007). For example, the SDSS led to the
discovery of one of the faintest known dSph galaxies, a new dwarf in
Ursa Major (called the UMa dSph) having $M_V$ $\approx$ $-$7 (Willman
et al. 2007). Most interesting is Leo T, which is an $M_V$ = $-$7.1
dSph galaxy with some recent star formation, providing one of the most
dramatic illustrations of the link between dSph and dI galaxies, and
the least luminous galaxy known to have recent star formation (Irwin et
al. 2007).

\subsubsection{Dwarf Spirals}

Sandage \& Binggeli (1984) described the classification of dwarf galaxies
based on the Virgo Cluster, and concluded that there are ``no real dwarf
spirals." This refers mainly to dwarf spirals that might be classified
as types Sa, Sb, or Sc, i.e., having both a bulge and a disk. Dwarf
late-type spirals are already built into de Vaucouleurs's modified
Hubble sequence as Sd-Sm types and connect directly to Magellanic 
irregulars, as shown
in Figure 1 of Sandage \& Binggeli (1984). Thus, any genuine examples of
dwarf Sa, Sb, or Sc spirals would be of great interest as they would
challenge the idea that for a galaxy to be able to make well-defined
spiral arms, it would have to be more massive than some lower limit
(estimated as 5$\times$10$^9$ $M_{\odot}$ by Sandage \& Binggeli).

Four of the best cases of genuine dwarf spirals are illustrated
in Figure~\ref{dwarfspirals}. The two rightmost frames
show IC 783 (BST type dS0,N) and IC 3328
(BST type dE1,N), both Virgo Cluster members having absolute
magnitudes $M_B^o$ $\approx$ $-$16 to $-$17, and found to have subtle
spiral structure by Barazza, Binggeli, \& Jerjen (2002) and Jerjen,
Kalnajs, \& Binggeli (2000), respectively. 
The patterns are hard to see in the direct SDSS color images
shown in Figure~\ref{dwarfspirals}, but
these authors use photometric models,
Fourier decomposition, and unsharp-masking to verify
the reality of the patterns. Barraza et al. conclude that many
of the bright early-type dwarfs in the Virgo Cluster have disks.
An example with a bar {\it and} spiral arms is NGC 4431 (shown as
dS0 in Figure~\ref{dwarfs}).

The leftmost panel of Figure~\ref{dwarfspirals} shows an HST wide
$V$-band image (Carollo et al. 1997) of NGC 3928, an absolute magnitude
$M_B^o$ = $-$18 galaxy which on small-scale, overexposed images looks
like an E0, but which harbors a miniature (2kpc diameter),
low-luminosity Sb spiral (van den Bergh 1980b).  Based on spectroscopic
analysis, Taniguchi \& Watanabe (1987) have suggested that NGC 3928 is
a spheroidal galaxy which experienced an accretion event that supplied
the gas for star formation in the miniature disk.

Schombert et al. (1995) brought attention to possible dwarf field spirals. 
One of their examples, D563$-$4, is shown in the second frame from the left in 
Figure~\ref{dwarfspirals}. This galaxy has $M_B^o$$\approx$$-$17. A few
other examples are given in the paper, and Schombert et al. find 
that they are not in general grand design spirals, are physically
small, and have low HI masses. However, van den Bergh (1998) considers
all of Schombert et al.'s examples as subgiant spirals rather then true 
dwarf spirals. A possible true dwarf spiral given by van den Bergh is
DDO 122 (type S V).

\subsection{Galaxy Zoo Project}

The Galaxy Zoo project (Lintott et al. 2008) has made extensive use of
SDSS color images. The project uses a website to enlist the help of
citizen scientists worldwide to classify a million galaxies as well as
note interesting and unusual cases in various forum threads. With such
a large database to work from, and the potential for discovery being
real, the project has attracted many competent amateur galaxy
morphologists.  One such discovery was a new class of galaxies called
``green peas," which are star-like objects that appear green in the
SDSS composite color images (Figure~\ref{zoofig}, left).  Cardemone et
al. (2009) used auxiliary SDSS data to show that peas are galaxies that
are green because of a high equivalent width of [OIII] 5007 emission.
They are sufficiently distinct from normal galaxies and quasars in a
two-color $g-r$ vs. $r-i$ plot that such a plot can be used to identify
more examples. Other characteristics noted are that peas are rare, no
bigger than 5 kpc in radius, lie in lower density environments than
normal galaxies, but may still have morphological characteristics
driven by mergers, are relatively low in mass and metallicity, and have
a high star formation rate. Cardemone et al.  conclude that peas are a
distinct class of galaxies that share some properties with luminous
blue compact galaxies and UV-luminous high redshift galaxies.

Another prominent colorful galactic-sized object identified by Galaxy
Zoo is ``Hanny's Voorwerp" (Figure~\ref{zoofig}), a peculiar collection
of blue clumps just south of IC 2497. Lintott et al. (2009) found that
the Voorwerp is mostly ionized gas, and after ruling out some possible
sources of ionizing radiation, concluded that the object could be the
first identified case of a quasar light echo. The implication is that
the companion galaxy underwent a temporary quasar phase.

Masters et al. (2010) examine the properties of face-on late-type
spiral galaxies whose colors are much redder than is typical
(Figure~\ref{zoofig}, right panels), suggesting that they are passive
objects where star formation has largely ceased or is lower than
normal.  These authors showed that red spirals are not excessively
dusty and tend to be near the high end of the mass spectrum. A range of
environmental densities was found, implying that environment alone is
not sufficient to make a spiral red. A significantly higher bar
fraction was found for red spirals as compared to blue spirals, which
Masters et al. suggest could mean the bars themselves may have acted to
shut down the star formation in these galaxies.

A major philosophical aspect of Galaxy Zoo is the value of human visual
interpretation of galaxy morphology. That is, the human eye can
integrate the detail in an image more reliably than a computer program
can, in spite of the latter's ability to classify numbers of galaxies
well beyond the capability of a single individual. This philosophy was
used to compile a major catalogue of galaxy merger pairs described by
Darg et al. (2010). In the initial set-up of Galaxy Zoo, a single
button allowed a classifier to select whether an object was a ``merger"
based simply on the appearance of peculiarities.  For each galaxy, a
weighted average number, $f_m$, was derived that characterized the
fraction of classifiers who interpreted a pair of galaxies as a merger,
essentially ``morphology by vote" and in a way reminiscent of RC3 where
morphological types were in many cases based on a weighted average from
a small number of classifiers (Buta et al. 1994). Taking $f_m$ $>$ 0.4,
Darg et al. (2004) identified 3003 pairs and groups of merging
galaxies, and also showed that the spiral to elliptical galaxy ratio in
merger pairs is a factor of two higher than for the global galaxy
population, suggesting that mergers involving spirals are detectable
for a longer period than those that do not involve spirals.

A similar philosophy to Galaxy Zoo was used by Buta (1995) to compile
the Catalogue of Southern Ringed Galaxies, and also by Schawinski et
al. (2007), who visually classified 48,023 SDSS galaxies to identify a
significant-sized sample of early-type galaxies for a study of the
connection between nuclear activity and star formation. In the latter
study, visual interpretation was argued to be needed to avoid bias
against star-forming early-type galaxies which would be excluded
from color-selected samples. A major result of this study was the
identification of a time sequence whereby an early-type galaxy has
its star formation suppressed by nuclear activity, a manifestation
of AGN feedback.

\subsection{Isolated Galaxies}

If interactions and mergers can have profound effects on
galaxy morphology, then the morphology of {\it isolated} galaxies
clearly is of great interest. Such galaxies allow us to see how
internal evolution alone affects morphology, i. e., what pure
``nature" morphologies look like. Karachentseva (1973) compiled
a large catalogue (the Catalogue of Isolated Galaxies, or
CIG) of 1050 isolated galaxies that has proven very useful
for examining this issue. A galaxy of diameter $D$ is sugggested
to be isolated if it
has no comparable-sized companions of diameter $d$ between $D$/4 and 
4$D$ within a distance of 20$d$. Verdes-Montenegro et al. (2005)
show that this means that an isolated galaxy 25 kpc in diameter,
in the presence of a typical field galaxy velocity of 150 km s$^{-1}$,
has not been visited by a comparable mass companion during the
past 3Gyr. These authors discuss the limitations of the CIG (e. g., 
the isolation criteria do not always work),
but in general it is the best source of isolated galaxies available.

Sulentic et al. (2006) examined all 1050 CIG galaxies on Palomar II
sky suvey charts in order to refine the sample and found that isolated 
galaxies cover all Hubble
types. Of these, 14\% were found to be E/S0 types, while 63\% were
Sb-Sc types, with the spiral population more luminous than the E/S0
population. Over the type range Sa-Sd, the proportion 
rises to 82\%. Thus, an isolated galaxy sample is very spiral-rich.
Nevertheless, the presence of
early-type galaxies in the sample implies that these are not likely 
to be ``nurture" formed, as such galaxies might be in denser environments. 
The refinement of the CIG sample forms the basis of the Analysis of the
Interstellar Medium of Isolated Galaxies (AMIGA) project (Verdes-Montenegro
et al. 2005).

Figure~\ref{isolateds} shows six examples of isolated Sb-Sc galaxies
from the AMIGA sample, based on SDSS color images. All of these look
relatively normal, but it is interesting how nonbarred galaxies like
NGC 2649 and 5622 (upper left frames in Figure~\ref{isolateds}) show
such conspicuous global spirals, which argues that the spirals in these
galaxies have not been excited by an interaction. 

Durbala et al. (2008)
analyzed the photometric properties of 100 isolated Sb-Sc AMIGA galaxies,
and found that a majority have pseudobulges rather than classical bulges.
In comparing the properties of isolated galaxies with a sample of Sb-Sc
galaxies selected without an isolation criterion, Durbala et al. found
that isolated spirals have longer bars and, using CAS parameters,
also less asymmetry, central concentration, and clumpiness, 

Durbala et al. (2009) analyzed the Fourier properties of the same set of
100 isolated spirals, and estimated bar lengths and strengths. Earlier
types in the sample were found to have longer and higher contrast bars
than later types. Spiral arm multiplicities were investigated also,
and it was shown that cases having an inner $m$=2 pattern and an outer
$m$=3 pattern occurred in 28\% of the sample. Elmegreen, Elmegreen,
\& Montenegro (1992) argued that in such morphologies, the $m$=3
pattern is driven internally by the $m$=2 pattern, and that three-armed
patterns measure the time elapsed since an interaction.

\subsection{Deep Field Color Imaging}

Particularly interesting in the domain of color galaxy morphology are
the various deep field surveys that have been at the heart of high
redshift studies. The most recent is the HST Wide Field Camera 3 Early
Release Science (ERS) data (Windhorst et al. 2010), which provides very deep panchromatic
images based on 10 filters ranging from 0.2$\mu$m to 2$\mu$m in wavlength. Figure~\ref{ers}
shows three subsections of the main ERS field, which coincides with the
GOODS south field (Giavalisco et al. 2004). The galaxies seen range in redshift
from $z$=0.08 to at least $z$=3. Windhorst et al. (2010) argue that images
like the ERS field are deep enough to allow probing of galaxy evolution in the
crucial redshift range $z$$\approx$1-3 where the galaxies assembled into their
massive shapes. By $z$$\approx$1, the Hubble sequence was largely in place.
Figure~\ref{ers} shows a variety of interesting nearby as
well as high $z$ morphologies, including some of those illustrated in Figure~\ref{highz}.
Images like the ERS field allow us to connect the local and distant galaxy populations
in unprecedented detail.

\section{Large-Scale Automated Galaxy Classification}

The process of galaxy classification thus far described is a manual
exercise where an observer attempts to sort a galaxy into its
appropriate stage, family, variety, outer ring classification, etc., by
visual inspection alone. For a small number of observers, this has been
done for as many as 14,000 galaxies by Nair \& Abraham (2010) and
48,000 galaxies by Schawinski et al. (2007), while for a large number
of observers working in concert (e.g., the Galaxy Zoo project) it has
been done for a million galaxies.  But critical to such ventures is the
preparation of images for classification, and a need for a homogeneous,
objective approach to very large numbers of galaxies. This has led to
extensive application of automated methods for classifying galaxies.
For example, Nair \& Abraham (2010) visually classified galaxies into
coded bins for the purpose of training an automatic classification
algorithm.

(The remainder of this part will be found in the printed publication.)

%A variety of approaches have been used, ranging from ``artificial
%intelligence" (neural network) techniques (Odewahn et al. 1992; most
%recently Banerji et al. 2010), to the CAS approach (Conselice 2003) and
%the Gini coefficient (Abraham et al. 2003), and to algorithms often
%used in other disciplines such as the automated cell morphology code
%described by Shamir (2009). The idea is to first classify a sample of
%galaxies by eye, use the algorithm to also classify them, and then
%compare how well the algorithm results agree with the visual results.
%Often the goal with automatic classification is not to automatically
%derive classifications such as ``(R$^{\prime}$)SB(r,nr)ab,", but simple
%distinctions of galaxies such as ellipticals, spirals, S0s, or
%edge-ons. Conselice et al. (2004) in fact argue that automated
%classification should not focus on reproducing detailed de Vaucouleurs
%types, for example, but instead focus on more tangible properties like
%the degree of central concentration, the degree of asymmetry, and the
%degree of patchiness, all of which in a way can distinguish late-type
%galaxies from early-type galaxies (but not necessarily ringed or barred
%galaxies from those lacking such features).  The reason for such a view
%is that high redshift galaxy morphology usually does not have the
%benefit of the detail seen in nearby galaxies.  van den Bergh (1998)
%discusses other issues connected with computer classifications,
%including the limitations of artificial neural networks and the
%usefulness of objectively-measured parameters.

\section{The Status and Future of Morphological Studies}

The large number of illustrations in this article attests to the
richness of the diversity of galaxy morphology. It is, of course,
not possible to illustrate all aspects of morphology that might
be worth discussing, but most interesting is how far {\it physical
galaxy morphology} has come in the 35 years since Allan Sandage
wrote his review of galaxy morphology in Volume IX of {\it Stars
and Stellar Systems}. Galaxy morphology is no longer the purely
descriptive subject it once used to be.

Internal perturbations such as bars are apparently capable of
generating a great deal of the interesting structure we see in disk
galaxies, and more theoretical and observational studies should
elucidate this further. The impact of bars on morphology seems well
understood, as summarized in detail by KK04 in their monumental review
article on pseudobulges. Bars redistribute angular momentum and
reorganize gas clouds to flow into resonance regions and fuel star
formation. The gathering of gas into resonance regions can drive the
formation of rings, and indeed can also build up the central mass
concentration to the point of bar destruction. Even failing this, the
pile-up of gas into the nuclear region can lead to the formation of a
pseudobulge.  The richness of barred galaxy morphology attests to the
strong role secular evolution plays in structuring galaxies.

Progress in understanding the role of mergers and interactions
on galaxy morphology has also proceeded at a rapid pace. Great
success in numerical simulations and the theory of interacting galaxies has
made it possible to link a specific type of interaction to a specific
morphology (e. g., collisional ring galaxies). The complex structure
of early-type galaxies, with boxy and disky isophote shapes, shells
and ripples, and other features, shows that interactions and mergers
play an important role in molding galaxies (see the excellent review
by Schweizer 1998). With the advent of the
Hubble Space Telescope, this role has been elucidated even more
clearly because the merger rate was higher in the past.

In spite of the theoretical progress, it is interesting that classical
morphology has not lost its relevance or usefulness even after more
than 80 years since Hubble published his famous 1926 paper.  No matter
how much progress in understanding the physical basis for morphology is
made, there is still a need for the ordering and insights provided by
classical Hubble-Sandage-de Vaucouleurs galaxy classifications.
Morphology went through a low phase in the 1980s and 90s when it was
perceived that galaxy classification placed too much emphasis on
unimportant details and was too descriptive to be useful.  It was
thought that the Hubble classification had gone as far as it can go,
and that another approach needed to be tried to build a more physical
picture of galaxies. At that time, there was a sense that the focus
should be more on the component ``building blocks" of galaxies, or what
might be called galactic subsystems (e.g., Djorgovski 1992).
Quantification of morphology became more possible as advanced
instrumentation allowed more detailed physical measurements to be made.
In the end, as morphology became better understood, it also became
clear what a type such as ``(R)SB(r)ab" might really mean, which
enhanced the value of classification (KK04). In addition, numerical
simulations became sophisticated enough to make predictions about
morphology (e.g., the R$_1$ and R$_2$ subclasses of outer rings and
pseudorings).  These types of things, as well as the movement of
morphology from the photographic domain to the digital imaging domain,
the broadening of the wavelength coverage available to morphological
studies from the optical to the ultraviolet and infrared domains, the
Sloan Digital Sky Survey, and the accessiblity of high redshift
galaxies to unprecedentedly detailed morphological study, all played a
role in bringing galaxy morphology to the forefront of extragalactic
research.

Even so, the writing of this article has shown that many important
galaxies and classes of galaxies have not been studied well enough to
have much modern data available. For example, in spite of the
considerable interest in collisional ring galaxies the past 20 years or
so, Struck (2010) was forced to lament that ring galaxies ``are
underobserved." The same can be said for resonance ring galaxies, giant
low surface brightness galaxies, dwarf spirals, Magellanic barred
spirals, counter-winding spirals, and other classes of interacting
galaxies.  The most that can be said about this is that further studies
will likely be made, especially if instrumentation facilitates the
objects in question. Rotation and dynamics are far short of photometry
for most classes of galaxies, but would add a great deal of insight if
obtainable.

At the other extreme, early-type (E and S0 galaxies) continue to
be the focus of major photometric, kinematic, and theoretical research
projects. Important clues to the formation and evolution of such
galaxies are contained in their intrinsic shapes (oblate, prolate,
triaxial), in the ages, metallicities, and radial mass-to-light ratios
of their stellar populations, in their three-dimensional orbital
structure, and in the kinematic peculiarities often found in such
systems (de Zeeuw et al. 2002). Among the most recent studies are the
massive photometric analysis of early-types in the Virgo Cluster by
Kormendy et al. (2009), and the ATLAS$^{3D}$ project described by
Cappellari et al. (2011). ATLAS$^{3D}$ is the largest kinematic
database of high-quality two-dimensional velocity field information
ever obtained for early-type galaxies, including 260 such galaxies in a
well-defined and complete sample. This survey is simply the latest part
of the long-term effort by many researchers, beginning in the 1980s, to
understand early-type galaxies in terms of quantitative parameters that
can be tied to theoretical models. Early-types have been a persistent
enigma in morphological studies, and considerable evidence suggests
that the E, S0 sequence as defined by Hubble, Sandage, and de
Vaucouleurs hides a great deal of important physics associated with
these objects. The ATLAS$^{3D}$ project was designed to exploit the
$\lambda_R$ parameter described by Emsellem et al. (2007; see section
5.1), which separates early-types into fast and slow rotators and
discriminates galaxies along the red color sequence (section 15.1).

These advances for early-type galaxies do not mean that quantitative
analyses of later-type galaxies are lacking.  As codes for
two-dimensional photometric decomposition become ever more
sophisticated (e. g., Peng et al. 2010; Laurikainen et al. 2010 and
references therein), parameters that characterize the bulges, disks,
bars, lenses, rings, and spiral patterns are being derived for large
numbers of galaxies (especially barred galaxies) that were not reliably
decomposable with earlier one-dimensional approaches.

For the future, it is to be hoped that the Sloan Digital Sky
Survey will be extended to cover the whole sky, and provide
access to high quality morphological studies of several million
more galaxies, some of which might have new and exotic structures.
The {\it James Webb Space Telescope} should be able to carry
the HST's torch to greater depths and resolutions of high redshift
galaxies, to further enhance our understanding of galaxy evolution.

This article is dedicated to Allan Sandage (1926-2010), one of the 20th
century's greatest astronomers, who helped set the stage for galaxy
morphology to be one of the most active fields of modern extragalactic
research. It was Dr. Sandage's efforts that firmly cemented Hubble's
ideas on morphology into astronomy.  The author is grateful to Dr.
Sandage for the inspiration he provided for this article and for his
standard of excellence in astronomy.

The author also gratefully acknowledges the helpful comments and
suggestions from the following people that considerably improved this
article:  Martin Bureau, Adriana Durbala, Debra Elmegreen, William
Keel, Jeffrey Kenney, Johan H. Knapen, Rebecca Koopmann, John Kormendy,
Eija Laurikainen, Barry Madore, Karen L. Masters, Patrick M.
Treuthardt, Sidney van den Bergh, and Xiaolei Zhang. The author also
thanks Debra Elmegreen for providing the images of high redshift
galaxies shown in Figure~\ref{highz}, Masfumi Yagi for the
illustrations in Figure~\ref{coma}, and John Kormendy, Jason Surace,
Donald P.  Schneider, and Rogier Windhorst for the use of published
illustrations from specific papers.  This article uses images from many
sources too numerous to acknowledge here, but mainly drawn from the
dVA, the NASA/IPAC Extragalactic Database, the Ohio State University
Bright Spiral Galaxy Survey (OSUBSGS), the Sloan Digital Sky Survey,
and several published papers by other authors. The NASA/IPAC
Extragalactic Database (NED) is operated by the Jet Propulsion
Laboratory, California Institute of Technology, under contract with the
National Aeronautics and Space Administration.  Funding for the OSUBSGS
was provided by grants from the NSF (grants AST 92-17716 and AST
96-17006), with additional funding from the Ohio State University.
Funding for the creation and distribution of the SDSS Archive has been
provided by the Alfred P.  Sloan Foundation, the Participating
Institutions, NASA, NSF, the U.S.  Department of Energy, the Japanese
Monbukagakusho, and Max Planck Society.  Observations with the NASA/ESA
{\it Hubble Space Telescope} were obtained at the Space Telescope
Science Institute, which is operated by the Association of Universities
for Research in Astronomy, Inc., under contract NAS 5-26555. The {\it
Spitzer Space Telescope} is operated by the Jet Propulsion Laboratory,
California Institute of Technology, under NASA contract 1407.  The {\it
Two Micron All-Sky Survey} is a joint project of the University of
Massachusetts and the Infrared Processing and Analysis
Center/California Institute of Technology, funded by the National
Aeronautics and Space Administration and the National Science
Foundation. GALEX is a NASA mission operated by the Jet Propulsion
Laboratory. GALEX data is from the Multimission Archive at the Space
Telescope Science Institute (MAST).  Support for MAST for non-HST data
is provided by the NASA Office of Space Science via grant NNX09AF08G
and by other grants and contracts. This article has also made use of
THINGS, ``The HI Nearby Galaxy Survey" (Walter et al. 2008), and
BIMA-SONG, the Berkeley-Illinois-Maryland Survey of Nearby Galaxies
(Helfer et al. 2003).

\noindent
\centerline{REFERENCES}

\noindent
Abraham, R. G., van den Bergh, S., Glazebrook, K., Ellis, R. S., Santiago, B. X., Surma, P., \& Griffiths, R. E. 1996, \apjs, 107, 1

\noindent
Abraham, R. G., van den Bergh, S., \& Nair, P. 2003, \apj, 588, 218

\noindent
Aguerri, J. A. L., M\'endez-Abreu, J., \& Corsini, E. M. 2009, \aap, 495, 491

\noindent
Appleton, P. \& Struck-Marcell, C. 1996, Fundamentals of Cosmic Physics, 16, 111

\noindent
Arp, H. 1966, \apjs, 14, 1

\noindent
Arp, H. C. \& Madore, B. F. 1987, Catalogue of Southern Peculiar Galaxies and Associations, Cambridge, Cambridge Univ. Press

\noindent
Arribas, S., Bushouse, H., Lucas, R. A., Colina, L., \& Borne, K. D. 2004, \aj, 127, 2522

\noindent
Athanassoula, E. 1992, \mnras, 259, 328

\noindent
Athanassoula, E. 2003, \mnras, 341, 1179

\noindent
Athanassoula, E. 2005, \mnras, 358, 1477

\noindent
Athanassoula, E. \& Bosma, A. 1985, \araa, 23, 147

\noindent
Athanassoula, E. et al. 1990, \mnras, 245, 130

\noindent
Athanassoula, E., Romero-G\'omez, M., \& Masdemont, J.
J. 2009a, \mnras, 394, 67

\noindent
Athanassoula, E., Romero-G\'omez, M., Bosma, A., \& Masdemont, J.
J. 2009b, \mnras, 400, 1706

\noindent
Bacon, R. et al. 2001, \mnras, 326, 23

\noindent
Bagetakos, I. Brinks, E., Walter, F., de Blok, W. J. G., Rich, J, W.,
Usero, A., \& Kennicutt, R. C. 2009, in The Evolving ISM in the Milky
Way and Other Galaxies, K. Sheth, A.  Noriega-Crespo, J. Ingalls, and
R. Paladini, eds., http://ssc.spitzer.caltech.edu/mtgs/ismevol, E17

\noindent
Bagley, M., Minchev, I., \& Quillen, A. C. 2009, \mnras, 395, 537

\noindent
Bahcall, J. N., Kirhakos, S., Saxe, D. H., \& Schneider, D. P. 1997, \apj, 479, 642

\noindent
Baldry, I. K., Glazebrook, K., Brinkmann, J., Ivezi\'c, Z.,
Lupton, R. H., Nichol, R. C., \& Szalay, A. S. 2004, \apj, 600, 681

\noindent
Banerji, M. et al. 2010, \mnras, 406, 342

\noindent
Barazza, F. D., Binggeli, B., \& Jerjen, H. 2002, \aj, 124, 1954

\noindent
Barazza, F. D., Jogee, S., \& Marinova, I. 2008, \apj, 675, 1194

\noindent
Barway, S., Khembavi, A., Wadadekar, Y., Ravikumar, C. D., \& Mayya, Y. D. 
2007, \apj, 661, L37

\noindent
Barway, S., Wadadekar, Y., \& Khembavi, A. K. 2011, \mnras, 410, L18

\noindent
Beckwith, S. V. W. et al. 2006, \aj, 132, 1729

\noindent
Bell, E. et al. 2004, \apj, 608, 752

\noindent
Belokurov, V. et al. 2007, \apj, 654, 897

\noindent
Bershady, M., Jangren, A., \& Conselice, C. J. 2000, \aj, 119, 2645

\noindent
Bertin, G., Lin, C. C., Lowe, S. A., \& Thurstans, R. P. 1989, \apj, 338, 78

\noindent
Bertola, F. 1987, in IAU Symp. 127, p. 135

\noindent
Bertin, G. \& Amorisco, N. C. 2010, \aap, 512, 17

\noindent
Binggeli, B., Sandage, A., \& Tammann, G. A. 1985, \aj, 90, 1681

\noindent
Binney, J. 1992, ARA\& A, 30, 51

\noindent
Blakeslee, J. P. 1999, \aj, 118, 1506

\noindent
Block, D. L., Bertin, G., Stockton, A., Grosbol, P., Moorwood, A. F. M., Peletier, R. F. 1994, \aap, 288, 365

\noindent
Block, D., Freeman, K. C., Puerari, I., Combes, F., Buta, R., Jarrett, T., \& Worthey, G. 2004a, in Penetrating Bars Through Masks of Cosmic Dust, eds. D. L. Block, I. Puerari, K. C. Freeman, R. Groess, \& E. Block, Springer, Kluwer, p. 15

\noindent
Block, D. L., Freeman, K. C., Jarrett, T. H., Puerari, I., Worthy, G., Combes, F., \& Groess, R. 2004b, \aap, 425, L37

\noindent
Block, D. L., Puerari, I., Elmegreen, B. G., Elmegreen, D. M., Fazio, G. G., \& Gehrz, R. D. 2009, \apj, 694,
115

\noindent
Book, L. G. \& Benson, A. J. 2010, \apj, 716, 810

\noindent
Bothun, G., Impey, C. D., Malin, D. F., \& Mould, J. R. 1987, \aj, 94, 23

\noindent
Bournaud, F. \& Combes, F. 2002, \aap, 392, 83

\noindent
Brook, C., Governato, F., Quinn, T., Wadsley, J., Brooks, A. M., Willman, B., Stilp, A., \& Jonsson, P. 2008, \apj, 689, 678

\noindent
Bureau, M., Aronica, G., Athanassoula, E., Dettmar, R.-J., Bosma, A., Freeman, K. C. 2006, \mnras, 370, 753

\noindent
Bush, S. J. \& Wilcots, E. M. 2004, \aj, 128, 2789

\noindent
Buta, R. 1987, ApJS, 64, 1

\noindent
Buta, R. 1995, \apjs, 96, 39 

\noindent
Buta, R. \& Block, D. L. 2001, \apj, 550, 243

\noindent
Buta, R. \& Combes, F. 1996, Fund. Cosmic Phys. 17, 95

\noindent
Buta R. \& Crocker D. A. 1993, \aj, 106, 939

\noindent
Buta, R., Byrd, G., \& Freeman, T. 2003, \aj, 125, 634

\noindent
Buta, R. J., Corwin, H. G., \& Odewahn, S. C. 2007, The de Vaucouleurs Atlas of Galaxies, Cambridge: Cambridge U. Press (dVA)

\noindent
Buta, R. J., Knapen, J. K., Elmegreen, B. G., Salo, H., Laurikainen, E.,
Elmegreen, D. M., Puerari, I., \& Block, D. L. 2009, \aj, 137, 4487

\noindent
Buta, R. et al. 2010a, \apjs, 190, 147

\noindent
Buta, R., Laurikainen, E., Salo, H., \& Knapen, J. H. 2010b, \apj, 721, 259

\noindent
Buta, R., Mitra, S., de Vaucouleurs, G., \& Corwin, H. G. 1994, \aj, 107, 118

\noindent
Buta, R., Laurikainen, E., Salo, H., Block, D. L., \& Knapen, J. H.
2006, \aj, 132, 1859

\noindent
Buta, R. \& Zhang, X, 2009, \apjs, 182, 559

\noindent
Butcher, H. \& Oemler, A. 1978, \apj, 219, 18

\noindent
Byrd, G. G., Rautiainen, P. Salo, H., Buta, R., \& Crocker, D. A.  1994, AJ, 108, 476

\noindent
Caon, N., Capaccioli, M., \& D'Onofrio, M. 1993, \mnras, 265, 1013

\noindent
Cappellari, M. et al. 2011, astro-ph 1012.1551

\noindent
Cardemone, C. et al. 2009, \mnras, 399, 1191

\noindent
Carignan, C. \& Purton, C. 1998, \apj, 506, 125

\noindent
Carollo, C. M., Stiavelli, M., de Zeeuw, P. T., \& Mack, J. 1997, \aj, 114, 2366

\noindent
Cayatte, V., Kotanyi, C., Balkowski, C., \& van Gorkom, J. H. 1994, \aj,
107, 1003

\noindent 
Chapman, S. C., Windhorst, R., Odewahn, S., et al. 2003, \apj, 599, 92

\noindent
Chung, A., van Gorkom, J. H., Kenney, J. D. P., Crowl, Hugh, \&
Vollmer, B. 2009, \aj, 138, 1741

\noindent
Cole, A. A. et al. 2007, \apj, 659, L17

\noindent
Comer\'on, S., Knapen, J. H., Beckman, J. E., Laurikainen, E., Salo, H., Martinez-Valpuesta, I., \& Buta, R. J. 2010, \mnras, 402, 2462

\noindent
Comer\'on, S., Martinez-Valpuesta, I., Knapen, J. H., \& Beckman, J.  2009, \apj, 706, L25

\noindent
Conselice, C. J. 1997, \pasp, 109, 1251

\noindent
Conselice, C. J. 2003, \apjs, 147, 1

\noindent
Conselice, C. J. et al. 2004, \apj, 600, L139

\noindent
Conselice, C. J., Rajgor, S., \& Myers, R. 2008, \mnras, 386, 909

\noindent
Conselice, C. J. 2009, \mnras, 399, L16

\noindent
Conselice, C. J. \& Arnold, J. 2009, \mnras, 397, 208

\noindent
Contopoulos, G. 1996, ASP Conf. Ser. 91, p. 454

\noindent
Contopoulos, G. \& Grosbol, P. 1989, Astr. Astrophys. Rev. 1, 261

\noindent
Cowie, L., Hu, E., \& Songaila, A. 1995, \aj, 110, 1576

\noindent
Crocker, D. A., Baugus, P. D., \& Buta, R. 1996, \apjs, 105, 353

\noindent
Cullen, H., Alexander, P., Green, D. A., \& Sheth, K. 2007, \mnras, 376, 98

\noindent
Darg, D. W. et al. 2010, \mnras, 401, 1043

\noindent
Davies, R. L., Efstathiou, G., Fall, S. M., Illingworth, G., \& Schechter, P. L. 1983, \apj, 266, 41

\noindent
de Vaucouleurs, G. 1956, Mem. Comm. Obs., 3, No. 13

\noindent
de Vaucouleurs, G. 1958, \apj, 127, 487

\noindent
de Vaucouleurs, G. 1959, Handbuch der Physik, 53, 275

\noindent
de Vaucouleurs, G. 1961, \apjs, 5, 233

\noindent
de Vaucouleurs, G. 1963, \apjs, 8, 31

\noindent
de Vaucouleurs, G. \& Freeman, K. C. 1972, Vistas in Astronomy 14, 163

\noindent
de Vaucouleurs, G. de Vaucouleurs, A., \& Corwin, H. G.  (1976), {\it Second Reference Catalogue of Bright Galaxies}, Univ. Texas Mono. Astr. No. 2 (RC2)

\noindent
de Vaucouleurs, G., de Vaucouleurs, A., Corwin, H. G., Buta, R., Paturel, G., \& Fouque\', P. 1991, Third Reference Catalogue of Bright Galaxies, New York, Springer (RC3)

\noindent
de Zeeuw, P. T. et al. 2002, \mnras, 329, 513

\noindent
Djorgovski, S. 1992, in Morphological and Physical Classification of Galaxies, G. Longo, M. Capaccioli, \& G. Busarello, eds., Dordrecht, Kluwer, p. 337

\noindent
Dressler, A. 1980, \apj, 236, 351

\noindent
Durbala, A., Buta, R., Sulentic, J. W., \& Verdes-Montenegro, L. 2009, \mnras, 397, 1756

\noindent
Durbala, A., Sulentic, J. W., Buta, R., \& Verdes=Montenegro, L. 2008, \mnras, 390, 881

\noindent
Elmegreen, D. M. 1981, \apjs, 47, 229

\noindent
Elmegreen, D. M. \& Elmegreen, B. G. 1987, \apj, 314, 3

\noindent
Elmegreen, D. \& Elmegreen, B. G. 2006, \apj, 651, 676 (EE06)

\noindent
Elmegreen, D., Elmegreen, B. G., \& Sheets, C. 2004, \apj, 603, 74 (EES04)

\noindent
Elmegreen, D., Elmegreen, B. G., \& Ferguson, T. E. 2005, \apj, 623, L71 

\noindent
Elmegreen, D., Elmegreen, B. G., Ferguson, T. E., \& Mullan, B. 2007a, \apj, 663, 734 (EEFM07)

\noindent
Elmegreen, D. M., Elmegreen, B. G., Ravindranath, S., \& Coe, D. A. 2007b,
\apj, 658, 763

\noindent
Elmegreen, D., Elmegreen, B. G., \& Hirst, A. 2004, \apj, 604, L21 (EEH04)

\noindent
Elmegreen, D. M., Elmegreen, B. G., Marcus, M. T., Shainyan, K., Yau, A., \& Petersen, M. 2009b, \apj, 701, 306

\noindent
Elmegreen, B.. G., Bournaud, F., \& Elmegreen, D. M. 2008, \apj, 688, 67

\noindent
Elmegreen, B. G., Elmegreen, D. M., \& Leitner, S. N. 2003, \apj, 590, 271 (EEL03)

\noindent
Elmegreen, B. G., Elmegreen, D. M., \& Montenegro, L. 1992, \apjs, 79, 37

\noindent
Elmegreen, B. G., Elmegreen, D. M., Fernandez, M. X., \& Lemonias, J. J. 2009a, \apj, 692, 12

\noindent
Emsellem, E. et al. 2007, \mnras, 379, 401

\noindent
Eskridge, P. B. et al. 2000, \aj, 119, 536

\noindent
Eskridge, P. B. et al. 2002, \apjs, 143, 73

\noindent
Fazio, G. G. et al. 2004, \apjs, 154, 10

\noindent
Ferguson, H. C. \& Binggeli, B. 1994, Astr. Ap. Rev. 6, 67

\noindent
Ferrarese, L. et al. 2006, \apjs, 164, 334

\noindent
Freeman, K. C. 1975, in Galaxies and the Universe, A. Sandage,
M. Sandage, \& J. Kristian, eds., Chicao, University of Chicago
Press, p. 409

\noindent
Garcia-Ruiz, I., Sancisi, R., \& Kuijken, K. 2002, \aap, 394, 769

\noindent
Giavalisco, M. et al. 2004, \apj, 600, L93

\noindent
Gil de Paz, A., Madore, B. F., \& Pevunova, O. 2003, \apjs, 147, 29

\noindent
Gil de Paz, A. et al. 2005, \apj, 627, L29

\noindent
Gillett, F. C., Forrest, W. J., \& Merrill, K. M. 1973, \apj, 183, 87

\noindent
Giovanelli, R. \& Haynes, M. P. 1985, \aj, 292, 404

\noindent
Graham, A. W. \& Guzm\'an, R. 2003, \aj, 125, 2936

\noindent
Grouchy, R. D., Buta, R., Salo, H., Laurikainen, E., \& Speltincx, T.  2008, \aj, 136, 980

\noindent
Grouchy, R. D., Buta, R. J., Salo, H., \& Laurikainen, E. 2010, \aj, 139, 2465

\noindent
Gunn, J. E. \& Gott, J. R. 1972, \apj, 176, 1

\noindent
Gunn, J. E. et al. 1998, \aj, 116, 3040

\noindent
Helfer, T. T., Thornley, M. D., Regan, M. W., Wong, T.,
Sheth, K., Vogel, S. N., Blitz, L., \& Bock, D. C.-J. 2003,
\apjs, 145, 259

\noindent
Helou, G. et al. 2004, \apjs, 154, 253

\noindent
Higdon, J. L. 1995, \apj, 455, 524

\noindent
Higdon, J. L., Buta, R. J., \& Purcell, G. B. 1998, \aj, 115, 80

\noindent
Holmberg, E. 1950, Medd. Lunds. Astron. Obs., Ser. II, No. 128

\noindent
Holwerda, B. W., Keel, W. C., Williams, B., Dalcanton, J. J., 
\& de Jong, R. S. 2009, \aj, 137, 3000

\noindent
Hubble, E. 1926, \apj, 64, 321

\noindent
Hubble, E. 1936, {\it The Realm of the Nebulae}, Yale Univ.  Press, Yale.

\noindent
Hubble, E. 1943, \apj, 97, 112

\noindent
Hunt, L. K. \& Malkan, M. A. 1999, \apj, 516, 660

\noindent
Hunter, D. A. 1997, \pasp, 109, 937

\noindent
Impey, C. \& Bothun, G. D. 1997, \araa, 35, 267

\noindent
Irwin, M. J. et al. 2007, \apj, 656, L13

\noindent
Jarrett, T. H. et al. 2003, \aj, 125, 525

\noindent
Jeans, J. 1929, Astronomy and Cosmogony, Cambridge, Cambridge University press

\noindent
Jerjen, H., Kalnajs, A., \& Binggeli, B. 2000, \aap, 358, 845

\noindent
Jogee, S. et al. 2009, \apj, 697, 1971

\noindent
Jokimaki, A., Orr, H., \& Russell, D. G. 2008, AP\&SS, 315, 249

\noindent
Karachentseva, V. E. 1973, Astrofiz. Issled-Izv. Spets. Astrofiz. Obs. 8, 3

\noindent
Karataeva, G. M., Tikhonov, N. A., Galazutdinova, O. A., Hagen-Thorn, V. A., \& Yakovleva, V. A. 2004, \aap, 421, 833

\noindent
Kennicutt, R. C., Tambln, P., \& Congdon, C. W. 1994, \apj, 435, 22

\noindent
Kerr, F. \& de Vaucouleurs, G. 1955, Australian Journal of Physics, 8, 508

\noindent
Knapen, J. H. 2005, \aap, 429, 141

\noindent
Knapen, J., Beckman, J. E., Shlosman, I., Peletier, R. F., Heller, C. H.,
\& de Jong, R. S. 1995a, \apj, 443, L73

\noindent
Knapen, J., Beckman, J. E., Heller, C. H., Shlosman, I., \& de Jong, R.
S. 1995b, \apj, 454, 623

\noindent
Knapen, J. H., Shlosman, I., \& Peletier, R. F. 2000, \apj, 529, 93

\noindent
Knapen, J. H. 2010, in Galaxies and Their Masks, D. L. Block, K. C. Freeman, \& I. Puerari, eds., New York, Spinger, in press

\noindent
Knapen, J. H. \& James, P. A. 2009, \apj, 698, 1437

\noindent
Koopmann, R. \& Kenney, J. D. P. 2004, \apj, 613, 866

\noindent
Kormendy, J. 1979, \apj, 227, 714

\noindent
Kormendy, J. 1985, \apj, 295, 73

\noindent
Kormendy, J. 1999, ASPC, 182, 124

\noindent
Kormendy, J. \& Bender, R. 1996, \apj, 464, L119

\noindent
Kormendy, J. \& Djorgovski, S. 1989, \araa, 27, 235

\noindent
Kormendy, J. \& Kennicutt, R. C. 2004, \araa, 42, 603 (KK04)

\noindent
Kormendy, J. \& Norman, C. A. 1979, \apj, 233, 539

\noindent
Kormendy, J., Fisher, D. B., Cornell, M. E., \& Bender, R. 2009,
\apjs, 182, 216

\noindent
Kuijken, K. \& Merrifield, M. R. 1995, \apj, 443, L13

\noindent
Laine, S., Shlosman, I., Knapen, J., \& Peletier, R. F. 2002, \apj,
567, 97

\noindent
Laurikainen, E. \& Salo, H. 2002, \mnras, 337, 1118

\noindent
Laurikainen, E., Salo, H., \& Buta, R. 2004, \apj, 607, 103

\noindent
Laurikainen, E., Salo, H., \& Buta, R. 2005, \mnras, 362, 1319

\noindent
Laurikainen, E., Salo, H., Buta, R., Knapen, J., Speltincx, T.,
\& Block, D. L.  2006, \mnras, 132, 2634

\noindent
Laurikainen, E., Salo, H., Buta, R., \& Knapen, J. 2007, \mnras, 381, 401

\noindent
Laurikainen, E., Salo, H., Buta, R., \& Knapen, J. 2009, \apj, 692, L34

\noindent
Laurikainen, E., Salo, H., Buta, R., Knapen, J. H., \& Comer\'on, S. 2010, \mnras, 405, 1089

\noindent
Lelli, F., Fraternali, F., \& Sancisi, R. 2010, \aap, 516, 11

\noindent
Lintott, C. et al. 2008, \mnras, 389, 1179

\noindent
Lintott, C. et al. 2009, \mnras, 399, 129

\noindent
Lisker, T., Debattista, V. P., Ferreras, I., \& Erwin, P. 2006, \mnras, 370, 477

\noindent
Lotz, J. M., Madau, P., Giavalisco, M., Primack, J., \& Ferguson, H. C. 2006, \apj, 636, 592

\noindent
Lupton, R., Blanton, M. R., Fekete, G., Hogg, D. W., O'Mullane, W., Szalay, A., \& Wherry, N. 2004, \pasp, 116, 133

\noindent
Lynden-Bell, D. \& Kalnajs, A. J. 1972, \mnras, 157, 1

\noindent
Madore, B. F., Nelson, E., \& Petrillo, K. 2009, \apjs, 181, 572

\noindent
Malin, D. F. \& Carter, D. 1980, Nature, 285, 643

\noindent
Malin, D. F. \& Carter, D. 1983, \apj, 274, 534

\noindent
Martin, D. C. et al. 2005, \apj, 619, L1

\noindent
Martin, D. C. et al. 2007, \apjs, 173, 342

\noindent
Martin, P. \& Friedli, D. 1997, \aap, 326, 449

\noindent
Martinez-Valpuesta, I., Shlosman, I., \& Heller, C. 2006, \apj, 637, 214

\noindent
Martinez-Valpuesta, I., Knapen, J. H., \& Buta, R. 2007, \aj, 134, 1863

\noindent
Masters, K. L. et al. 2010, \mnras, 405, 783

\noindent
Mateo, M. 1998, \araa, 36, 435

\noindent
Matthews, T. A., Morgan, W. W., \& Schmidt, M. 1964, \apj, 140, 35

\noindent
Mazzuca, L. M., Swaters, R. A., Veilleux, S., \& Knapen, J. H. 2009, BAAS, 41, 693

\noindent
McGaugh, S., Schombert, J. M., \& Bothun, G. D. 1995, \aj, 109, 2019

\noindent
McKernan, B., Ford, K. E. S., \& Reynolds, C. S. 2010, astro-ph 1005.4907

\noindent
Menanteau, F., Abraham, R. G., \&  Ellis, R. S. 2001, \mnras, 322, 1

\noindent
Menanteau, F. et al. 2004, \apj, 612, 202

\noindent
Meurer, G. R., Carignan, C., Beaulieu, S. F., \& Freeman, K. C. 1996, \aj, 111, 1551

\noindent
Moore, B., Katz, N., Lake, G., Dressler, A., \& Oemler, A. 1996, Nature, 379, 613

\noindent
Morgan, W. W. 1958, PASP, 70, 364

\noindent
Nair, P. B. \& Abraham, R. G. 2010, \apjs, 186, 427

\noindent
Odewahn, S. C. 1991, \aj, 101, 829

\noindent
Odewahn, S. C., Stockwell, E. B., Pennington, R. L., Humphreys, R. M.,
\& Zumach, W. A. 1992, \aj, 103, 318 

\noindent
Oosterloo, T. A., Morganti, R., Sadler, E. M., Vergani, D., \& Caldwell, N.  2002, \aj, 123, 729

\noindent
Pahre, M., Ashby, M. L. N., Fazio, G. G., \& Willner, S. P. 2004, \apjs, 154, 235

\noindent
Pence, W. D., Taylor, K., Freeman, K. C., de Vaucouleurs, G., \&
Atherton, P. 1988, \apj, 326, 564

\noindent
Peng, C. Y., Ho, L. C., Impey, C. D., \& Rix, H-W. 2010, \aj, 139,
2097

\noindent
Pereira-Santaella, M., Alonso-Herrero, A., Rieke, G. H.,
Colina, L. D\'iaz-Santos, T., Smith, J.-D. T.,
P\'erez-Gonz\'alez, P. G., \& Engelbracht, C. W. 2010, \apjs, 188, 447

\noindent
Quillen, A. C., Frogel, J. A., \& Gonz\'alez, R. A. 1994, \apj, 437, 162

\noindent
Rahman, N., Howell, J. H., Helou, G., Mazzarella, J. M., \& Buckalew, B. 2007, \apj, 663, 908

\noindent
Rautiainen, P. \& Salo, H. 2000, \aap, 362, 465

\noindent
Rautiainen, P., Salo, H., \& Buta, R. 2004, \mnras, 349, 933

\noindent
Ravindranath, S. et al. 2006, \apj, 652, 963

\noindent
Regan, M. \& Teuben, P. 2003, \apj, 582, 723

\noindent
Regan, M. \& Teuben, P. 2004, \apj, 600, 595

\noindent
Regan, M. W., Thornley, M. D., Helfer, T. T.; Sheth,
K., Wong, T., Vogel, S. N., Blitz, L., \& Bock, D. C.-J.
2001, \apj, 561, 218

\noindent
Regan, M. et al. 2004, \apjs, 154, 204

\noindent
Reynolds, J. H. 1927, Observatory, 50, 185

\noindent
Revaz, Y. \& Pfenniger, D. 2007, in Island Universes, Astrophysics \& Space Science Proceedings, p. 149

\noindent
Rieke, G. H. \& Low, F. J. 1972, \apj, 176, L95

\noindent
Rix, H.-W. \& Rieke, M. J. 1993, \apj, 418, 123

\noindent
Rix, H.-W. et al. 2004, \apjs, 152, 163

\noindent
Romano, R., Mayya, Y. D., \& Vorobyov, E. I.  2008, \aj, 136, 1259

\noindent
Romero-G\'omez, M., Masdemont, J. J., Athanassoula, E., \&
Garc\'ia-G\'omez, C. 2006, \aap, 453, 39

\noindent
Romero-G\'omez, M., Athanassoual, E., Masdemont, J. J., \&
Garc\'ia-G\'omez, C. 2007, \aap, 472, 63

\noindent
Ryder, S. D., Buta, R. J., Toledo, H., Shukla, H. Staveley-Smith, L.,
\& Walsh, W. 1996, \apj, 460, 665

\noindent
Sackett, P. D., Rix, H.-W., Jarvis, B. J., \& Freeman, K. C.  1994, \apj, 436, 629

\noindent
Saha, K., de Jong, R., \& Holwerda, B. 2009, \mnras, 396, 409

\noindent
Salo, H. \& Laurikainen, E. 2000a, \mnras, 319, 377

\noindent
Salo, H. \& Laurikainen, E. 2000b, \mnras, 319, 393

\noindent
Salo, H., Laurikainen, E., Buta, R., \& Knapen, J. H. 2010, \apj, 715, L56

\noindent
Sandage, A. 1961, The Hubble Atlas of Galaxies, Carnegie Inst. of Wash. Publ. No. 618

\noindent
Sandage, A. 1975, in {\it Galaxies and the Universe, Stars and Stellar Systems, Vol. IX}, A. Sandage, M. Sandage, \& J. Kristian, eds., p. 1.

\noindent
Sandage, A. \& Bedke, J. 1994, The Carnegie Atlas of Galaxies, Carnegie Inst. of Wash. Pub. No. 638

\noindent
Sandage, A. \& Binggeli, B. 1984, \aj, 89, 919

\noindent
Sandage, A. \& Tammann, G. A. 1981, {\it A Revised Shapley-Ames Catalog of Bright Galaxies}, Carnegie Institute of Washington Publ. No. 635 (first edition) (RSA)

\noindent
Sanders, D. B. \& Mirabel, I. F. 1996, \araa, 34, 749

\noindent
Schawinski, K., Thomas, D., Sarzi, M., Maraston, C., Kavaraj, S., Joo,
S., Yi, S. K., \& Silk, J. 2007, \mnras, 382, 1415

\noindent
Schombert, J. 1986, \apjs, 60, 603

\noindent
Schombert, J. 1987, \apjs, 64, 643

\noindent
Schombert, J. 1988, \apj, 328, 475

\noindent
Schombert, J., Pildis, R. A., Eder, J. A., \& Oemler, A. 1995, \aj, 110, 2067

\noindent
Schwarz, M. P. 1981, \apj, 247, 77

\noindent
Schwarz, M. P. 1984, \mnras, 209, 93

\noindent
Schweizer, F. 1998, in Galaxies: Interactions and Induced Star Formation, R. C. Kennicutt, et al., eds., Berlin, Springer, p. 105

\noindent
Schweizer, F. \& Seitzer, P. 1988, \apj, 328, 88

\noindent
Schweizer, F., Ford, W. K., Jedrzejewski, R., \& Giovanelli, R. 1987, ApJ, 320, 454

\noindent
Schweizer, F., van Gorkom, J., \& Seitzer, P. 1989, \apj, 338, 770

\noindent
Scoville, N. et a. 2007, \apjs, 172, 38

\noindent
Seiden, P. E. \& Gerola, H. 1982, Fund. Cosmic Phys. 7, 241

\noindent
Seigar, M. S., Block, D. L., Puerari, I., Chorney, N. E., \& James,
P. A. 2005, \mnras, 359, 1065

\noindent
Seigar, M. S., Kennefick, D., Kennefick, J., \& Lacy, C. H. 2008, \apj,
678, L93

\noindent
Sellwood, J. A. 2010, in Planets, Stars, and Stellar Systems, Vol. 5, in production

\noindent
Sellwood, J. A. \& Wilkinson, A. 1993, Reports on Progress in Physics, 56, 173

\noindent
Shamir, L. 2009, \mnras, 399, 1367

\noindent
Sheth, K. et al. 2008, \apj, 675, 1141

\noindent
Sheth, K. et al. 2010, \pasp, 122, 1397

\noindent
Sil'chenko, O. K. \& Afanasiev, V. L. 2004, \aj, 127, 2641

\noindent
Simkin, S. M., Su, H. J., \& Schwarz, M. P. 1980, \apj, 237, 404

\noindent
Skrutskie, M. F. et al. 2006, \aj, 131, 1163

\noindent
Sparke, L. S., van Moorsel, G., Erwin, P., \& Wehner, E. M. H. 2008, \aj,
135, 99

\noindent
Spitzer, L. \& Baade, W. 1951, \apj, 113, 413

\noindent
Steidel, C. \& Hamilton, D. 1992, \aj, 104, 941

\noindent
Steidel, C., Giavalisco, M., Dickinson, M., \& Adelberger, K. 1996, \aj, 112, 352

\noindent
Strateva, I. et al. 2001, \aj, 122, 1861

\noindent
Struck, C. 2010, \mnras, 403, 1516

\noindent
Struck, C., Appleton, P. N., Borne, K. D., \& Lucas, R. A.  1996, \aj, 112, 1868

\noindent
Sulentic, J. et al. 2006, \aap, 449, 937

\noindent
Surace, J. A., Sanders, D. B., Vacca, W. D., Veilleux, S., \& Mazzarella, J. M.  1998, \apj, 492, 116

\noindent
Taniguchi, Y. \& Watanabe, M. 1987, \apj, 313, 89

\noindent
Taylor-Mager, V. A., Conselice, C. J., Windhorst, R. A., \& Jansen, R. A.
2007, \apj, 659, 162

\noindent
Theis, C., Sparke, L., \& Gallagher, J. 2006, \aap, 446, 905

\noindent
Theys, J. C. \& Spiegel, J. C. 1976, \apj, 208, 650

\noindent
Thilker, D. A. et al. 2010, \apj, 714, L171

\noindent
Thornley, M. 1996, \apj, 469, 45

\noindent
Thuan, T. X. \& Martin, G. E. 1981, \apj, 247, 823

\noindent
Thuan, T. X., Izotov, Y. I., Lipovetsky, V. A. 1997, \apj, 477, 661

\noindent
Tolstoy, E., Hill, V., \& Tosi, M. 2009,\araa, 47, 371

\noindent
Treuthardt, P., Salo, H., Rautinainen, P., \& Buta, R. 2008, \aj, 136, 300

\noindent
Vaisanen, P., Ryder, S., Mattila, S., \& Kotilainen, J.  2008, \apj, 689, L37

\noindent
van den Bergh, S. 1980a, \pasp, 92, 122

\noindent
van den Bergh, S. 1980b, \pasp, 92, 409

\noindent
van den Bergh, S. 1995, \aj, 110, 613

\noindent
van den Bergh, S. 1998, Galaxy Morphology and Classification, Cambridge, Cambridge University Press

\noindent
van den Bergh, S. 2009a, \apj, 694, L120

\noindent
van den Bergh, S. 2009b, \apj, 702, 1502

\noindent
van den Bergh, S., Abraham, R. G., Ellis, R. S., Tanvir, N. R., Santiago, B.  X., \& Glazebrook, K. G 1996, \aj, 112, 359

\noindent
van den Bergh, S., Cohen, J. G., Hogg, D. W., \& Blandford, R. 2000, \aj, 120, 2190

\noindent
van den Bergh, S., Pierce, M., \& Tully, R. B. 1990, \apj, 359, 4

\noindent
van der Wel, A., Bell, E. F., Holden, B. P., Skibba, R. A., \& Rix, H.-W.
2010, \apj, 714, 1779

\noindent
van Driel, W. et al. 1995, \aj, 109, 942

\noindent
Vanzi, L. \& Sauvage, M. 2004, \aap, 415, 509

\noindent
Verdes-Montenegro, L., Sulentic, J., Lisenfeld, U., Leon, S., Espada, D., Garcia, E., Sabater, J., \& Verley, S. 2005, \aap, 436, 443

\noindent
Veron-Cetty, M. \& Veron, P. 2006, \aap, 455, 773

\noindent
Volonteri, M., Saracco, P., Chincarini, G., \& Bolzonella, M. 2000, \aap, 362, 487

\noindent
Walter, F., Brinks, E., de Blok, W. J. G., Bigiel, F., Kennicutt, R. C.,
Thornley. M. D., \& Leroy, A. 2008, \aj,136, 2563

\noindent
White, R. E. \& Keel, W. C. 1992, Nature, 359, 129

\noindent
Whitmore, B. C., Lucas, R. A., McElroy, D. B., Steiman-Cameron, T. Y., Sackett, P. D., \& Olling, R. P. 1990, \aj, 100, 1489

\noindent
Williams, R. E. et al. 1996, \apj, 112, 1335

\noindent
Willman, B. et al. 2005, \apj, 626, L85

\noindent
Willner, S. et al. 2004, \apjs, 154, 222

\noindent
Windhorst, R. et al. 2010, \apjs, May, 2010, in press

\noindent
Wirth, A. \& Gallagher, J. S. 1984, \apj, 282, 85

\noindent
Wyder, T. K. et al. 2007, \apjs, 173, 293

\noindent
Yagi, M., Yoshida, M., Komiyama, Y., Kashikawa, N.; Furusawa, H.,
Okamura, S., Graham, A. W.; Miller, N. A.; Carter, D., Mobasher, B.,
\& Jogee, S. 2010, \aj, 140, 1814

\noindent
York, D. G. et al. 2000, \aj, 120, 1579

\noindent
Zhang, X. 1996, \apj, 457, 125

\noindent
Zhang, X. 1998, \apj, 499, 93

\noindent
Zhang, X. 1999, \apj, 518, 613

\noindent
Zhang, X. \& Buta, R. 2007, \aj, 133, 2584

\clearpage

\clearpage

\clearpage

\clearpage

\clearpage

\clearpage

\clearpage

\clearpage

\clearpage

\clearpage

\clearpage

\clearpage

\clearpage

\clearpage

\clearpage

\clearpage
\begin{figure}
\figurenum{16. cont.}
\caption{Examples showing the continuity of inner rings (r) and lenses
(l), for barred and nonbarred galaxies. The galaxies are
(left to right): Row 1 
- NGC 7187, 1553, and 4909; Row 2 - NGC 1326, 2859, and 1291;
Row 3 - ESO 426$-$2, NGC 1211, NGC 1543. All images are $B$ or $g$-band
from the dVA.}
\end{figure}

\clearpage
\begin{figure}
\plotone{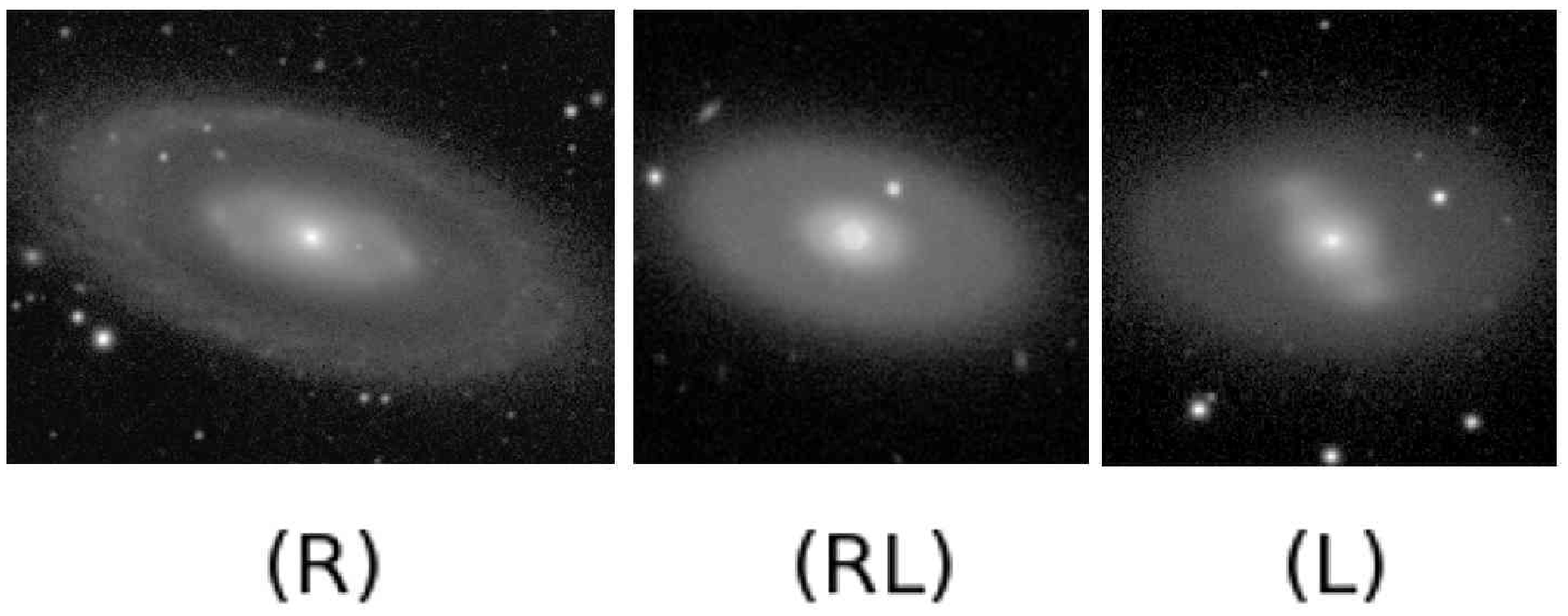}
\caption{Examples showing the continuity of outer rings (R) and lenses
(L). The galaxies are (left to right): - NGC 7020 (dVA $B$), NGC 5602 (SDSS image),
and 2983 (dVA $B$).}
\label{outer_lenses}
\end{figure}

\clearpage
\begin{figure}
\plotone{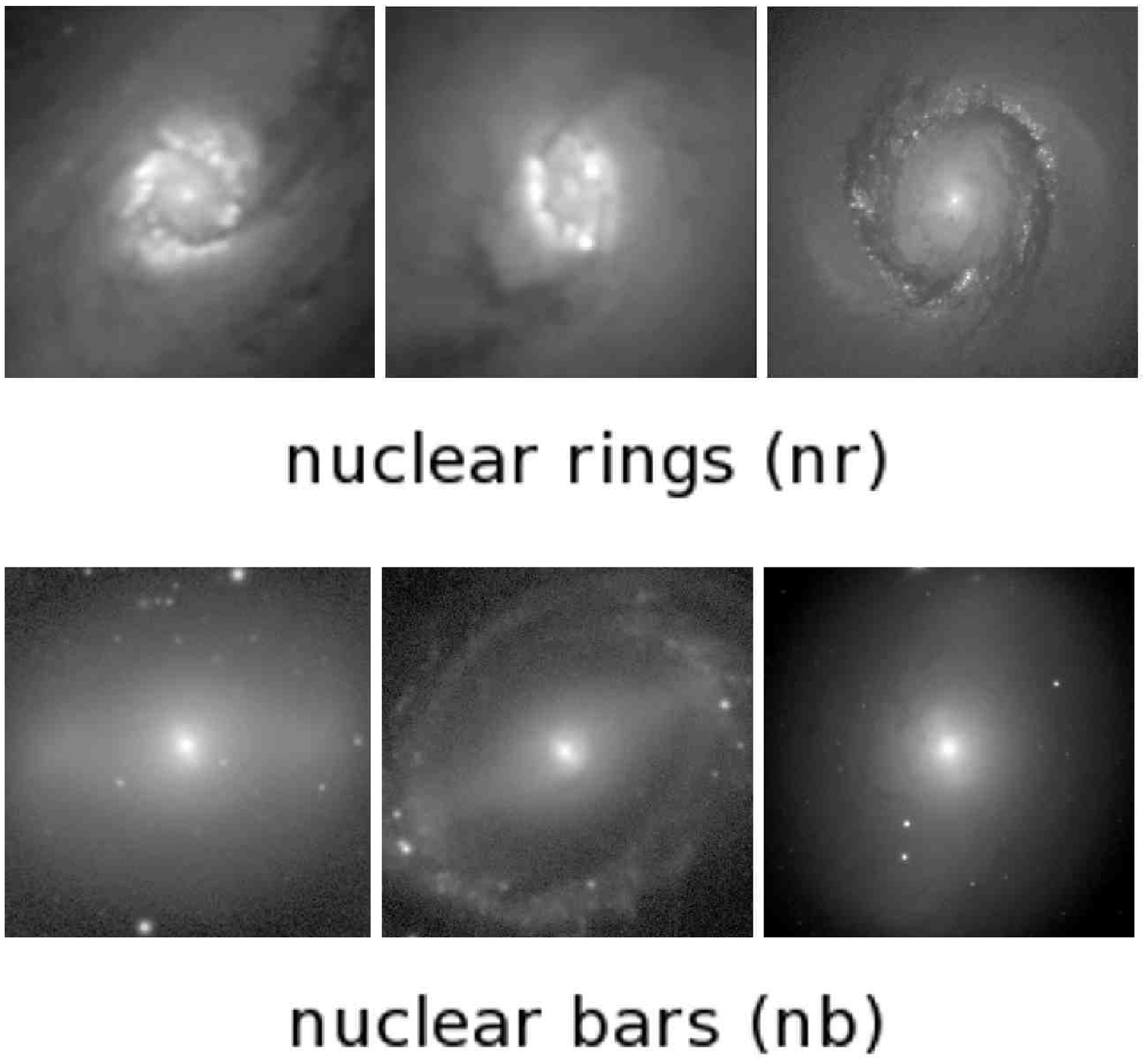}
\caption{Examples of nuclear rings and secondary (nuclear) bars. The
galaxies are (left to right): Row 1 - NGC 1097, 3351, and 4314; Row 2
- NGC 1543, 5850, and 1291. All images are $B$-band from the dVA.}
\label{nrnb}
\end{figure}

\clearpage
\begin{figure}
\plotone{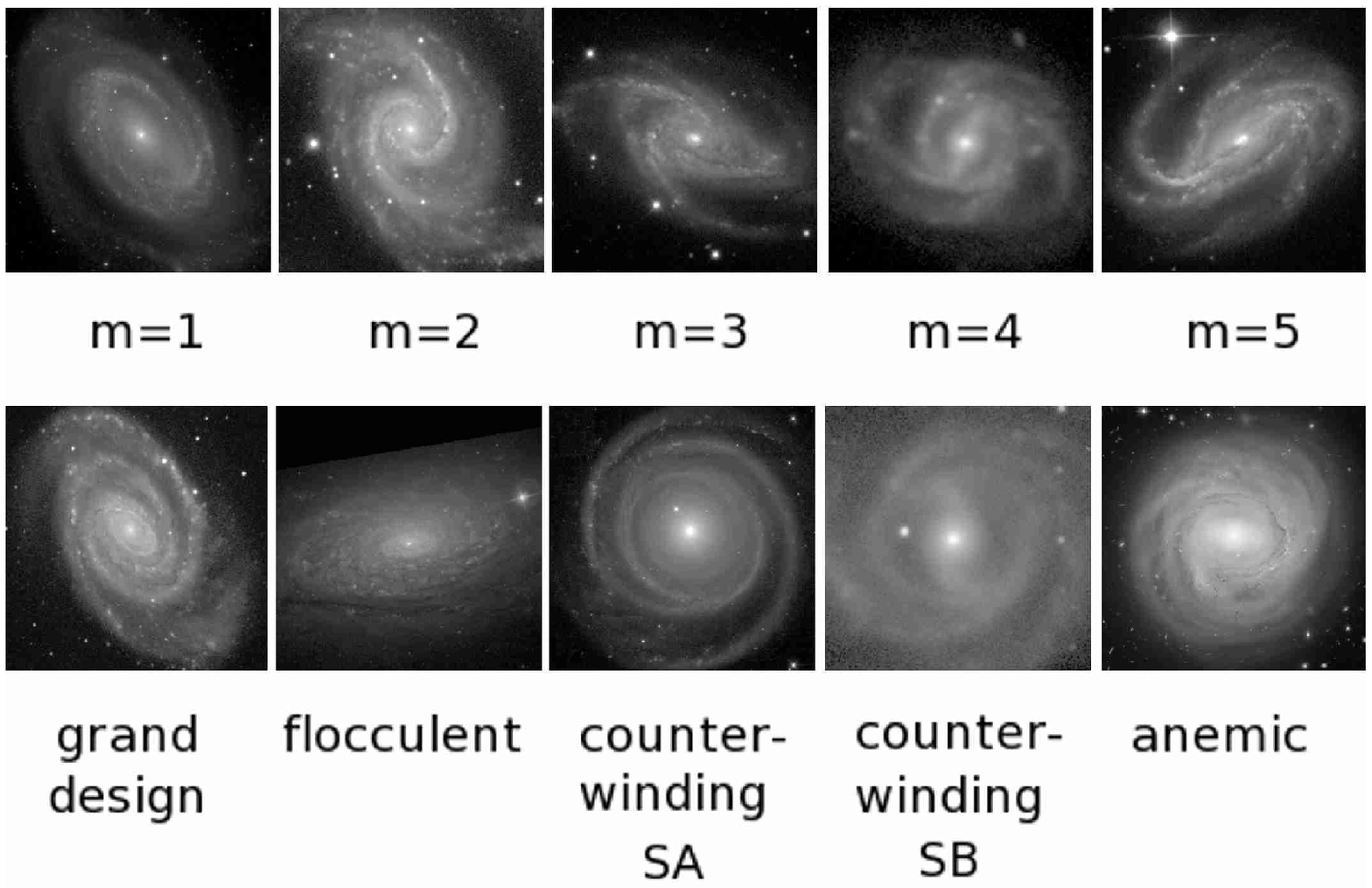}
\caption{Examples showing spiral arm character differences in the
form of arm multiplicity, grand design and flocculent spirals,
counter-winding spirals, and an anemic spiral.
The galaxies are (left to right): Row 1 - NGC 4725, 1566, 5054, ESO 566$-$24,
and NGC 613; Row 2 - NGC 5364, 5055, 4622, 3124, and 4921. All images are
$B$-band from
the dVA except NGC 5055, which is SDSS $g$-band, and NGC 4921, 
which is from Hubble Heritage.}
\label{spiraltypes}
\end{figure}

\clearpage
\begin{figure}
\plotone{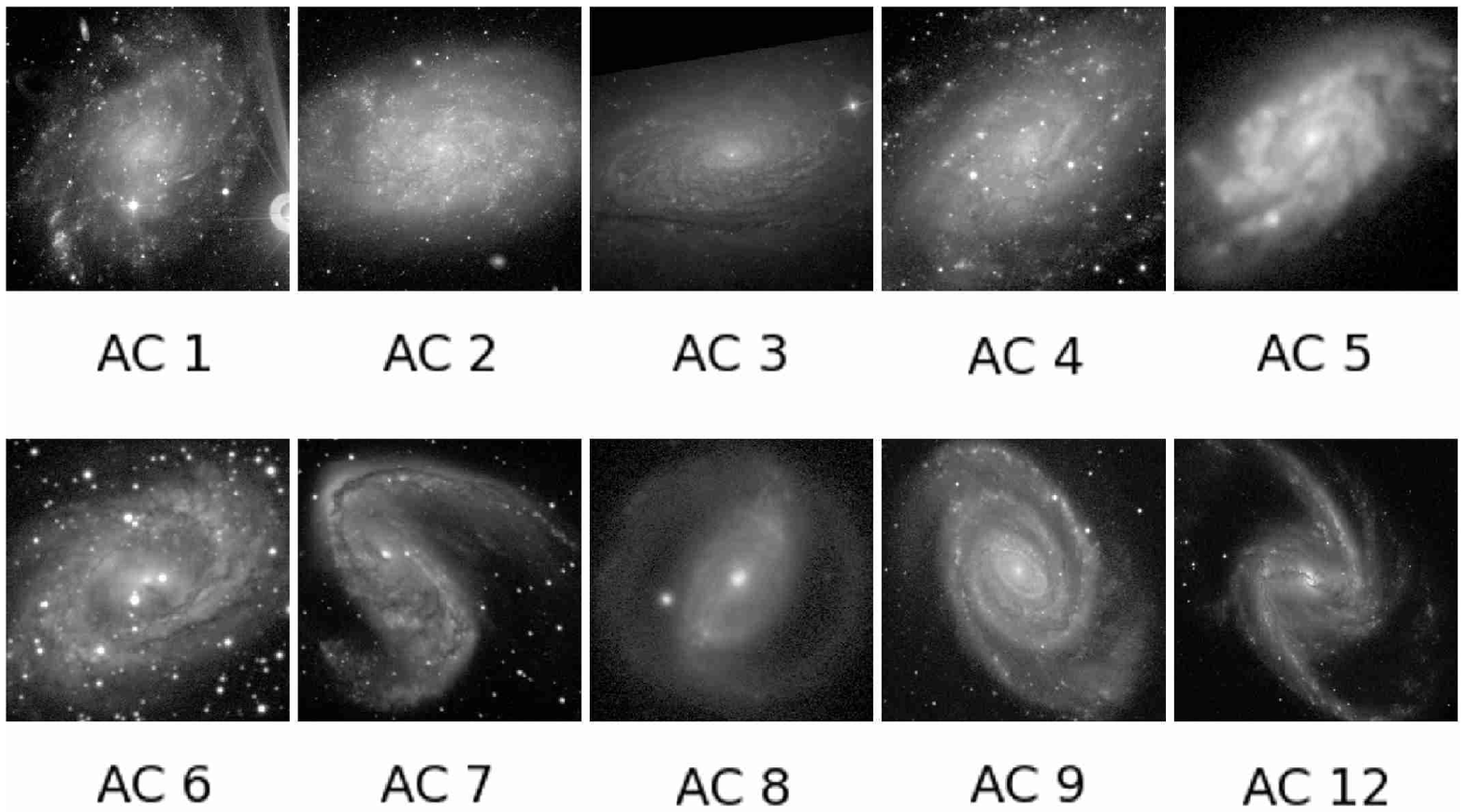}
\caption{Examples showing the spiral arm classes of Elmegeen \& Elmegreen
(1987).
The galaxies are (left to right): Row 1 - NGC 45, 7793, 5055, 2403, and 1084.
Row 2: NGC 6300, 2442, 3504, 5364, and 1365. All images are $B$-band 
from the dVA, except NGC 5055 which is SDSS $g$-band.}
\label{arm_classes}
\end{figure}

\clearpage
\begin{figure}
\plotone{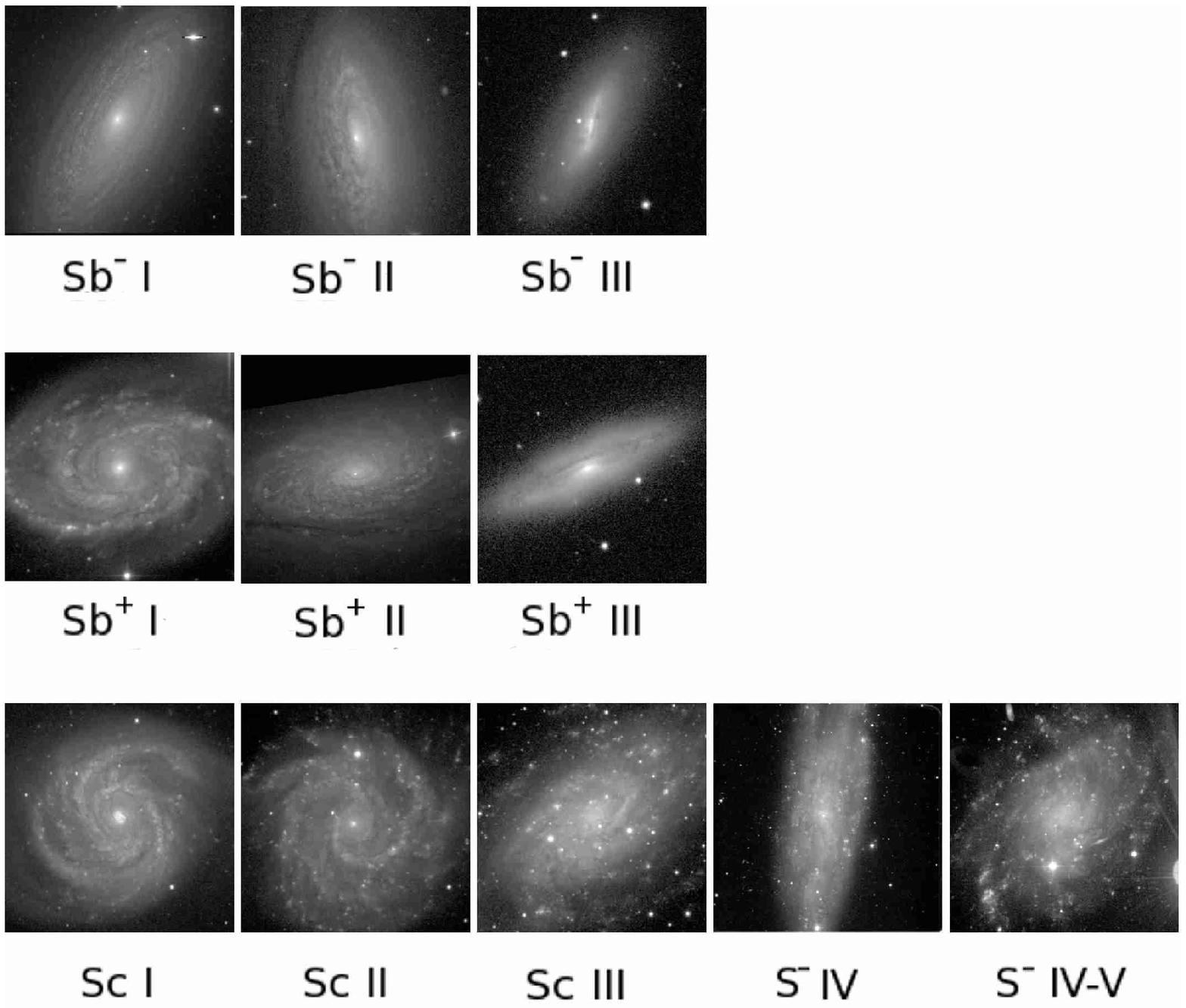}
\caption{Examples showing van den Bergh luminosity classes.
The galaxies are (left to right): Row 1 - NGC 2841 (dVA $B$), 3675 (SDSS $g$), and 4064 (SDSS $g$).
Row 2: NGC 5371 (dVA $B$), 5055 (SDSS $g$), and 4586 (SDSS $g$). Row 3:
NGC 4321, 3184, 2403, 247, and 45 (all dVA $B$). The classifications are
in the van den Bergh system (see van den Bergh 1998).}
\label{lclasses}
\end{figure}

\clearpage
\begin{figure}
\plotone{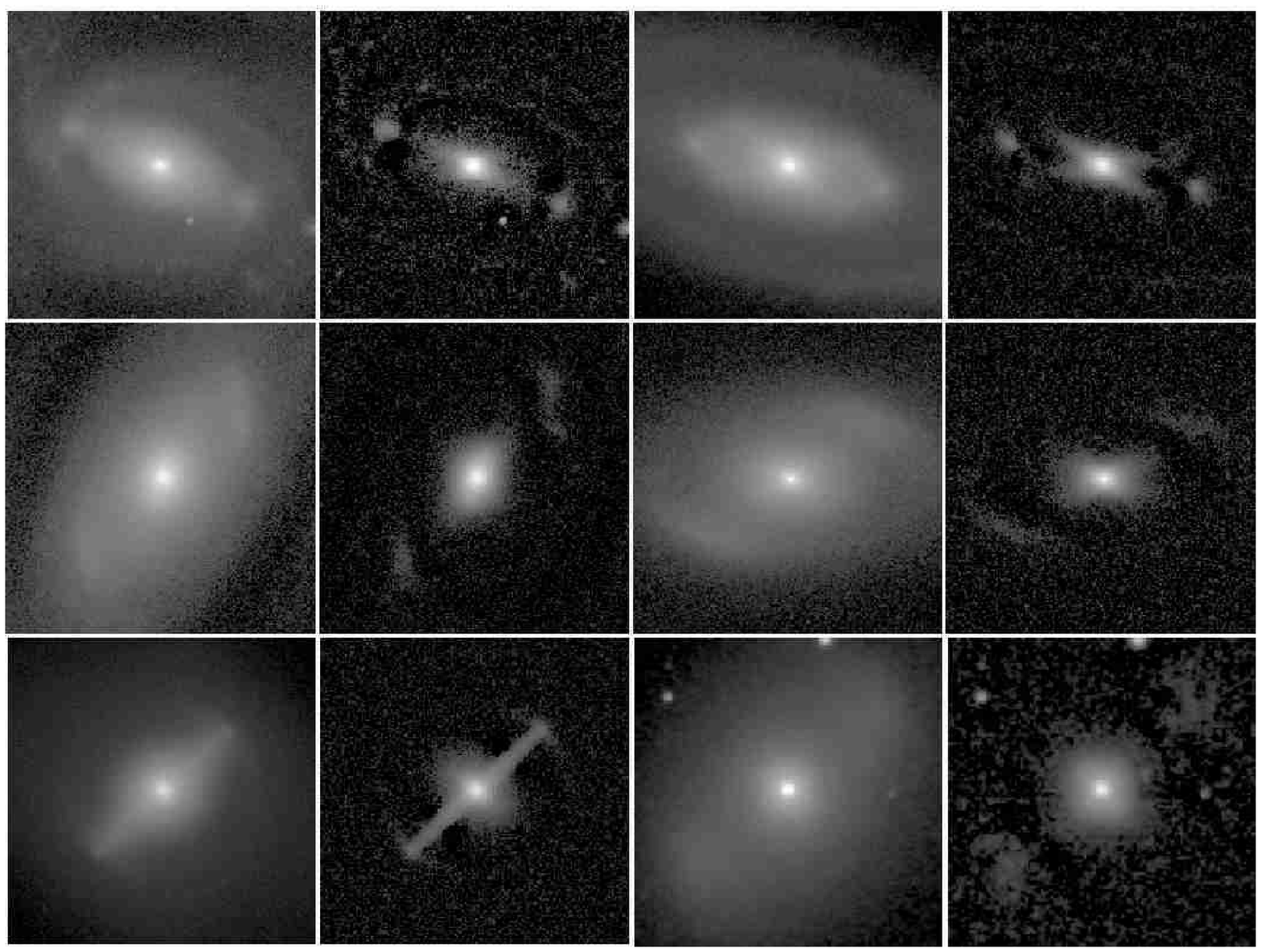}
\caption{Examples showing ansae bar morphologies as compared to one
mostly non-ansae bar. For each galaxy, the left frame is the full
image while the right frame is an unsharp masked image, both in units
of mag arcsec$^{-2}$. The galaxies are:
(left to right): Row 1 - NGC 5375 (SDSS $g$) and 7020 ($I$) (both
round ansae type);
Row 2 - NGC 7098 ($I$, linear, partly wavy ansae) and NGC 1079 ($K_s$, curved ansae);
Row 3 - NGC 4643 ($I$, mostly non-ansae type with trace of ring arcs at
bar ends) and NGC 4151 (OSUBSGS $H$, irregular ansae).}
\label{ansae}
\end{figure}

\clearpage
\begin{figure}
\plotone{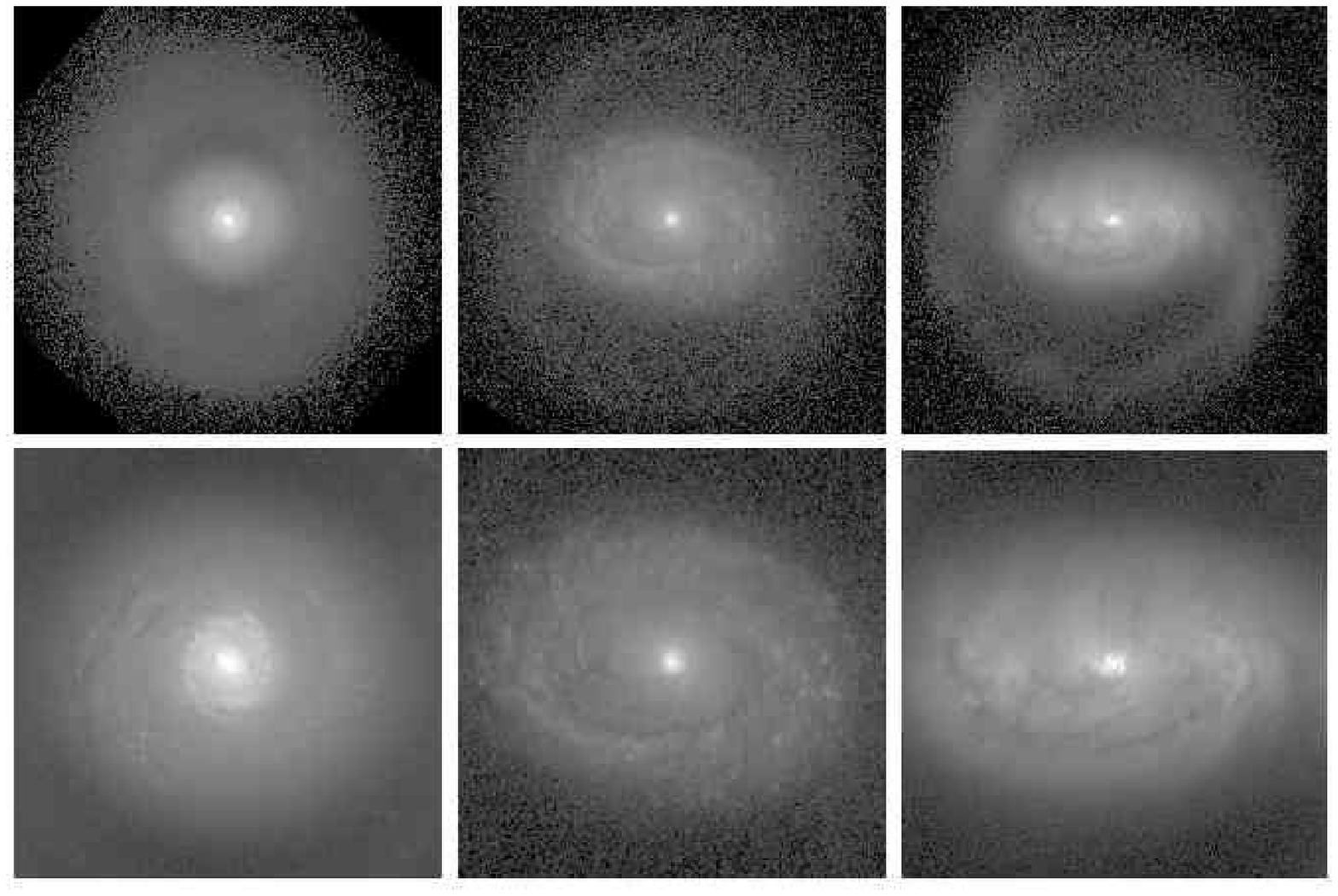}
\caption{Examples of oval disk galaxies. Left to right: NGC 4736 ($B$),
NGC 4941 ($B$), and NGC 1808 ($V$). The images are cleaned of foreground
and background objects and have been deprojected and rotated so that the
major axis of the oval is horizontal. The upper panels show the ovals
imbedded within outer rings, while the lower panels focus on the ovals
alone.}
\label{ovals}
\end{figure}

\clearpage
\begin{figure}
\plotone{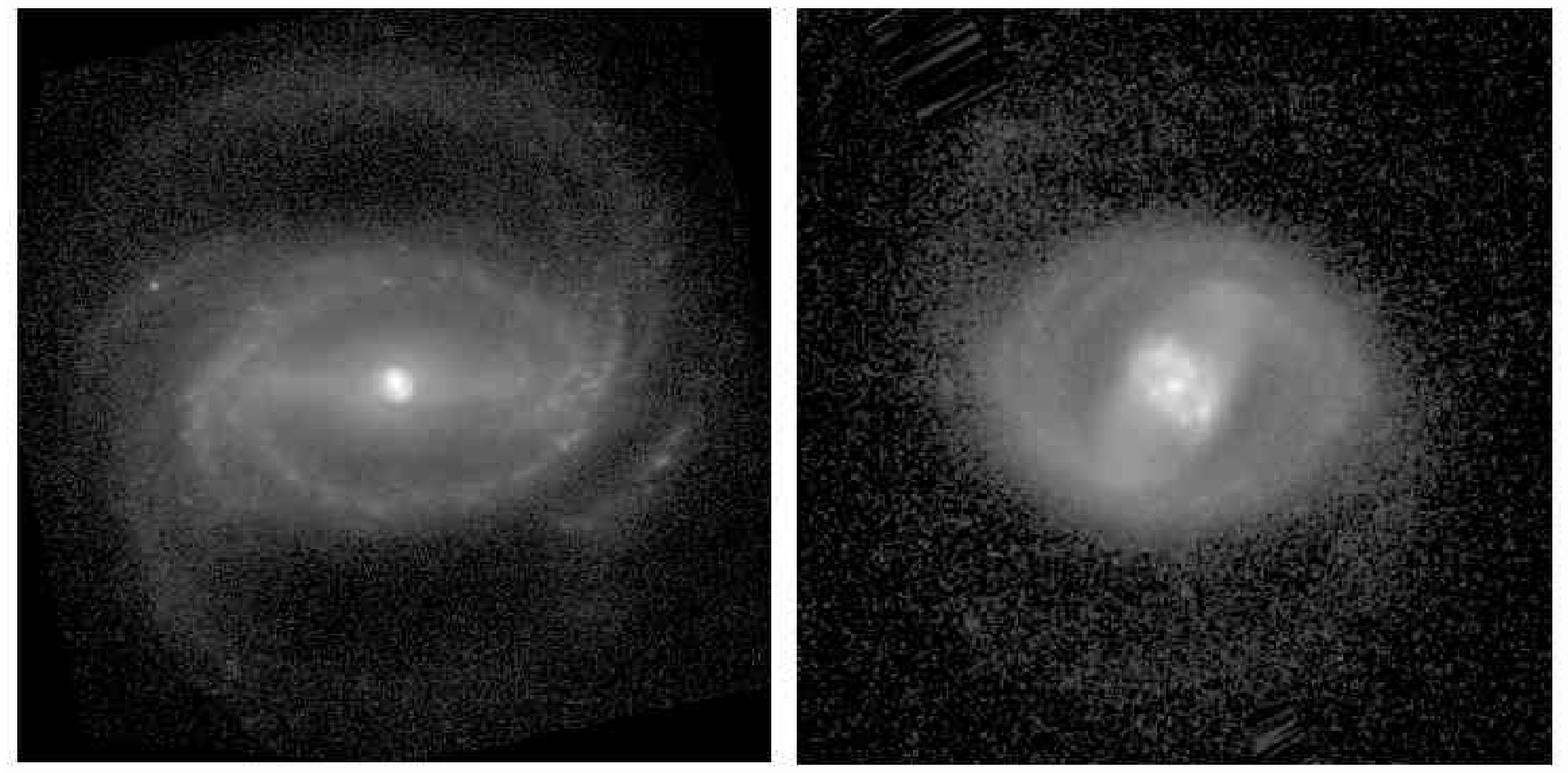}
\caption{Examples of two galaxies having highly a elongated inner ring
at the boundary of a broad oval. Left to right: NGC 1433, ESO 565$-$11
(both $B$-band). Each galaxy also has a prominent bar which is aligned
with the oval and inner ring in NGC 1433, but misaligned with these
features in ESO 565$-$11. ESO 565$-$11 also has a highly elongated,
large nuclear ring of star-forming regions.}
\label{inner_rings}
\end{figure}

\clearpage
\begin{figure}
\plotone{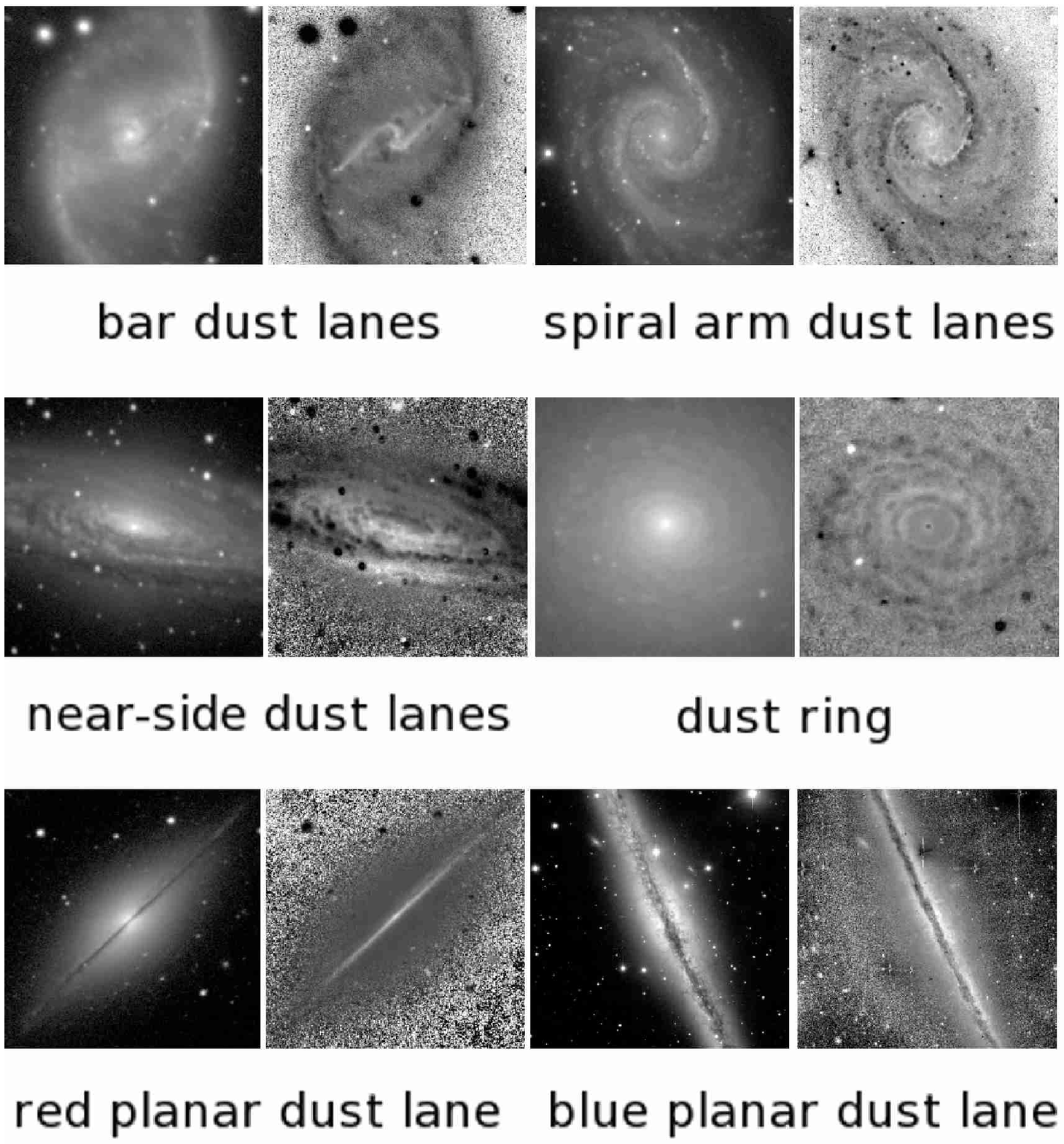}
\caption{Examples showing different classes of dust lanes (left to right) -
Row 1 NGC 1530, $V$-band image and $V-K_s$ color index map; NGC 1566,
$B$-band and $B-K_s$ color index map; Row 2 - NGC 7331, $B$-band image
and $B-I$ color index map; NGC 7217, $B$-band image and $B-I$ color
index map; Row 3 - NGC 7814, $B$-band image and $B-I$ color index map;
NGC 891, $B$-band and $B-V$ color index map.}
\label{dust_types}
\end{figure}

\clearpage
\begin{figure}
\plotone{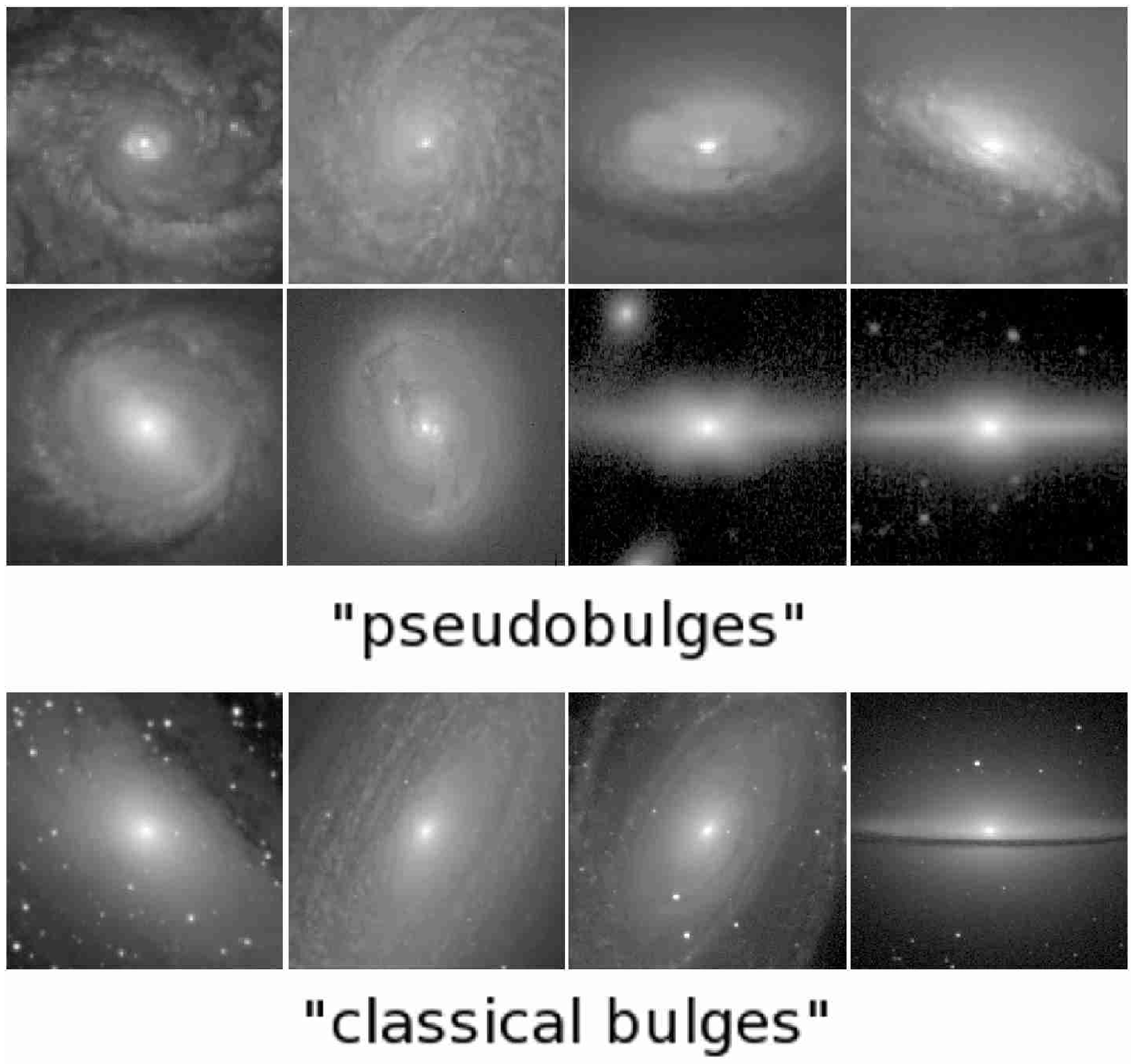}
\caption{Examples of pseudobulges and classical bulges in spiral
galaxies (left to right): Row 1: NGC 3177, 4030, 5377, and 1353 (all HST
wide $V$-band filter F606W; KK04);
Row 2: NGC 6782 ($I$-band, F814W), 3081 (wide $B$, F450W), 128 ($K_s$), and 1381 ($K_s$); Row 3: NGC 224 (M31), 2841, 3031 (M81),
and 4594 (M104) (all $B$-band). The images of NGC 128 and 1381 are from Bureau
et al. (2006).}
\label{bulges}
\end{figure}

\clearpage
\begin{figure}
\plotone{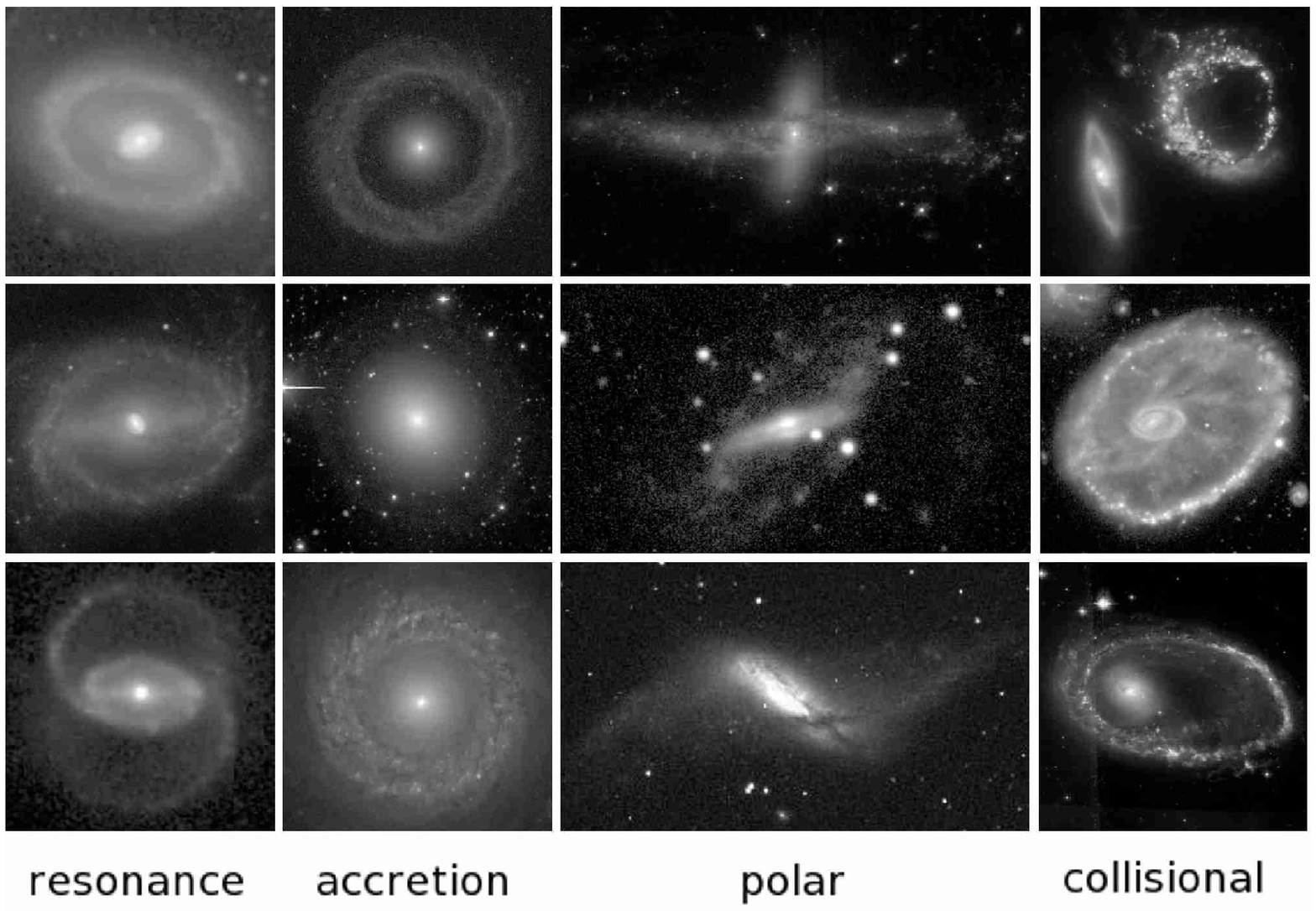}
\caption{Different classes of ring phenomena seen in galaxies (top
to bottom): Column 1 - NGC 3081, NGC 1433, and UGC 12646. Column 2
- Hoag's Object, IC 2006, and NGC 7742; Column 3 - NGC 4650A, ESO
235$-$58, and NGC 660; Column 4 - Arp 147, the Cartwheel, and the
Lindsay-Shapley ring. All images are from the dVA except Arp 147.}
\label{ringtypes}
\end{figure}

\clearpage
\begin{figure}
\plotone{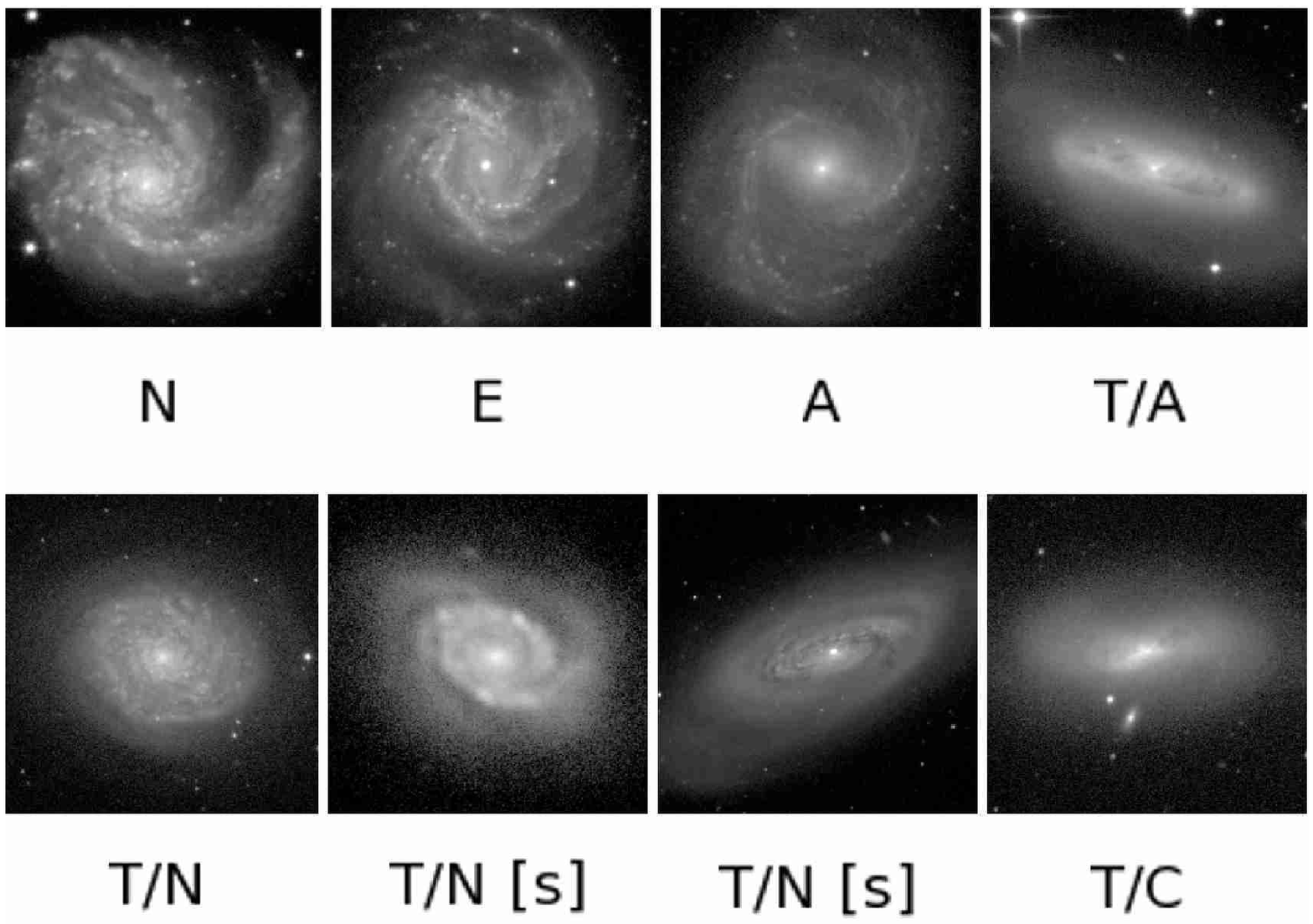}
\caption{Blue light morphologies of eight Virgo Cluster spirals having 
different Koopmann \& Kenney (2004) H$\alpha$ star formation morphologies.
The galaxies are (left to right): Row 1: NGC 4254 (normal N);
NGC 4303 (Enhanced E); NGC 4548 (Anemic A); NGC 4293 (Truncated/Anemic
T/A); Row 2: NGC 4689 (Truncated/Normal T/N); NGC 4580 and 4569
(Truncated/Normal (severe) T/N [s]); NGC 4424 (Truncated/Compact
T/C). Images are dVA $B$, except for NGC 4424 which is SDSS $g$.}
\label{virgotypes}
\end{figure}

\clearpage
\begin{figure}
\plotone{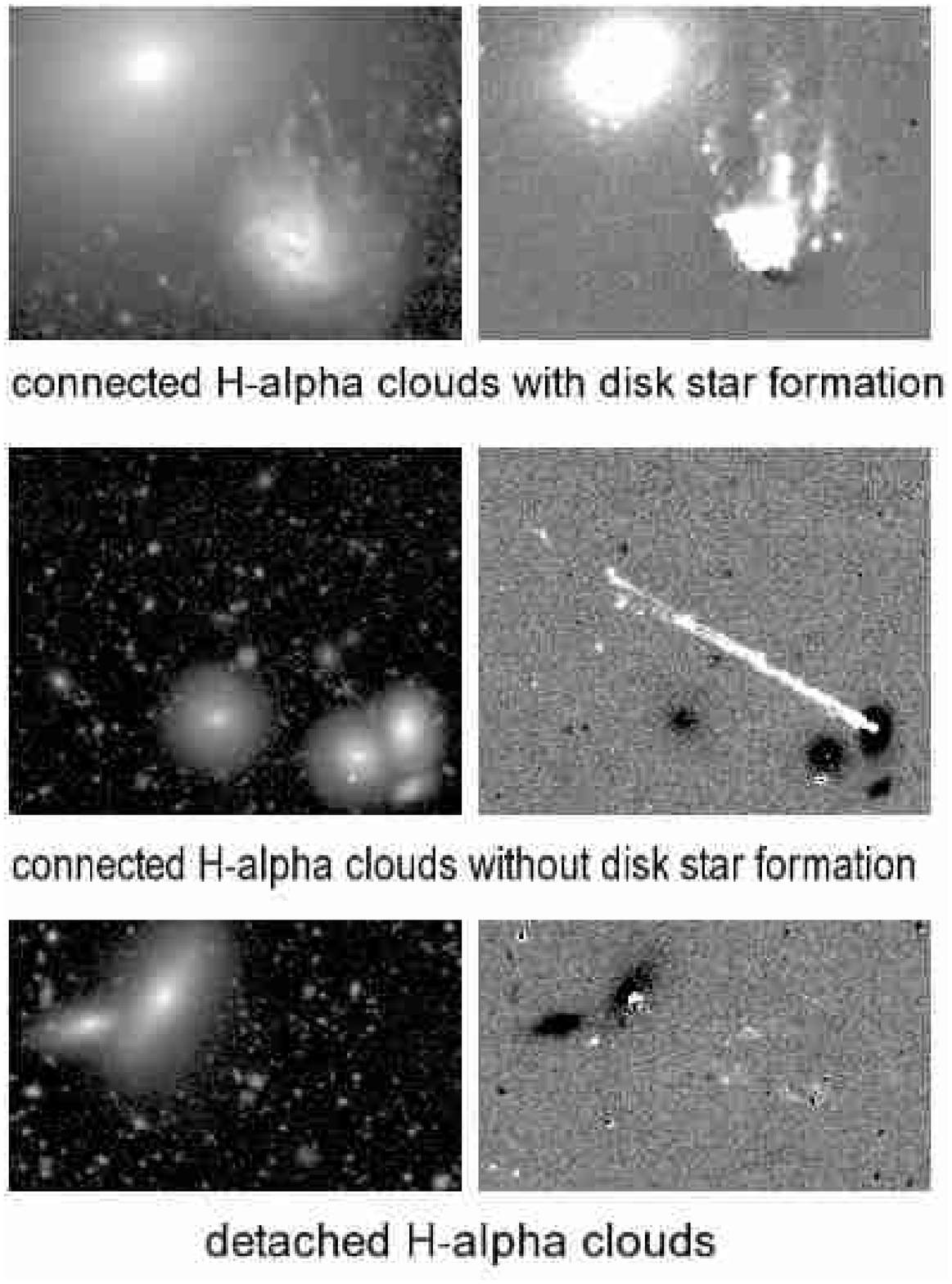}
\caption{}
\label{coma}
\end{figure}
\begin{figure}
\figurenum{29. (cont.)}
\caption{Three galaxies in a possible evolutionary stripping sequence in the
Coma Cluster. The images and categories are from Yagi et al. (2010). The
left frames are $B$-band images in units of mag arcsec$^{-2}$, while the
right frames are net H$\alpha$ images in linear intensity units (called
NB$-$R by Yagi et al.). From top to bottom, the galaxies are GMP 3816,
GMP 2910, and GMP 2923. The idea is that GMP 3816 is in an earlier phase
of stripping, such that there is still considerable disk ionized gas;
GMP 2910 is in a more advanced phase with still connected clouds but an
absence of disk emission; and finally GMP 2923 is in the most advanced
phase of the three, showing only scattered HII regions.}
\end{figure}

\clearpage
\begin{figure}
\plotone{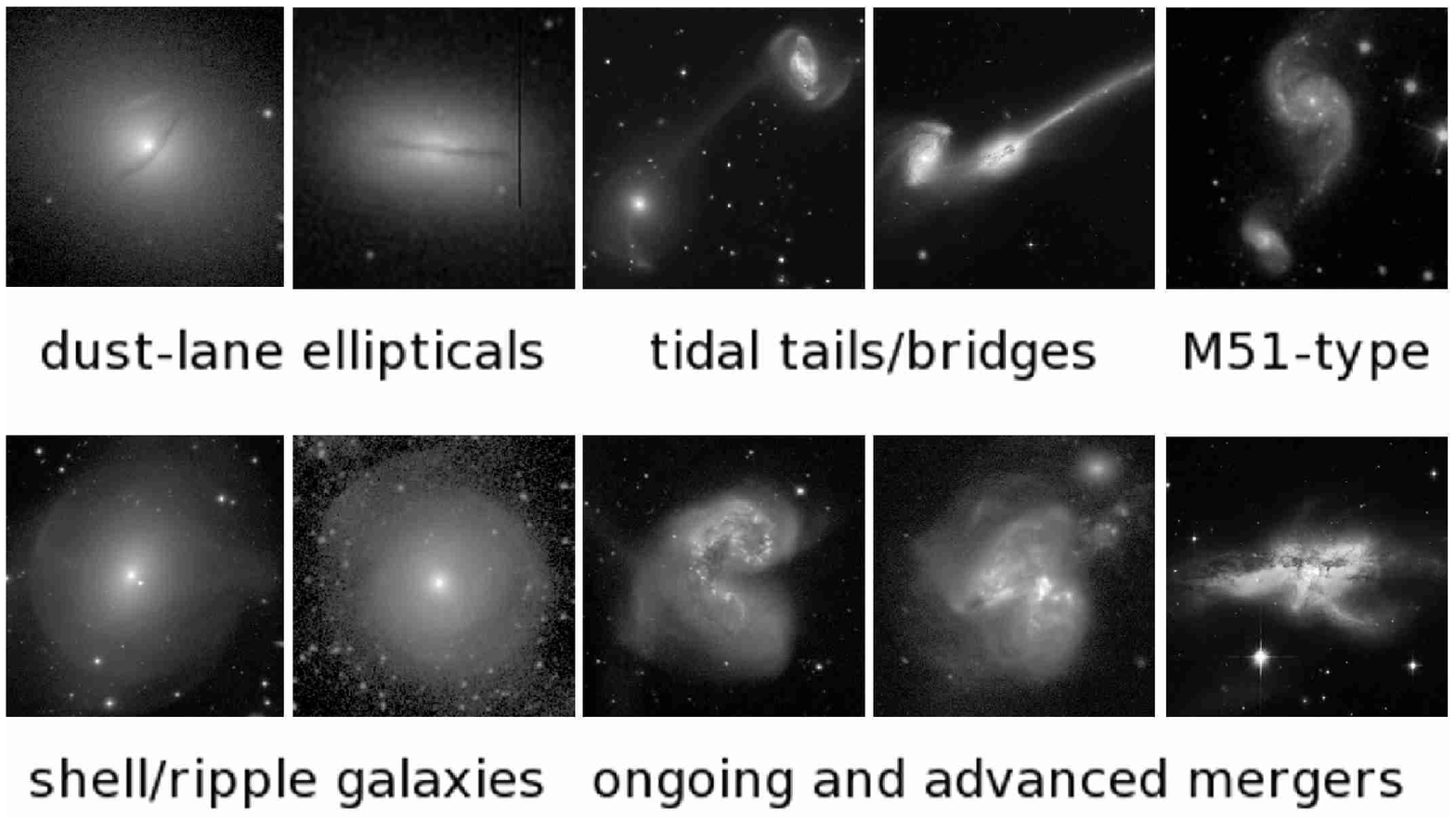}
\caption{Interacting and peculiar galaxies. The galaxies are
(left to right): Row 1: NGC 5485 (SDSS), 4370 (SDSS), 5216/18 (A.
Block), 4676 (Hubble Heritage), and 2535-6 (SDSS);
Row 2: NGC 2865, 474, 4038-9, 3690, and 6240, all $B$ except NGC 474,
which is a 3.6$\mu$m image (section 12).
}
\label{pec_Es}
\end{figure}

\clearpage
\begin{figure}
\plotone{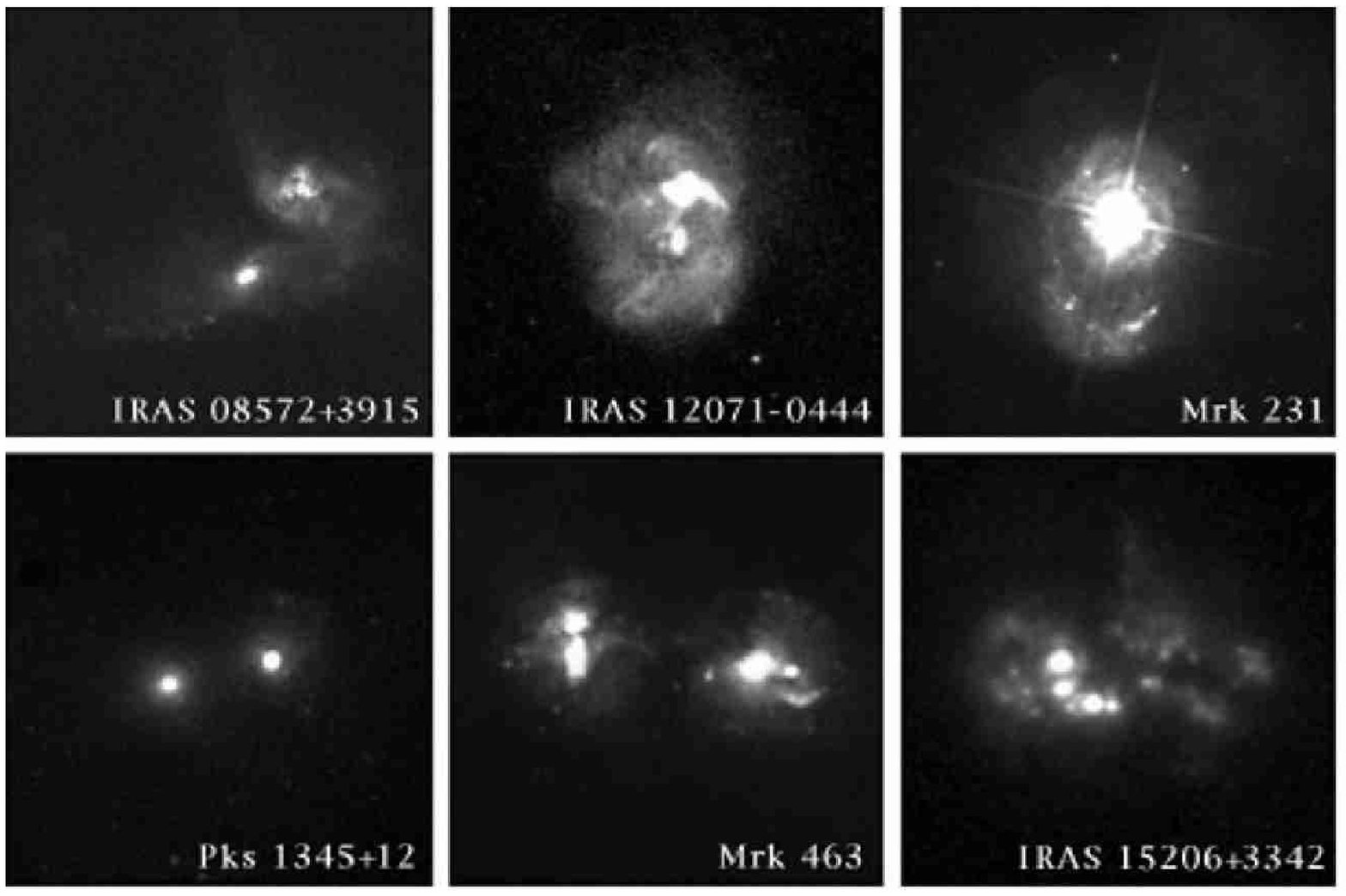}
\caption{The morphologies of six ``ultra-luminous infrared galaxies" from
HST optical/near-IR imaging (Surace et al. 1998). These images are not
in units of mag arcsec$^{-2}$.}
\label{ulirgs}
\end{figure}

\clearpage
\begin{figure}
\plotone{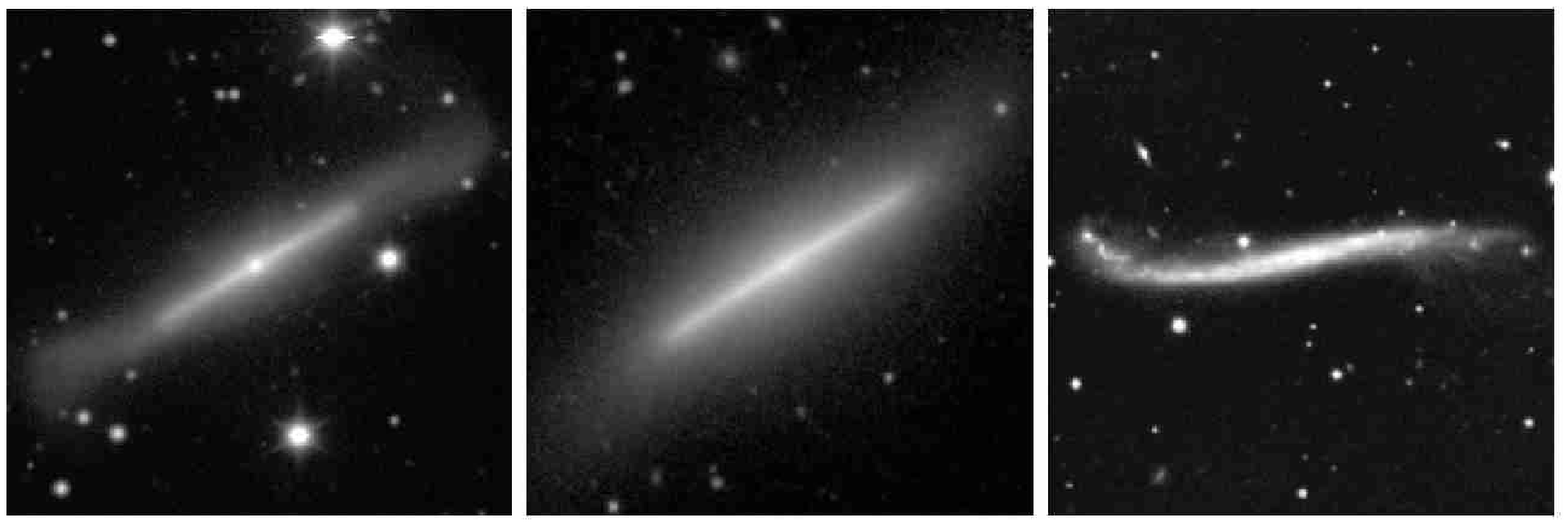}
\caption{Three galaxies showing strong optical disk warping. Left to right:
NGC 4762 ($B$), NGC 4452 (SDSS), UGC 3697 (Internet Encyc. of Science)}
\label{warps}
\end{figure}

\clearpage
\begin{figure}
\vspace{-0.5truein}
\plotone{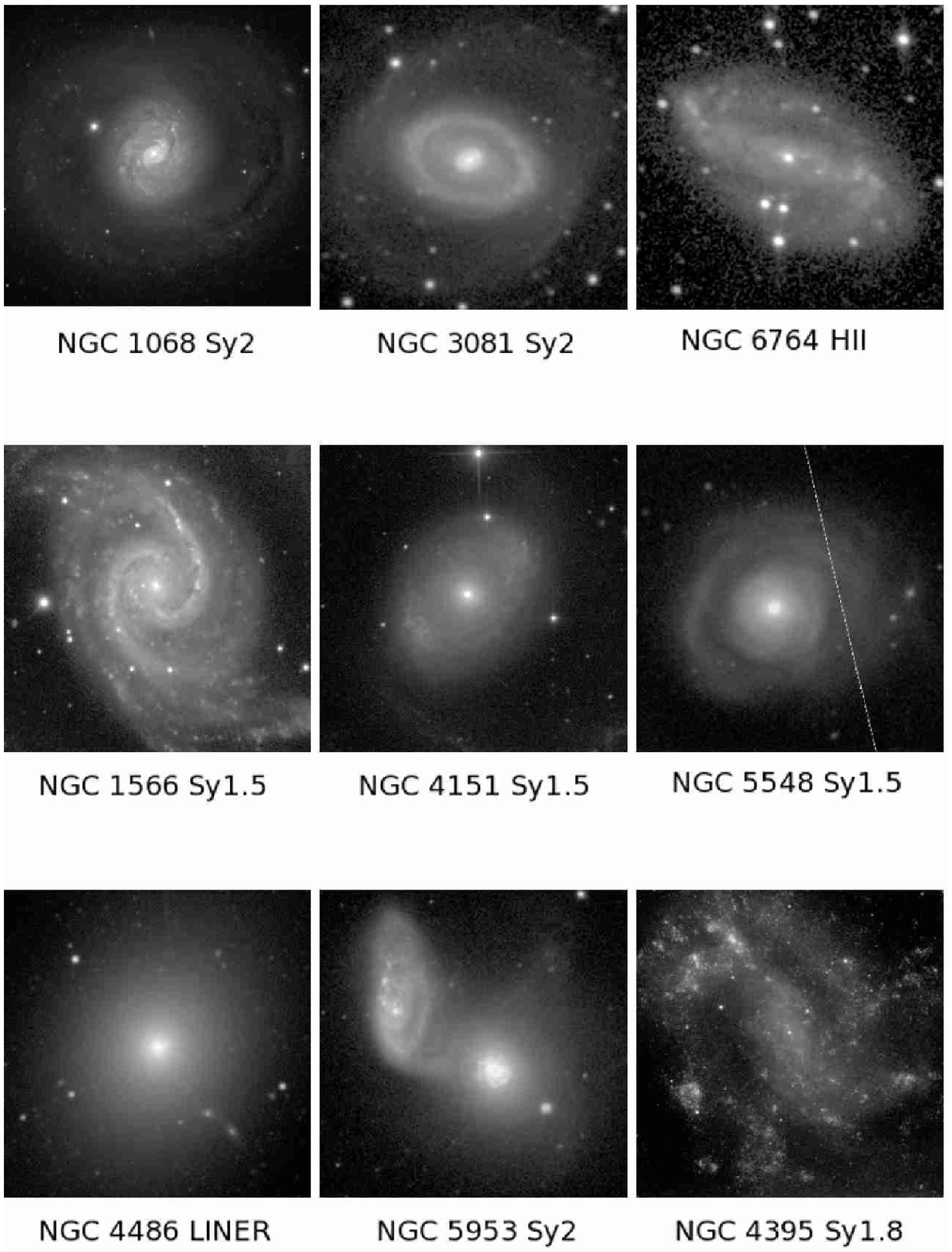}
\vspace{-0.5truein}
\caption{Images of nearby active galaxies, dVA $B$-band except for
NGC 5548 and 5953, which are SDSS. The activity classification
is from Veron-Cetty \& Veron (2006).}
\label{actives}
\end{figure}

\clearpage
\begin{figure}
\plotone{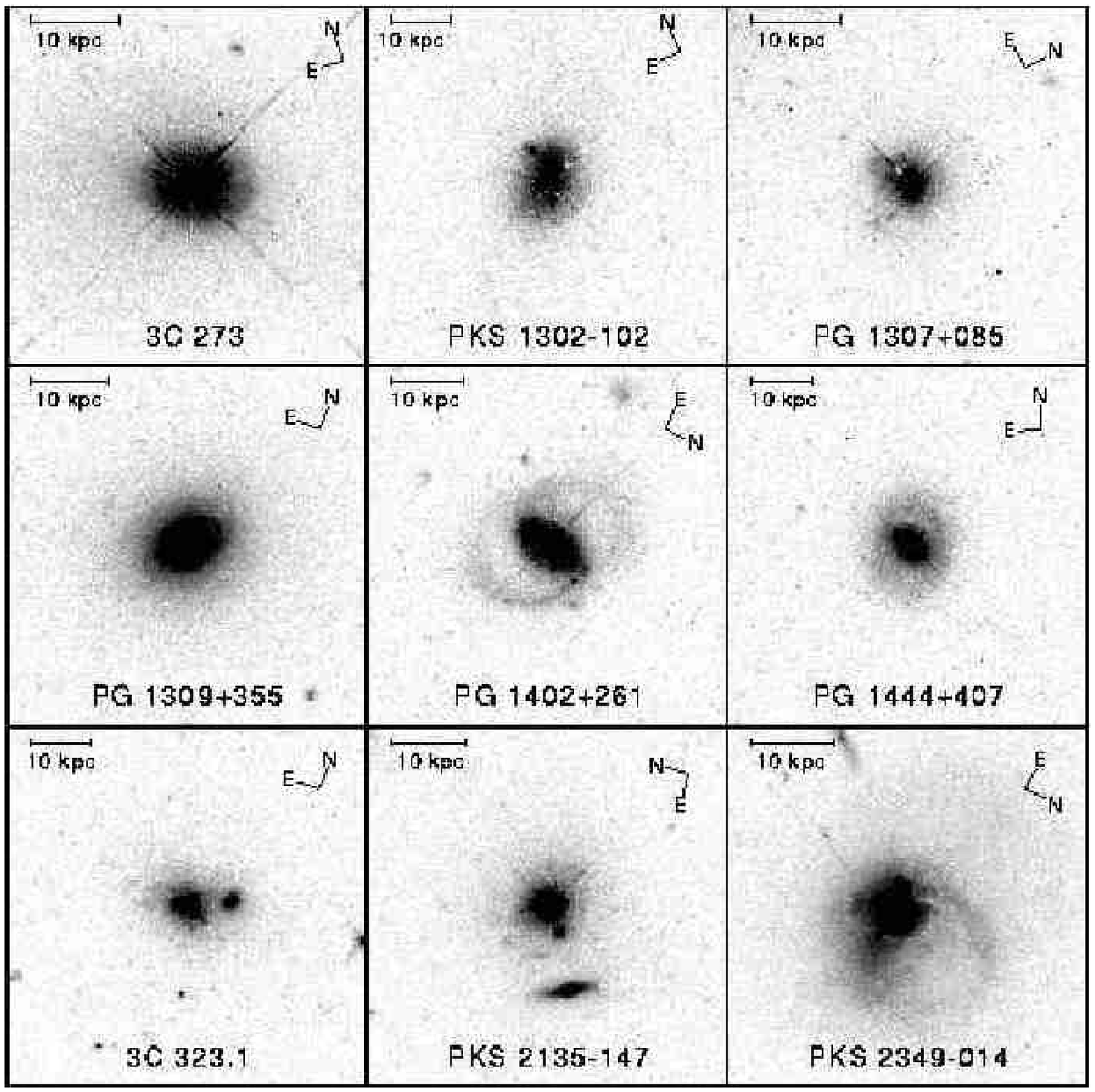}
\caption{The morphologies of the host galaxies of nine nearby quasars,
from Bahcall et al. (1997). These images are not in units of mag arcsec$^{-2}$.}
\label{quasars}
\end{figure}

\clearpage
\begin{figure}
\plotone{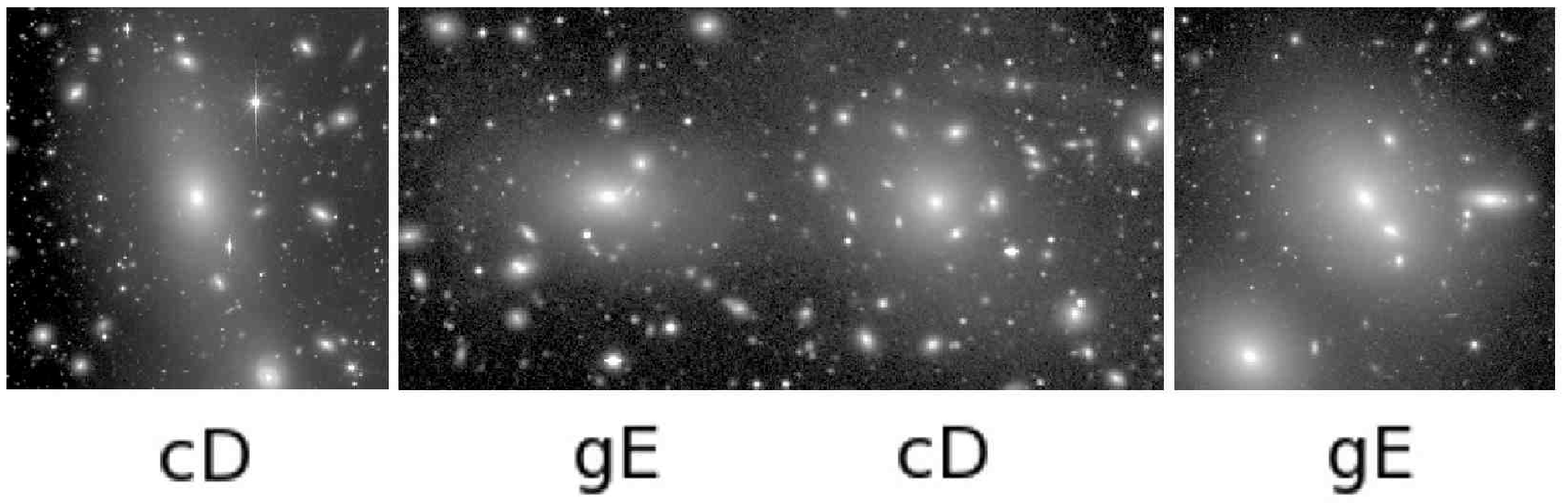}
\caption{Deep images of the brightest members of three rich galaxy clusters
(left to right): UGC 10143 in A2152 ($R$-band), NGC 4889 (left) and 4874 
(right) in A1656 ($B$-band), and NGC 6041 ($R$-band) in A2151. The images
of UGC 10143 and NGC 6041 are from Blakeslee (1999), while that of NGC 4874-89
is from the dVA.}
\label{bcms}
\end{figure}

\clearpage
\begin{figure}
\plotone{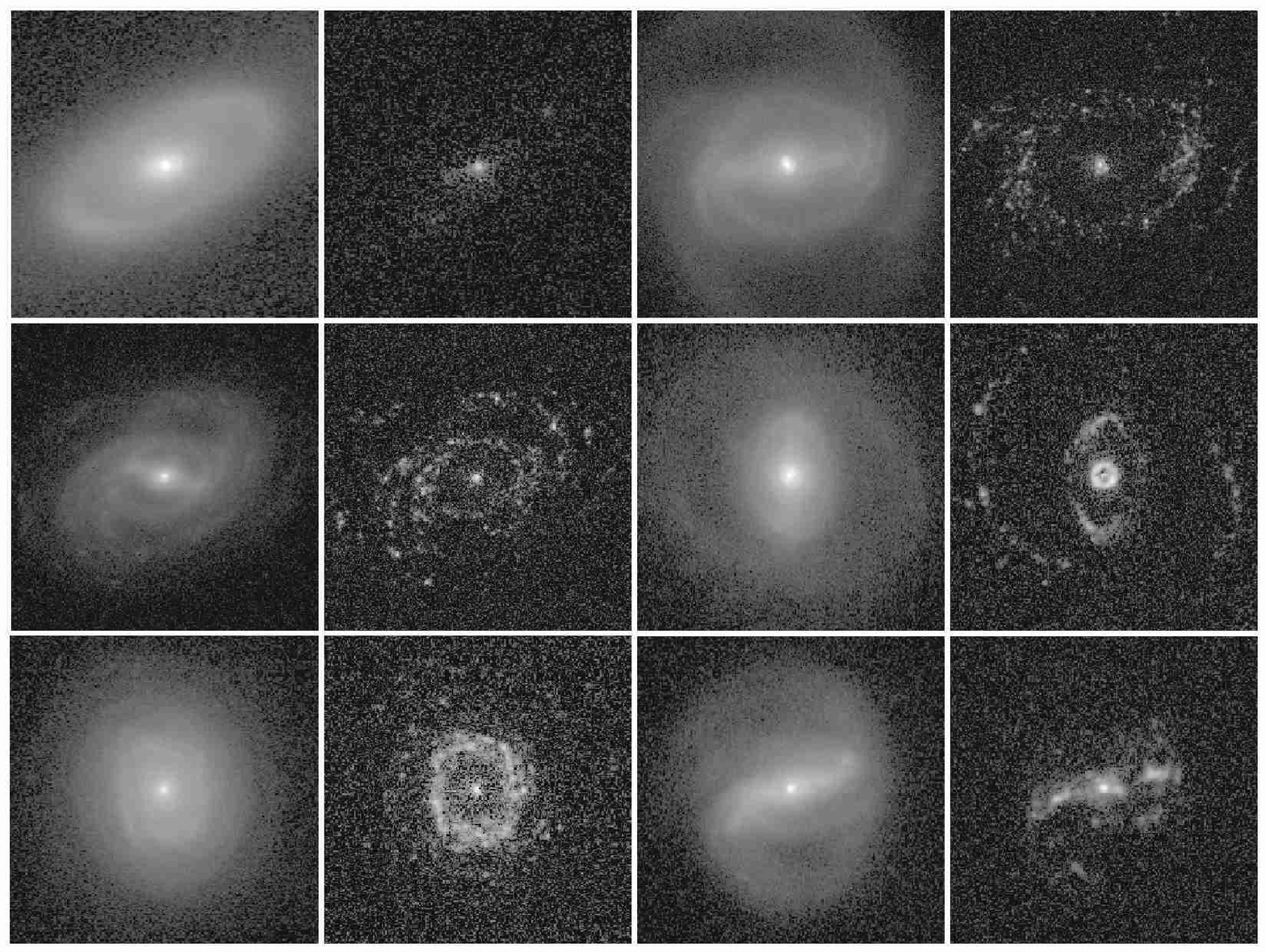}
\caption{Red continuum (left) and net H$\alpha$ (right) images of six
early-to-intermediate-type galaxies. The galaxies are (left to right):
top row: NGC 7702, 1433; middle row: NGC 7329, 6782; bottom row: NGC
6935, 7267. From Crocker, Baugus, \& Buta (1996).}
\label{halpha}
\end{figure}

\clearpage
\begin{figure}
\plotone{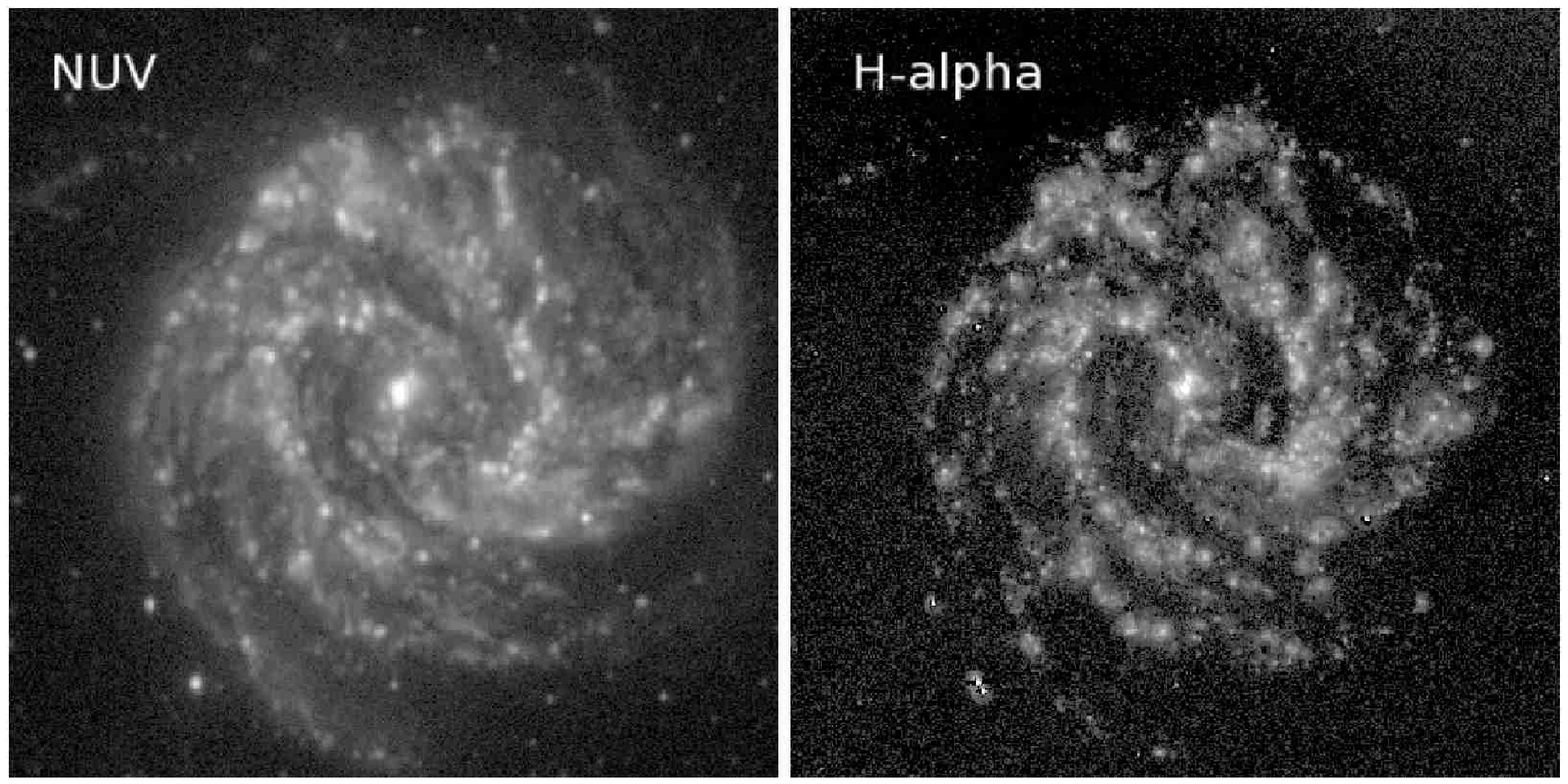}
\caption{Comparison of a near-UV image (0.225$\mu$m) with an H$\alpha$
image for the nearby spiral galaxy M83.}
\label{nuv-halpha}
\end{figure}

\clearpage
\begin{figure}
\plotone{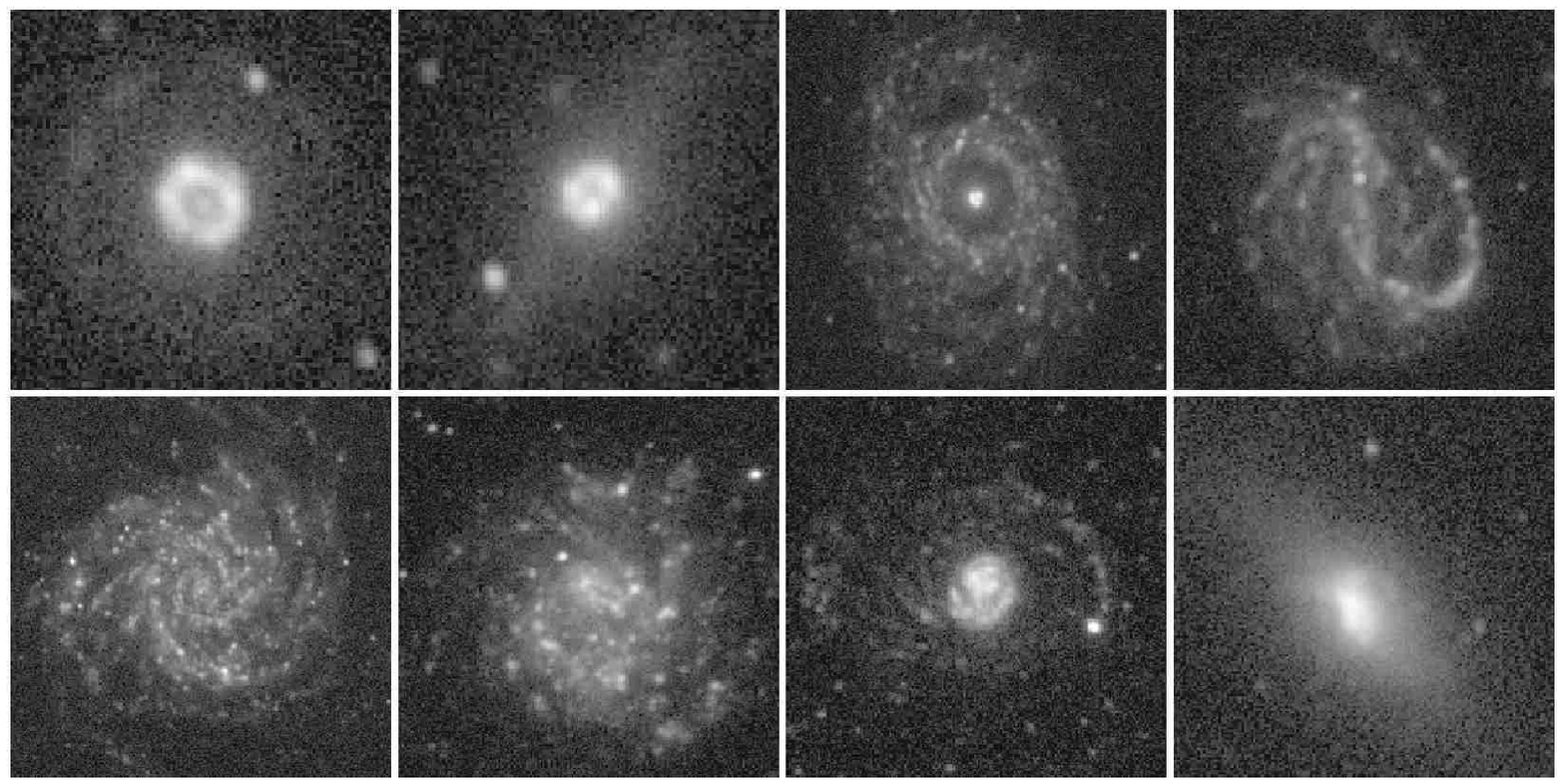}
\caption{GALEX near-UV images of eight galaxies: (left to right)
Top row: NGC 1317, 4314, 3351, and 7479; bottom row: NGC 628, 5474,
4625, and 5253}
\label{nuv}
\end{figure}

\clearpage
\begin{figure}
\vspace{-0.5truein}
\plotone{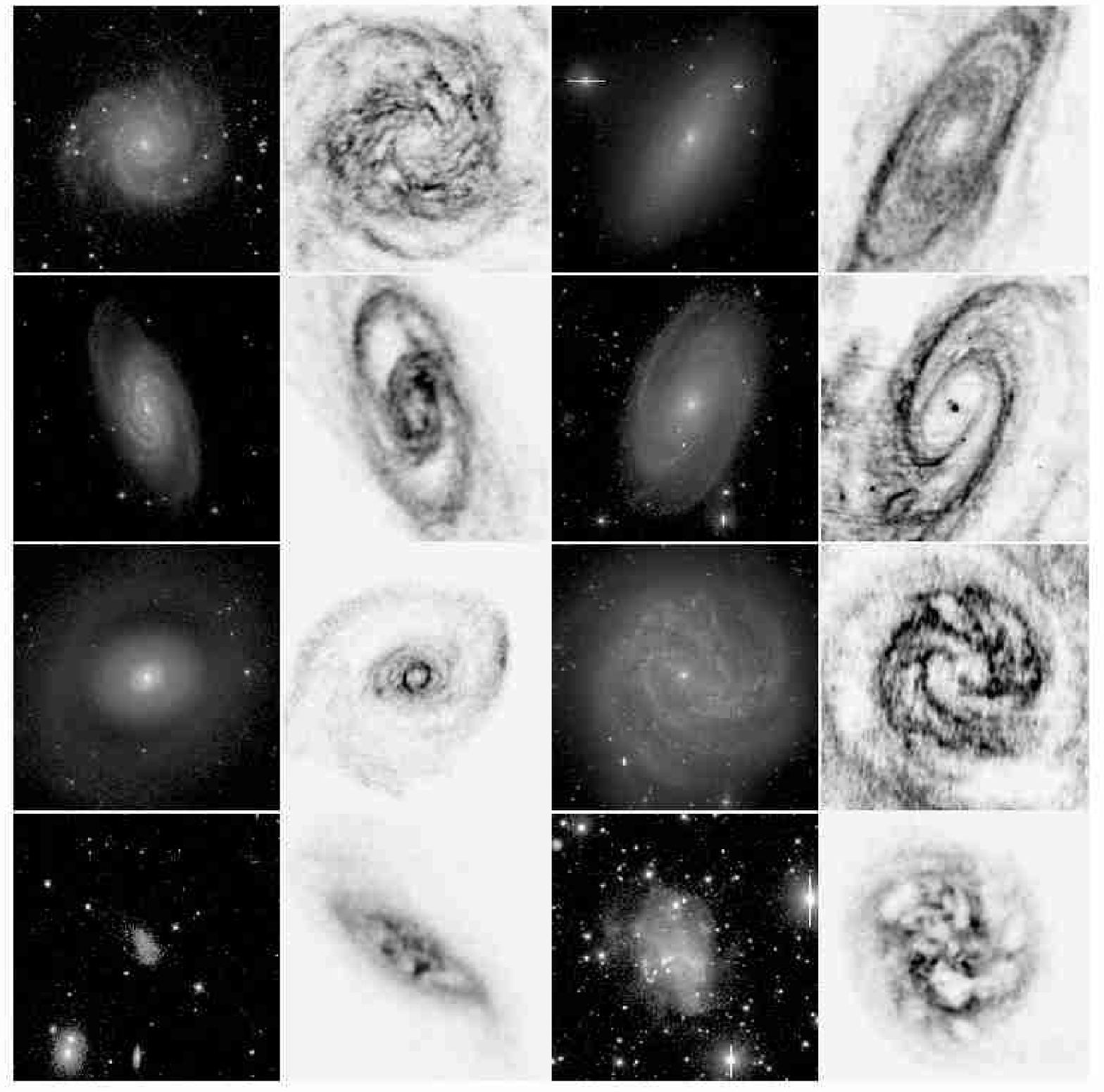}
\vspace{-0.5truein}
\caption{HI morphologies (Walter et al. 2008) 
of eight galaxies as compared to optical
$B$-band morphologies. Left panels: NGC 628 (M74), NGC 4258 (M106),
NGC 4736 (M94), and DDO 154. Right panels: NGC 2841, NGC 3031 (M81);
NGC 5236 (M83), and DDO 50 (Ho II).}
\label{himaps}
\end{figure}

\clearpage
\begin{figure}
\vspace{-0.5truein}
\plotone{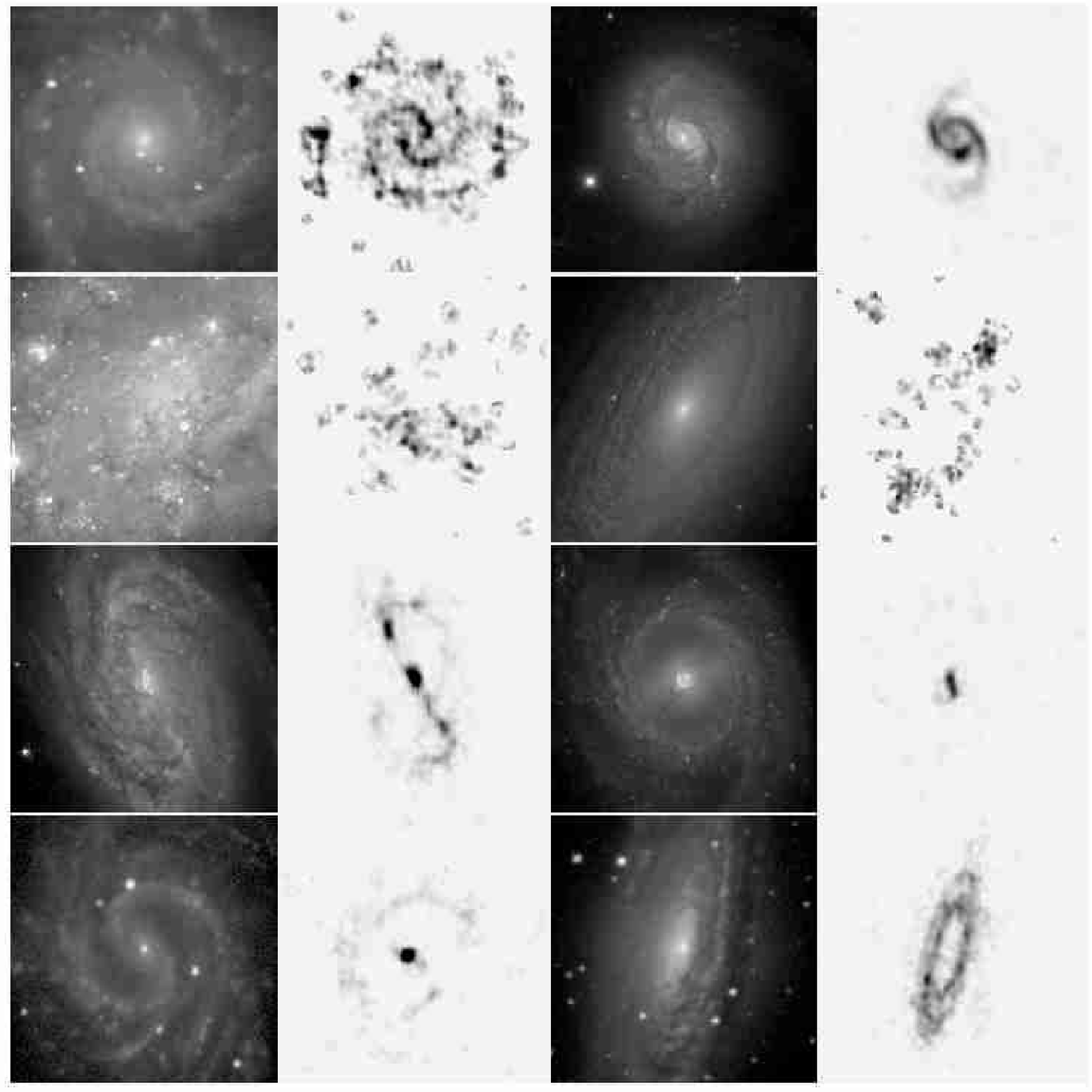}
\vspace{-0.5truein}
\caption{CO morphologies (Helfer et al. 2003)
of eight galaxies as compared to optical
$B$-band morphologies. Left panels: NGC 628 (M74), NGC 2403,
NGC 2903, and NGC 4535. Right panels: NGC 1068 (M77), NGC 2841;
NGC 3351 (M95), and NGC 7331.}
\label{comaps}
\end{figure}

\clearpage
\begin{figure}
\vspace{-0.5truein}
\plotone{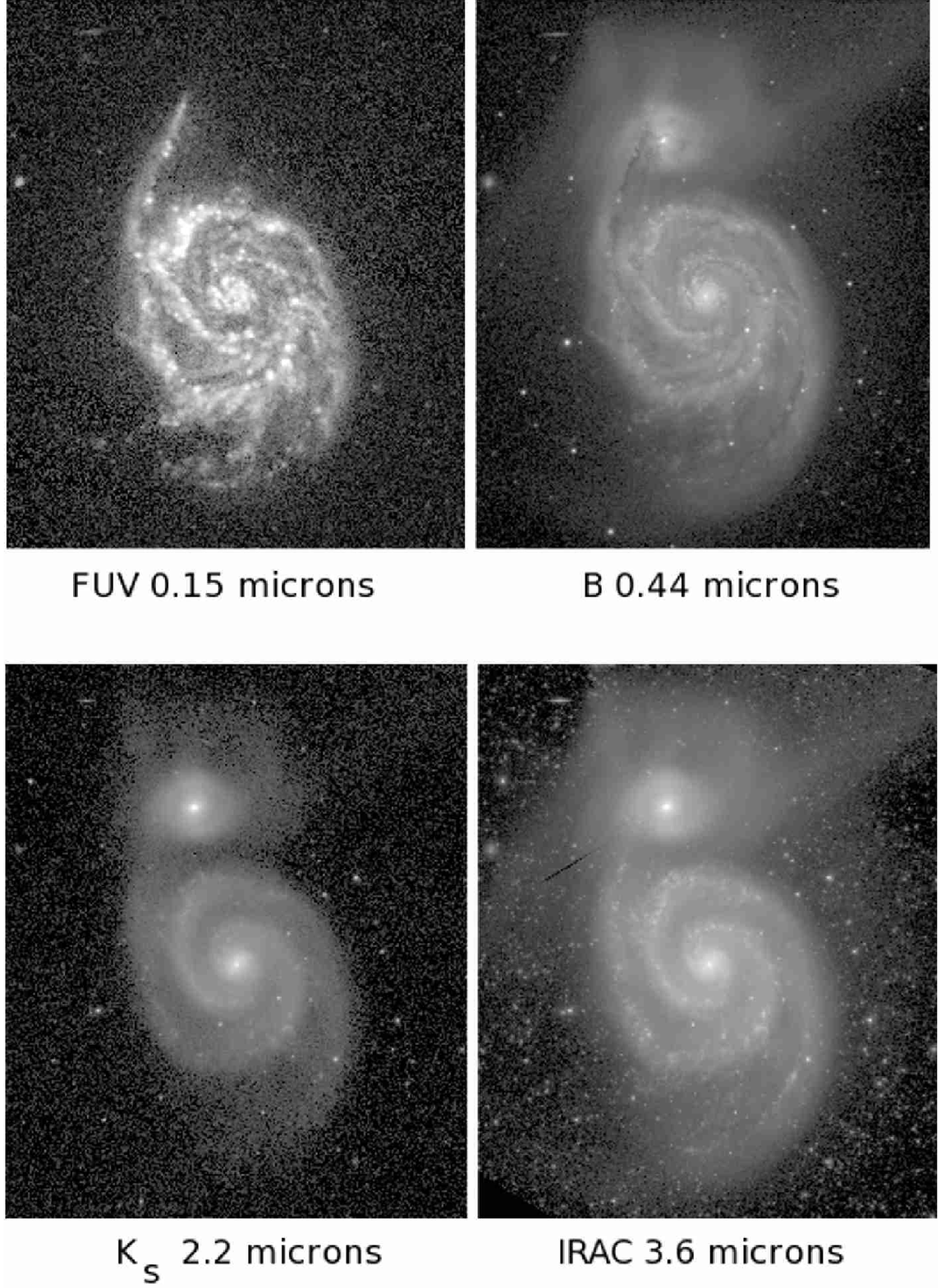}
\vspace{-0.5truein}
\caption{Multi-wavelength images of M51.}
\label{m51_frames}
\end{figure}

\clearpage
\begin{figure}
\plotone{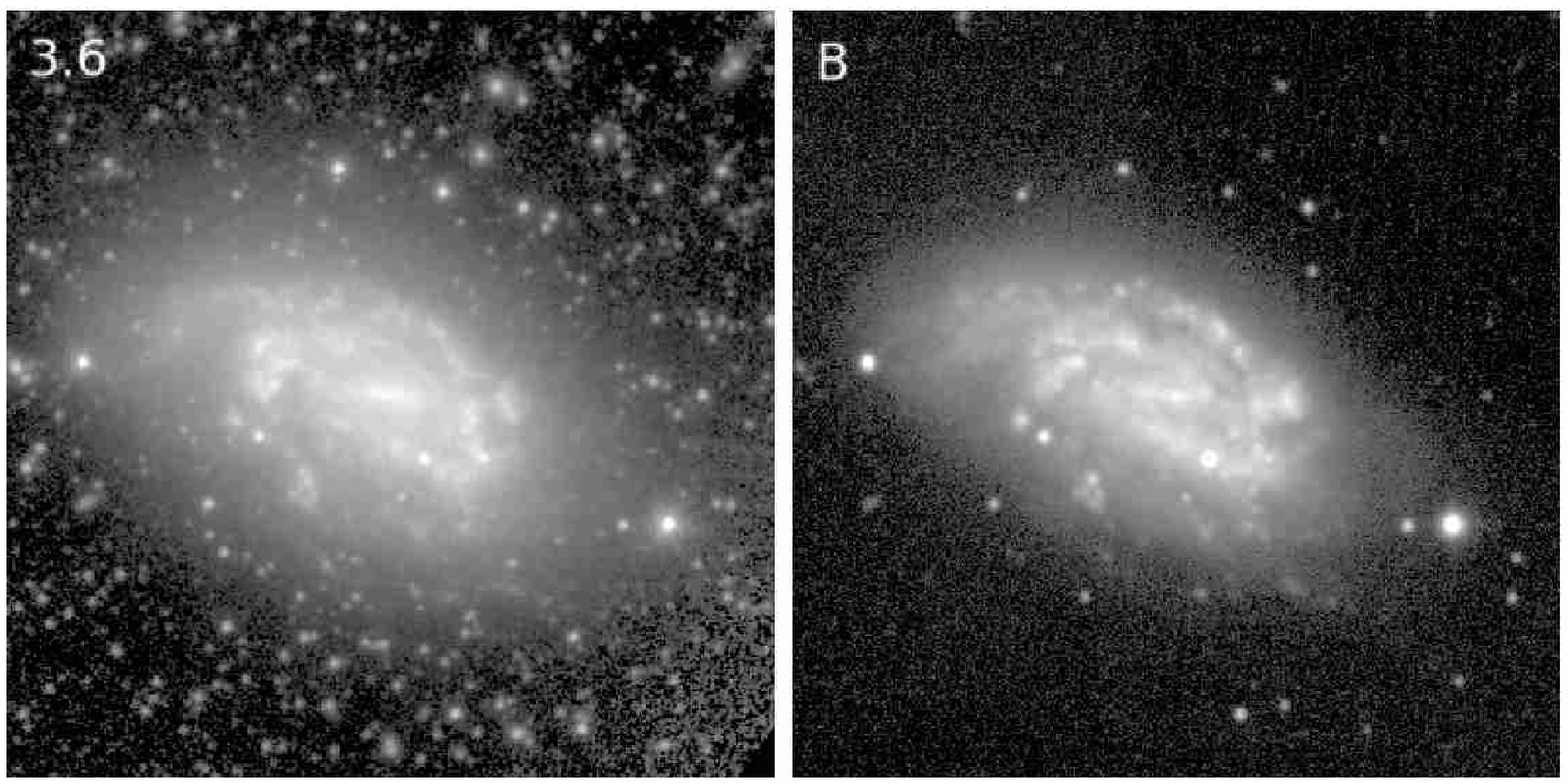}
\caption{Comparison of IRAC 3.6$\mu$m image of NGC 1559 (left) with a ground-based
$B$-band image of the same galaxy at right. Note the significant
correspondence of features between the two very different wavelength
domains in this case.} 
\label{ngc1559}
\end{figure}

\clearpage
\begin{figure}
\vspace{-0.75truein}
\plotone{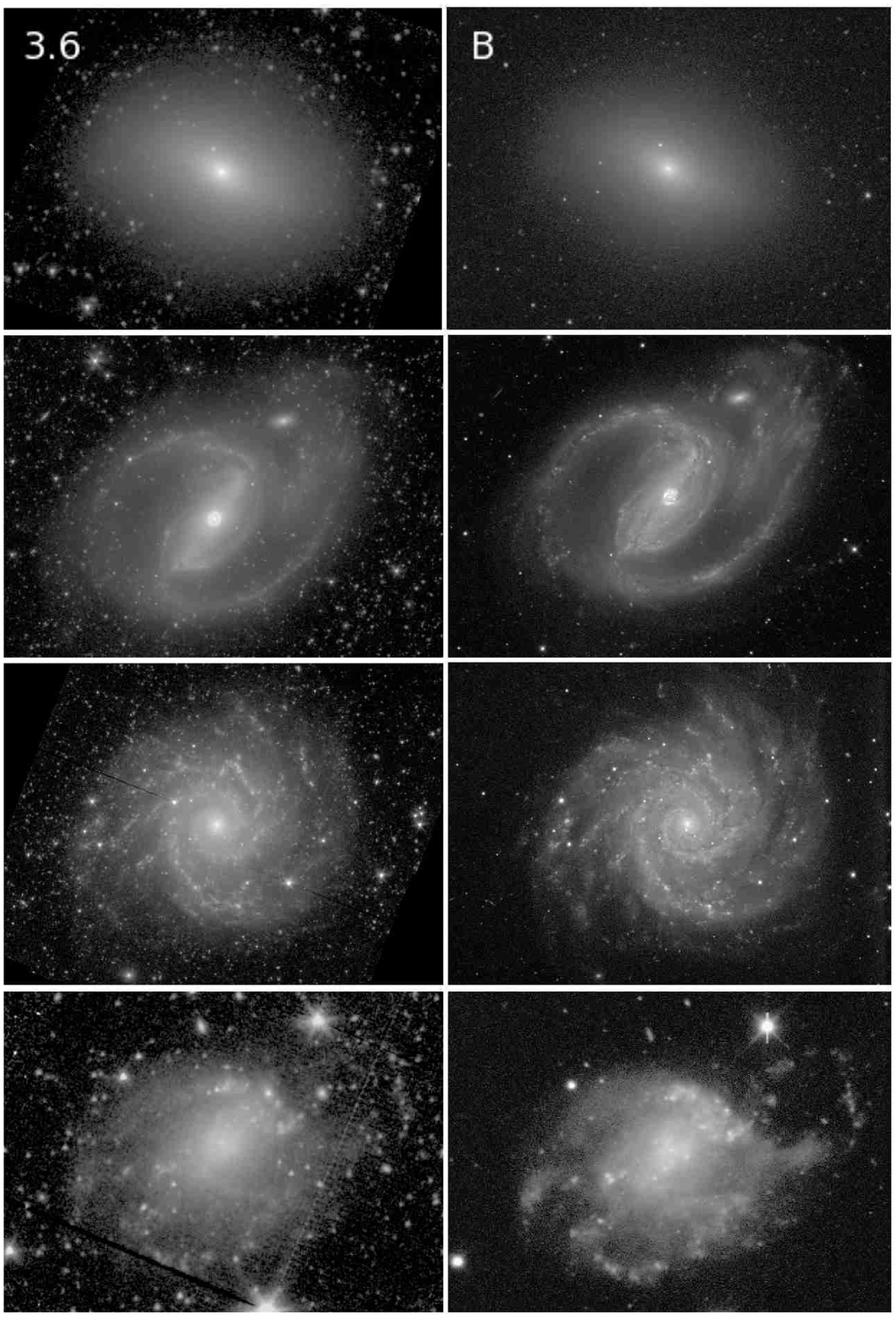}
\vspace{-0.65truein}
\caption{Comparison of IRAC 3.6$\mu$m images (left frames) with ground-based
$B$-band images for (top to bottom): NGC 584, 1097, 628, and 428.}
\label{nir01}
\end{figure}

\clearpage
\begin{figure}
\plotone{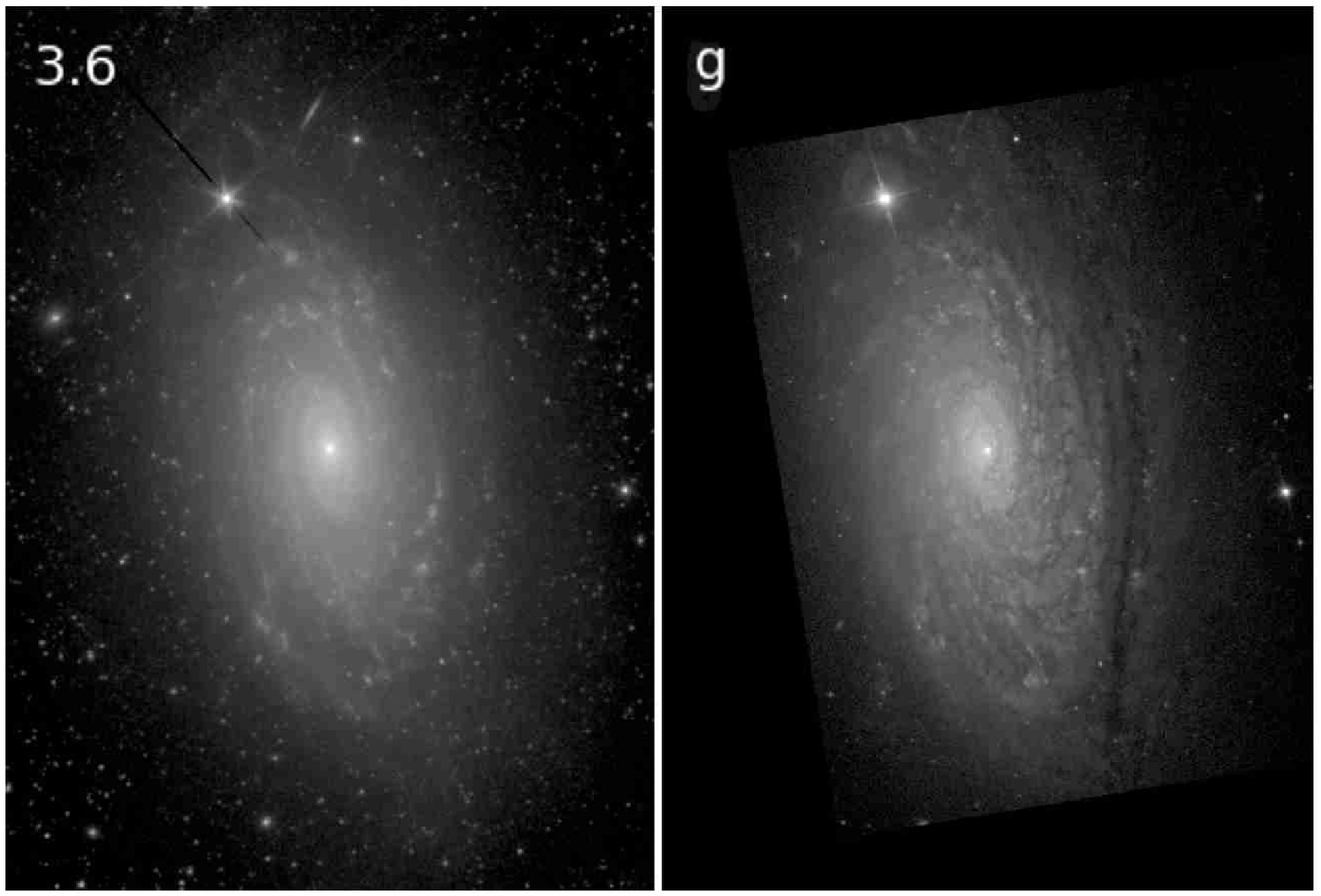}
\caption{Comparison of IRAC 3.6$\mu$m and SDSS $g$-band 
images of the flocculent spiral galaxy NGC 5055.}
\label{nir03}
\end{figure}

\clearpage
\begin{figure}
\plotone{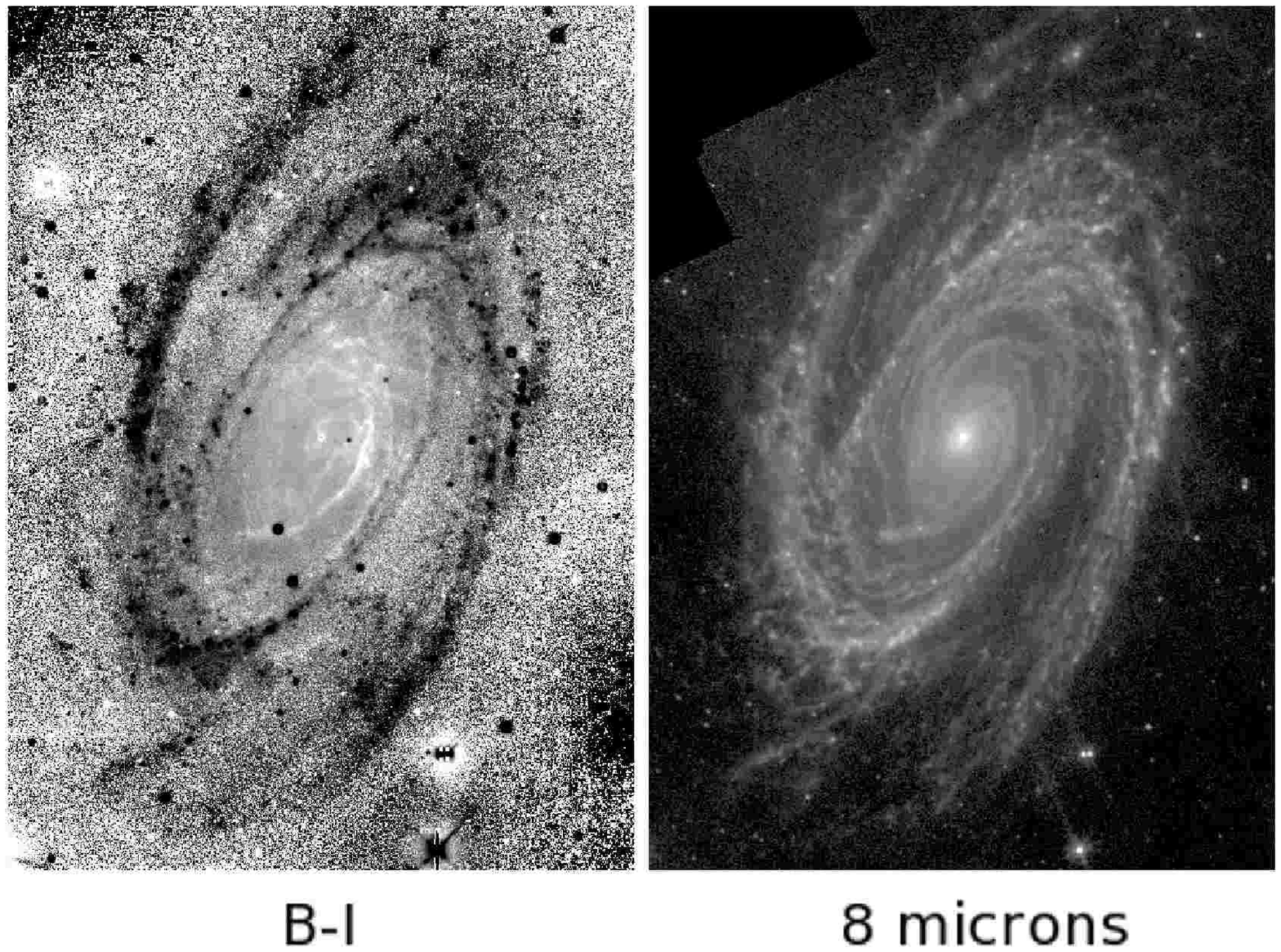}
\caption{A comparison between a $B-I$ color index map and an 8$\mu$m
dust emission map of the nearby spiral galaxy M81.}
\label{m81frames}
\end{figure}

\clearpage
\begin{figure}
\plotone{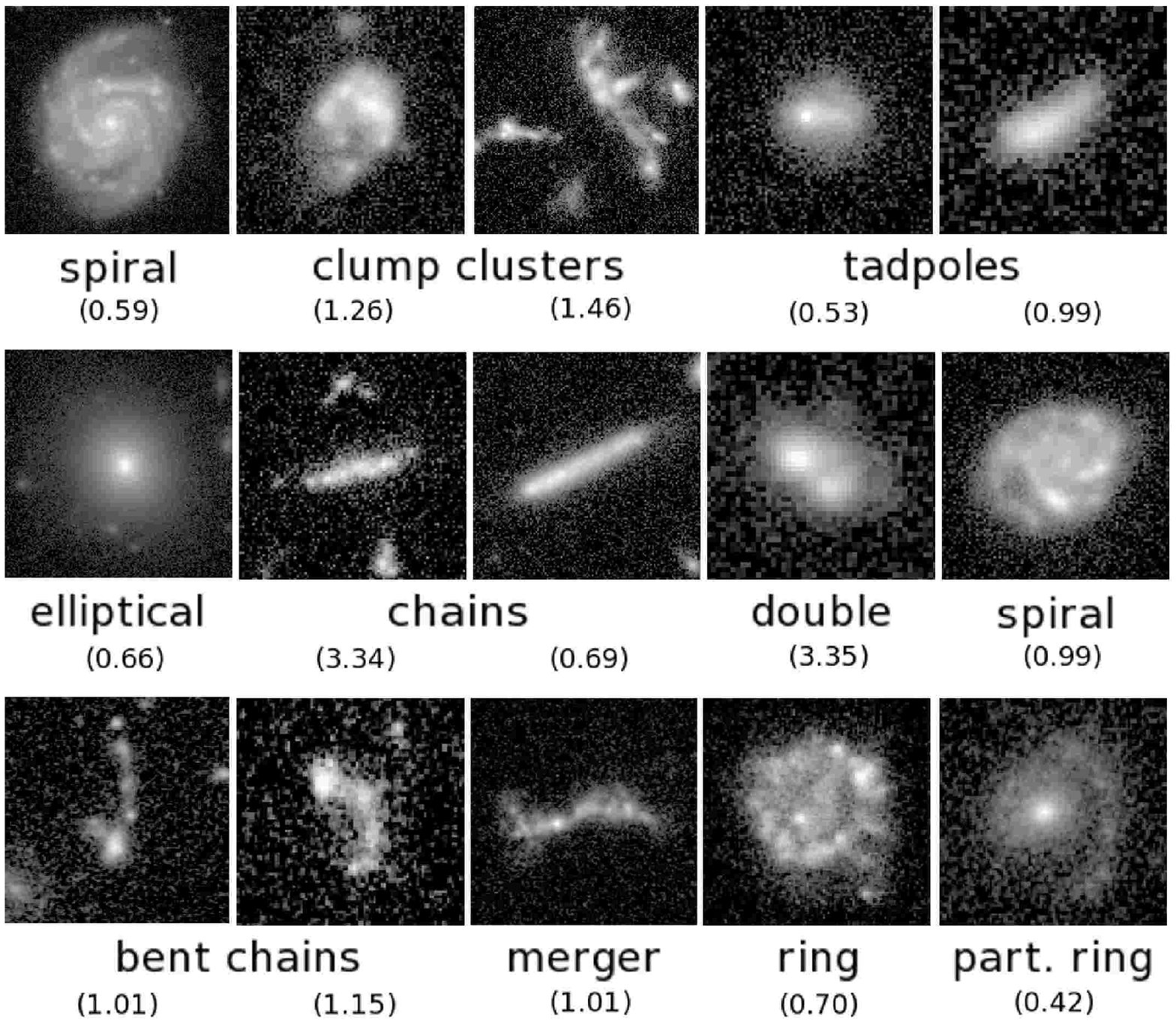}
\caption{Intermediate to high redshift galaxy morphology ($V$ and $i$-bands). The categories are due to
EES04 and EE06. The number in parentheses below each
frame is the redshift $z$ of the galaxy shown.}
\label{highz}
\end{figure}

\clearpage
\begin{figure}
\plotone{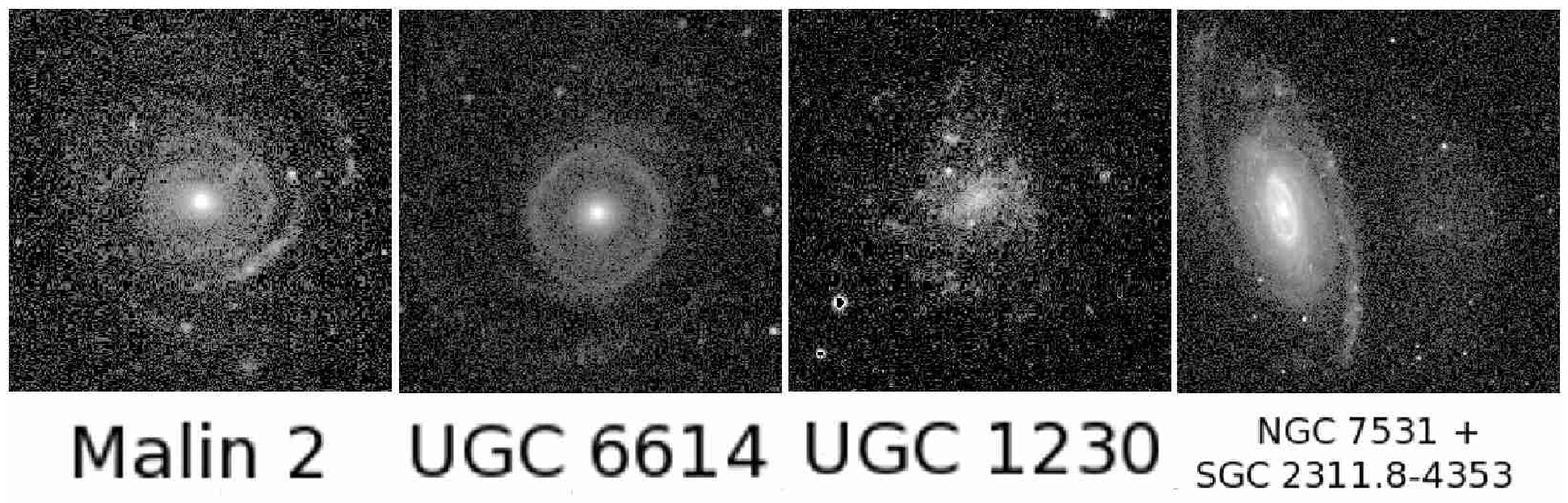}
\caption{Examples of giant or large low surface brightness galaxies.
In the far right panel, SGC 2311.8$-$4353 is the diffuse object to the
right of high surface brightness spiral NGC 7531. All of these images are
$B$-band.}
\label{glsbs}
\end{figure}

\clearpage
\begin{figure}
\plotone{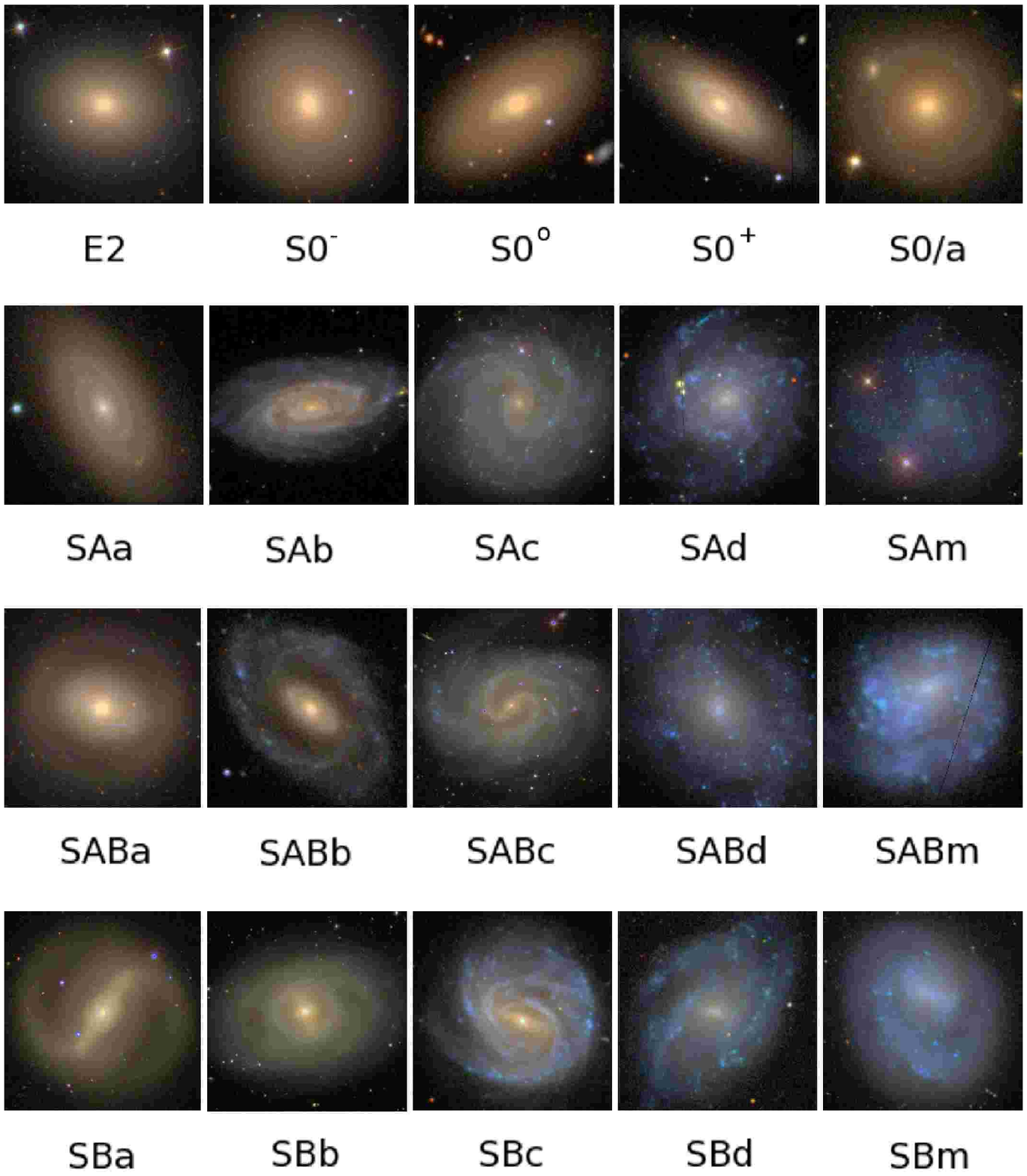}
\caption{The Hubble tuning fork of ellipticals, S0s, and spirals of different
bar classifications are shown here using SDSS color images. The galaxies are
(left to right): Row 1 - NGC 3608, 4203, 6278, 4324, and 932. Row 2 - NGC
4305, 5351, 3184, 5668, and IC 4182. Row 3 - NGC 4457, 5409, 4535, 5585, and 
3445. Row 4 - NGC 4314, 3351, 3367, 4519, and 4618.}
\label{colortf}
\end{figure}

\clearpage
\begin{figure}
\plotone{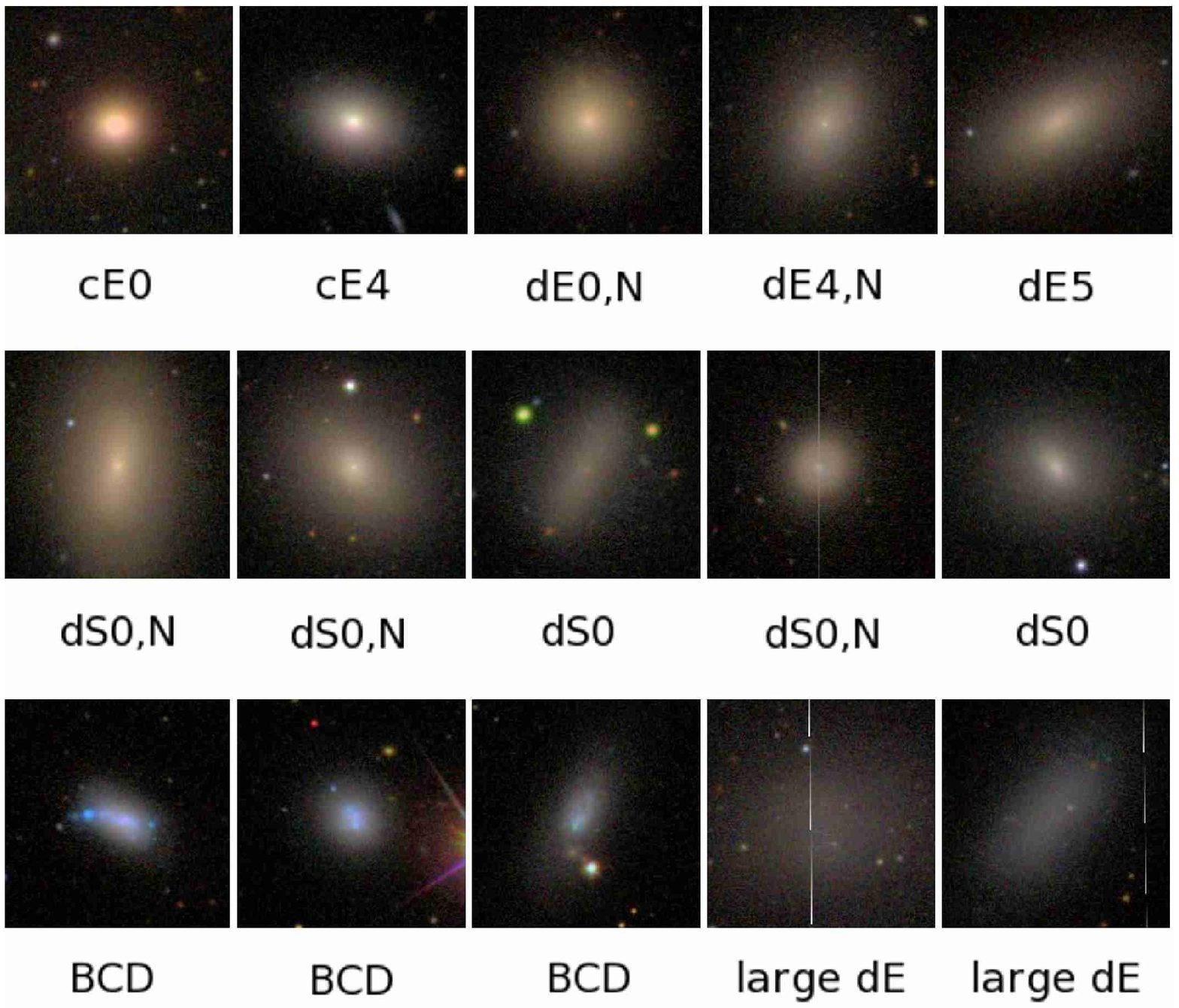}
\caption{Examples of dwarf galaxies in the Virgo Cluster, drawn from the
catalogue of Binggeli, Sandage, \& Tammann (1985) and highlighted using
SDSS color images. The classifications are from the BST catalogue and
the galaxies are (left to right): Row 1 - NGC 4486B, IC 767, IC 3470, IC
3735, and UGC 7436. Row 2 - NGC 4431, IC 781, IC 3292, VCC 278, IC 3586.
Row 3 - VCC 459, VCC 2033, VCC 841, IC 3475, and IC 3647.}
\label{dwarfs}
\end{figure}

\clearpage
\begin{figure}
\plotone{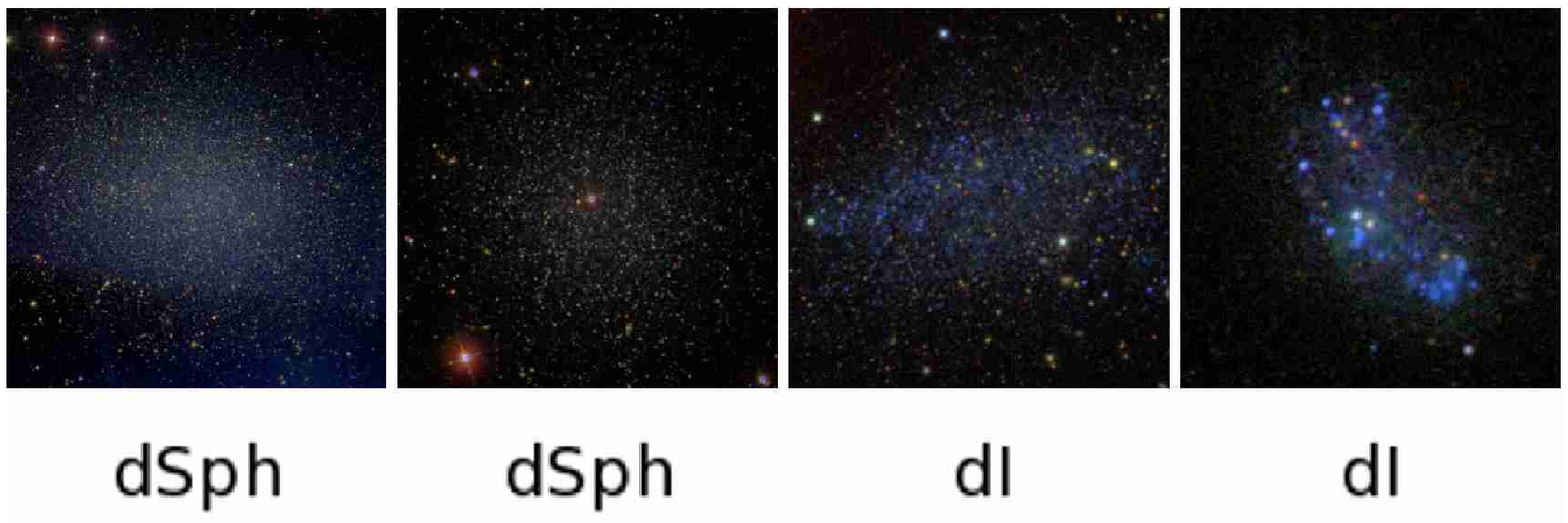}
\caption{Four Local Group dwarfs having $M_V$ $>$ $-$12 (left to right):
Leo I, Leo II, Leo A, and DDO 155 (all SDSS color images)}
\label{lgroup}
\end{figure}

\clearpage
\begin{figure}
\plotone{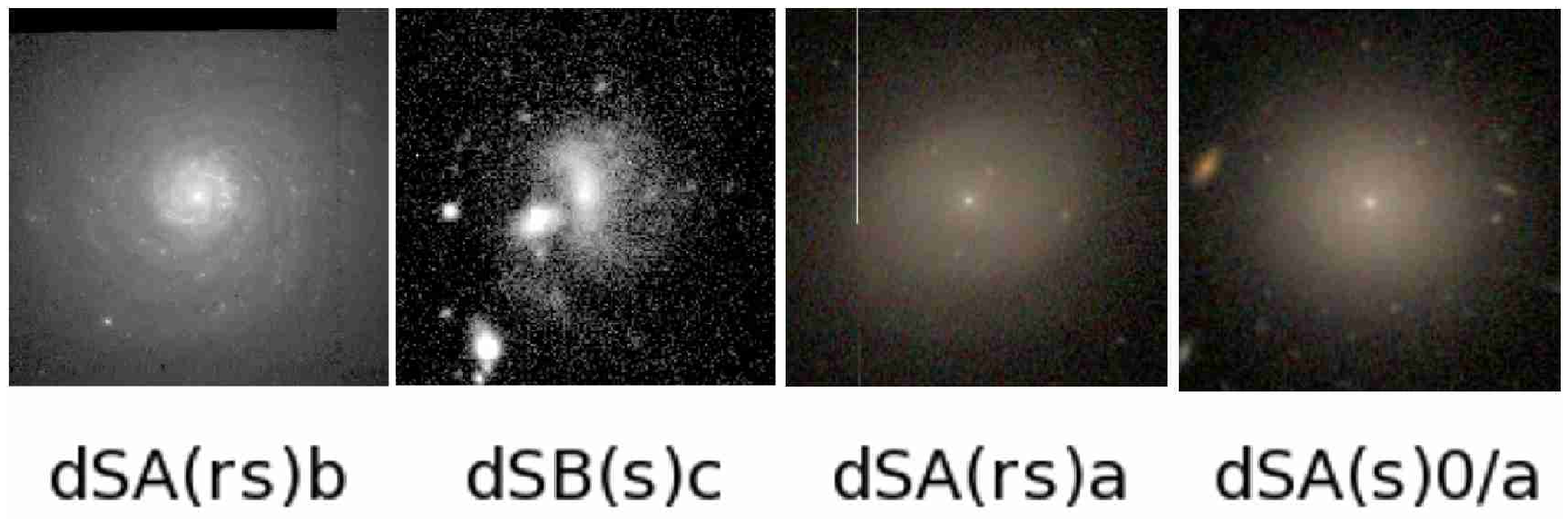}
\caption{Four dwarf spiral galaxies (left to right): NGC 3928, D563$-$4,
IC 783, and IC 3328}
\label{dwarfspirals}
\end{figure}

\clearpage
\begin{figure}
\plotone{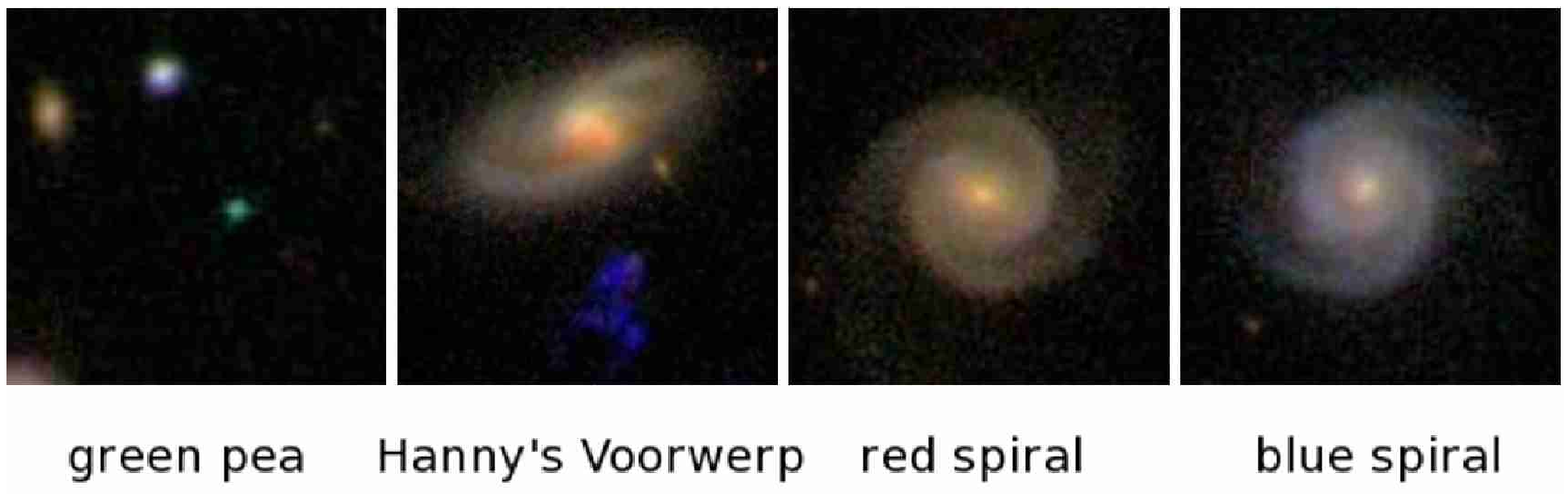}
\caption{Interesting morphologies and color characteristics found
by the Galaxy Zoo project team participants. A ``green pea" is a
star-like galaxy with a high flux in [OIII] 5007. ``Hanny's
Voorwerp" is a cloud of ionized gas 
that may be the light echo of a quasar outburst in the nucleus
of nearby IC 2497. Red spirals are morphologically similar to blue spirals,
but have a lower star formation rate. (All SDSS color images)}
\label{zoofig}
\end{figure}

\clearpage
\begin{figure}
\plotone{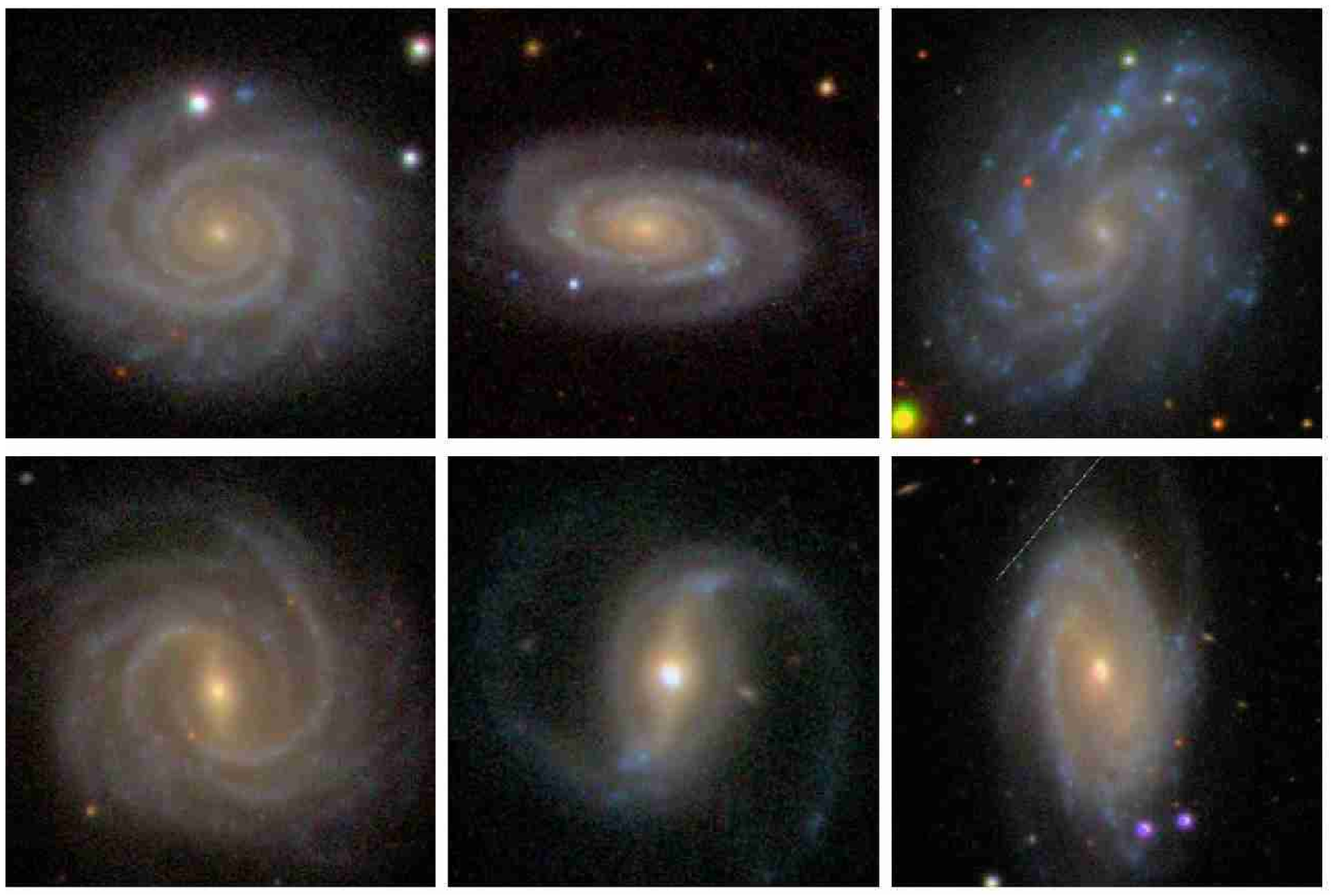}
\caption{SDSS color images of six isolated Sb-Sc galaxies, both barred and nonbarred, from the
AMIGA sample, a refined version of the Catalogue of Isolated Galaxies
(Karachentseva 1973). The galaxies are (left to right): Row 1: NGC 2649, 5622, and 5584; Row 2: NGC 4662, 4719, and 2712.}
\label{isolateds}
\end{figure}

\clearpage
\begin{figure}
\vspace{-1.25truein}
\plotone{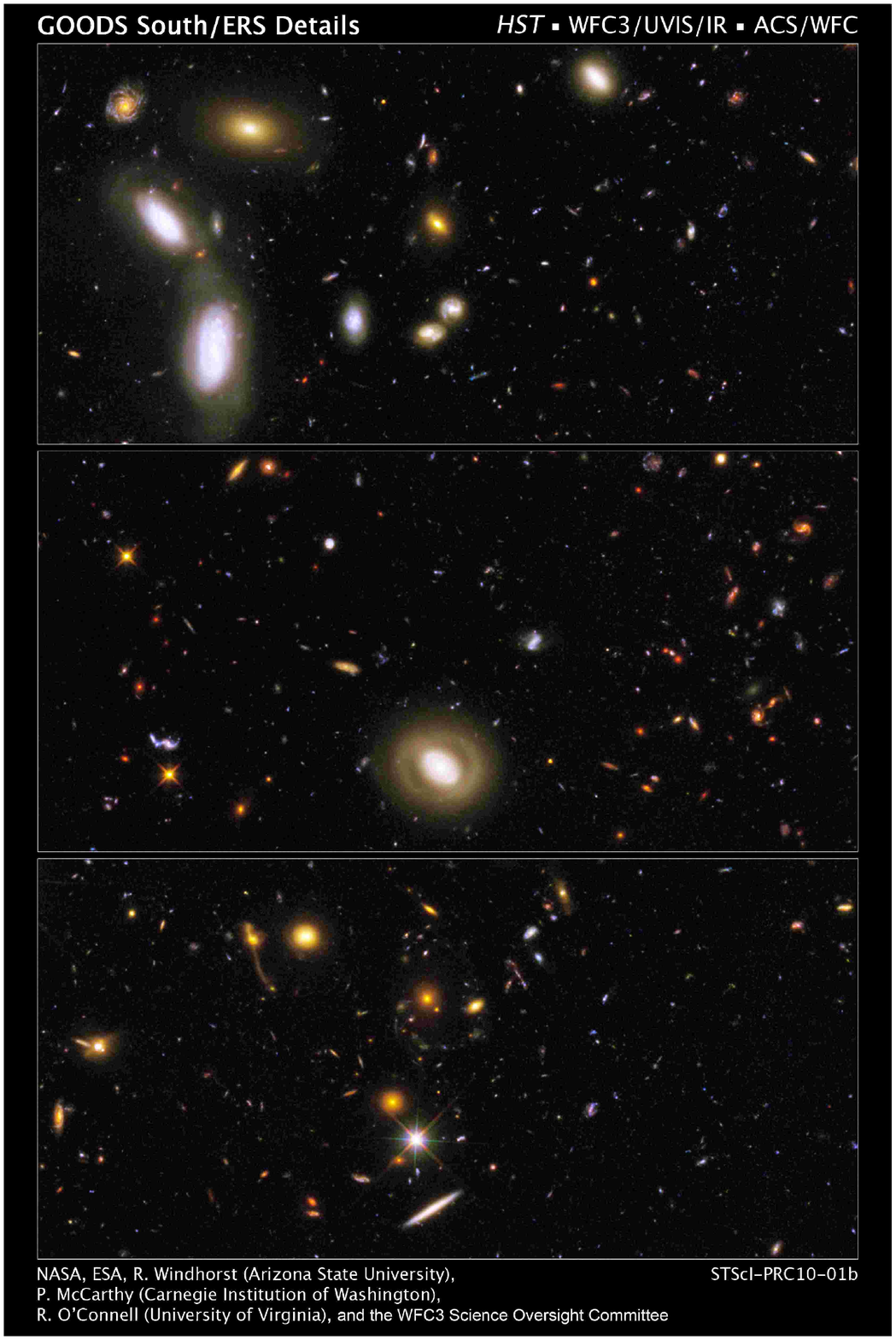}
\caption{}
\label{ers}
\end{figure}
\begin{figure}
\figurenum{54. (cont.)}
\caption{Three subsections from the WFC3-ERS survey of the GOODS-south
field (Windhorst et al. 2010). The colors are based on images obtained
with 10 filters ranging from 0.2$\mu$m to 2$\mu$m.}
\end{figure}

\end{document}